\documentclass[12pt]{article}
\pdfoutput=1
\usepackage{titling}
\usepackage{amsmath}
\usepackage{slashed}
\usepackage{amssymb}
\usepackage{epsfig}
\usepackage{graphicx}
\usepackage{multirow}
\usepackage{color}
\usepackage[normal]{subfigure}
\usepackage{rotating}
\usepackage{hyperref}
\usepackage[margin=0.9in]{geometry}
\usepackage[table]{xcolor}
\usepackage{enumitem}
\usepackage[utf8x]{inputenc}
\usepackage[compress,numbers,sort]{natbib}
\usepackage{authblk}
\usepackage{colortbl}
\usepackage{pdflscape}
\usepackage{color}

\definecolor{nicered}{rgb}{0.7,0.1,0.1}
\definecolor{nicegreen}{rgb}{0.1,0.5,0.1}
\definecolor{red}{rgb}{1.0, 0, 0}
\hypersetup{colorlinks,citecolor= nicegreen,linkcolor= nicered}

\def\eq#1{{Eq.~(\ref{#1})}}
\def\eqs#1#2{{Eqs.~(\ref{#1})--(\ref{#2})}}
\def\fig#1{{Fig.~\ref{#1}}}

\def\Table#1{{Table~\ref{#1}}}
\def\Tables#1#2{{Tables~\ref{#1}--\ref{#2}}}
\def\sect#1{{Sect.~\ref{#1}}}

\def\app#1{{Appendix~\ref{#1}}}

\def\vev#1{\left\langle #1\right\rangle}
\def\abs#1{\left| #1\right|}

\def\Im{\mbox{Im}\,}
\def\Re{\mbox{Re}\,}

\def\etc{\hbox{\it etc}{}}
\renewcommand{\bar}{\overline}

\definecolor{LightCyan}{rgb}{0.88,1,1}
\definecolor{piggypink}{rgb}{0.99, 0.87, 0.9}
\definecolor{applegreen}{rgb}{0.55, 0.71, 0.0}
\definecolor{darkpastelgreen}{rgb}{0.01, 0.75, 0.24}
\definecolor{green-yellow}{rgb}{0.68, 1.0, 0.18}

\newcommand{\beq}{\begin{equation}}
\newcommand{\eeq}{\end{equation}}
\newcommand{\bea}{\begin{eqnarray}}
\newcommand{\eea}{\end{eqnarray}}

\newcommand{\Red}[1]{{\color{nicered}{#1}}}

\newcommand{\published}[1]{%
\gdef\puB{#1}}
\newcommand{\puB}{}

\title{\bf{Accidental matter at the LHC}}

\author[1]{Luca Di Luzio\thanks{luca.di.luzio@ge.infn.it}}
\author[2]{Ramona Gr\"{o}ber\thanks{ramona.groeber@roma3.infn.it}}
\author[3a,b,c]{Jernej F. Kamenik\thanks{jernej.kamenik@cern.ch}}
\author[4]{Marco Nardecchia\thanks{M.nardecchia@damtp.cam.ac.uk}}

\affil[1]{\emph{Dipartimento di Fisica, Universit\`a di Genova and INFN, Sezione di Genova, via Dodecaneso 33, 16159 Genova, Italy}}
\affil[2]{\emph{INFN, Sezione di Roma Tre, 
via della Vasca Navale 84, 00146 Roma, Italy}}
\affil[3a]{\emph{Jo\v{z}ef Stefan Institute, Jamova 39, 1000 Ljubljana, Slovenia}}
\affil[3b]{\emph{Faculty of Mathematics and Physics, University of Ljubljana, Jadranska 19, 1000 Ljubljana, Slovenia}}
\affil[3c]{\emph{CERN TH-PH Division, Meyrin, Switzerland}}
\affil[4]{\emph{DAMTP, University of Cambridge, Wilberforce Road, Cambridge CB3 0WA, United Kingdom}}

\date{}

\published{\flushright RM3-TH/15-3\vskip1cm}

\begin{document}

\maketitle

\begin{abstract}
\normalsize
We classify weak-scale extensions of the Standard Model which automatically 
preserve its accidental and approximate symmetry structure at the renormalizable level 
and which are hence invisible to low-energy indirect probes. By requiring the 
consistency of the effective field theory up to scales of 
$\Lambda_{\text{eff}} \approx 10^{15}$ GeV and after applying cosmological constraints, 
we arrive at a finite set of possibilities that we analyze in detail. 
One of the most striking signatures of this framework is the presence of 
new charged and/or colored states which can be efficiently produced in high-energy particle 
colliders and which are stable on the scale of detectors. 
\end{abstract}

\clearpage

\tableofcontents

\clearpage

\section{Introduction}
\label{intro}

The electroweak (EW) hierarchy puzzle suggests that new physics (NP) degrees of freedom should appear around or not much above the EW scale. Hence the search for NP is a clear target being vigorously pursued by the LHC experiments. On the other hand, indirect searches for NP using flavor and CP violating observables have already probed NP scales up to $10^8$~GeV (c.f.~\cite{Isidori:2010kg,Cirigliano:2013lpa,Kamenik:2014xya} for recent reviews). Thus, the overall excellent agreement with the CKM paradigm predictions suggests a large mass gap above the EW scale. Perhaps even more strikingly,  searches for baryon (B) and lepton (L) number violating processes at low energies suggest that these accidental quantum numbers of the standard model (SM) are good symmetries of nature up to scales of the order of $10^{15}$ GeV (c.f.~\cite{Cirigliano:2013lpa}).

Explicit models of TeV scale NP need to resolve the apparent conflict between these two sets of expectations by postulating exact or approximate symmetries which in term forbid or sufficiently suppress the most dangerous contributions to flavor changing neutral currents, CP violation, as well as B and L changing processes. These include explicit B, L or their anomaly free combinations, discrete space-time symmetries including C, P, CP but also new internal symmetries like R-parity (or R-symmetry) in  supersymmetric (SUSY) extensions of the SM, KK-parity in extra-dimensional setups,  as well as abelian or non-abelian horizontal (flavor) symmetries.

At the heart of  these problems is the fact that generically, extending the SM particle content will either (i) break some of the SM accidental symmetries, and/or (ii) introduce new sources of breaking of the approximate SM symmetries, which in general will not be aligned with existing SM symmetry breaking directions. Examples of the first kind include B and L. Flavor, CP and custodial symmetry of the Higgs potential fall into the second category. 

Consider the SM as the renormalizable part of an effective field theory (EFT)
\begin{equation}
\mathcal L = \mathcal L_{\rm SM}^{(d\leq 4)} + \sum_{d>4} \frac{1}{\Lambda_{\rm eff}^{d-4}} \mathcal L^{(d)} \,,
\end{equation}
where only SM fields appear as dynamical degrees of freedom in $\mathcal L$, and $d$ denotes the canonical operator dimension. Assuming $\mathcal O(1)$ coefficients in the EFT operator expansion, currently all experimental evidence in particle physics can be accommodated by such a generic theory with a very large cut-off $\Lambda_{\rm eff} \approx 10^{15}$~GeV.\footnote{Cosmological observations suggest that most of the mass in the observable Universe cannot be accounted for by known forms of matter. The possibility of particle dark matter within our setup is briefly discussed below.} This particular scale is intriguing since it can account for both the observed neutrino masses suggesting the presence of L violating $\mathcal L^{(5)}$, as well as null results of all flavor, CP and B violation probes constraining $\mathcal L^{(d\geq 6)}$. One may thus ask the following well defined question. {\it Which extensions of the SM particle content with masses close to the EW scale (i) form consistent EFTs with a cut-off scale as high as $10^{15}$~{\rm GeV}, (ii) automatically preserve the accidental and approximate symmetry structure of the SM and thus do not require the introduction of additional protective mechanisms in order to remain viable LHC targets in light of negative search results of the numerous indirect probes, and (iii) are cosmologically viable?}

In the present paper we explore such possibilities by adding to the SM EW (and possibly color) multiplets, requiring that their SM gauge quantum numbers alone forbid all renormalizable interactions which would break any of the SM approximate or accidental symmetries. In particular, in the SM the global $\mathcal G_F \equiv {\rm U(3)}^5$ flavor symmetry of quarks and leptons is only broken by their respective Yukawas (and the gauging of hypercharge). In order not to introduce new sources of $\mathcal G_F$ breaking, one should only consider $\mathcal G_F$ singlet operator extensions of the SM.
In Sect.~\ref{safeext} we list all $d\leq 3$ operators involving quark and lepton fields that transform nontrivially under $\mathcal G_F$ and demand that the new degrees of freedom do not couple to any of these at the renormalizable level.  Since both B and L are subgroups of $\mathcal G_F$ the above prescription also automatically preserves these accidental symmetries of the SM.  Furthermore, in most cases this also ensures the absence of new sources of breaking for both custodial and CP symmetries at the renormalizable level. The exceptions where new breaking can arise in the scalar potential are discussed in \sect{scalpot}.

The relevant Lagrangian is restricted only by imposing the SM gauge and Lorentz invariance, not by new symmetries. Finally, such a theory is assumed to represent a consistent description of nature up to the cut-off scale $\Lambda_{\rm eff}$. In particular, we require that all of the marginal couplings (in particular the SM gauge couplings) remain perturbative up to $\Lambda_{\rm eff}$.  The physical idea behind this requirement is that a Landau pole might be associated 
with the emergence of some new, generic dynamics that will break the accidental 
symmetries of the SM at scales above the Landau pole. As we show in Sect.~\ref{landau}, this condition (together with the cosmological constraints on stable charged particles) 
limits the size of the new representations and leads to a finite list of possible SM extensions.

\setlength{\tabcolsep}{4pt}

\begin{table}[htbp]
\centering
\begin{tabular}{@{} |c|c|c|c|c|c| @{}}
\hline
Spin  & $\chi$ &  $Q_{\textrm{LP}}$ & $\mathcal{O}_{\rm decay}$ & dim$(\mathcal{O}_{\rm decay})$ & $\Lambda_{\rm{Landau}}^{\rm{2-loop}}$[GeV] \\ 
\hline
\hline
0 & $(1,1,0)$ & 0  & $\chi H H^{\dag}$  & 3 &$\gg m_{\rm Pl}$ ($g_1$)  \\ 
0 & $(1,3,0)^\ddag$ & 0,1 & $ \chi H H^{\dag} $ & 3 & $\gg m_{\rm Pl}$ ($g_1$)  \\ 
0 & $(1,4,1/2)^\ddag$ & -1,0,1,2 & $ \chi H H^{\dag} H^{\dag} $ & 4 & $\gg m_{\rm Pl}$ ($g_1$)  \\ 
0 & $(1,4,3/2)^\ddag$ & 0,1,2,3 & $ \chi H^{\dag} H^{\dag} H^{\dag} $ & 4 & $\gg m_{\rm Pl}$ ($g_1$)  \\ 
\hline
\rowcolor{LightCyan}
0 & $(1,2,3/2)$ & 1,2  & $\chi H^{\dag} \ell \ell ,~ \chi^{\dag} H^{\dag} e^c e^c ,~ D^\mu \chi^\dag \ell^\dag \overline{\sigma}_\mu e^c$  & 5 &$\gg m_{\rm Pl}$ ($g_1$)  \\ 
\rowcolor{LightCyan}
0 & $(1,2,5/2)$ & 2,3 & $ \chi^{\dag} H e^c e^c $ & 5 & $\gg m_{\rm Pl}$ ($g_1$)  \\ 
\rowcolor{LightCyan}
0 & $(1,5,0)$ &  0,1,2  & $ \chi H H H^{\dag} H^{\dag} ,~ \chi W^{\mu \nu} W_{\mu \nu} ,~ \chi^3  H^{\dagger} H $ & 5 & $\gg m_{\rm Pl}$ ($g_1$) \\ 
\rowcolor{LightCyan}
0 & $(1,5,1)$ &  -1,0,1,2,3 & $ \chi^{\dag} H H H H^{\dag},~\chi \chi \chi^{\dagger} H^{\dagger} H^{\dagger} $ & 5 & $\gg m_{\rm Pl}$ ($g_1$) \\ 
\rowcolor{LightCyan}
0 & $(1,5,2)$ & 0,1,2,3,4 & $ \chi^{\dag} H H H H$ & 5 &  $3.5 \times 10^{18}$ ($g_1$)  \\ 
\rowcolor{LightCyan}
0 & $(1,7,0)^\star$ & 0,1,2,3 & $\chi^3 H^{\dagger} H$
& 5 & $1.4 \times 10^{16}$ ($g_2$)  \\ 
\hline
\rowcolor{piggypink}
1/2 & $(1,4,1/2)$ &  -1 & $\chi^c \ell H H,~\chi \ell H^{\dag} H ,~ \chi \sigma^{\mu \nu} \ell W_{\mu \nu}$ & 5 & $8.1 \times 10^{18}$ ($g_2$)  \\ 
\rowcolor{piggypink}
1/2 & $(1,4,3/2)$ &  0 & $\chi \ell H^{\dag} H^{\dag}$  & 5 & $2.7 \times 10^{15}$ ($g_1$) \\ 
\rowcolor{piggypink}
1/2 & $(1,5,0)$ &  0 & $\chi \ell H H H^{\dag} ,~ \chi \sigma^{\mu \nu} \ell H W_{\mu \nu}$  & 6 & $8.3 \times 10^{17}$ ($g_2$) \\ 
  \hline
  \end{tabular}
  \caption{\label{summary1} 
  List of new weak-scale uncolored states $\chi$ which can couple to SM fields 
  at the renormalizable level
  without breaking $\mathcal{G_F}$, 
  and which are compatible with cosmology and  
  an EFT cut-off scale of $\Lambda_{\rm{eff}} \simeq 10^{15}$ GeV. 
  The possible electromagnetic charges of the LP in the multiplet are denoted by $Q_{\textrm{LP}}$, while 
  $\mathcal{O}_{\rm decay}$ denotes the lowest dimensional operators responsible for the decay of $\chi$. 
  States with $Y=0$ are understood to be real.
  In the last column, the Landau pole has been 
  estimated at two loops by integrating in the new multiplet at the scale of the $Z$ boson mass $m_Z$, 
  while the symbol in the bracket stands for the gauge coupling, $g_{1,2,3}$, 
  triggering the Landau pole 
  and $m_{\rm Pl} = 1.22 \times 10^{19}$~GeV is the Planck mass. The states marked with $\ddag$ and $\star$ are constrained by EW precision tests and BBN, respectively, to  lie possibly beyond the LHC reach.}
\end{table}

\setlength{\tabcolsep}{2pt}

\begin{table}[htbp]
\centering
\begin{tabular}{@{} |c|c|c|c|c|c| @{}}
\hline
Spin  & $\chi$ &  $Q_{\textrm{LP}}$ & $\mathcal{O}_{\rm decay}$ & dim$(\mathcal{O}_{\rm decay})$ & $\Lambda_{\rm{Landau}}^{\rm{2-loop}}$[GeV]  \\ 
\hline 
\hline
\rowcolor{LightCyan}
0 & $(3,1,5/3)$ & ${5}/{3}$ & 
$
\begin{array} {lcl}
& \chi^{\dag} H q e^c ,~ \chi H^{\dag} u^c \ell, & \\ 
&  D^\mu \chi^\dag u^{c\dag} \overline{\sigma}_\mu e^c & 
\end{array}
$ & 5 &$\gg m_{\rm Pl}$ ($g_1$)  \\ 
\hline
\rowcolor{LightCyan}
0 & $(\overline{3},2,5/6)$ & ${1}/{3}, {4}/{3}$ 
&  
$\begin{array} {lcl} 
&\chi^{\dag} H qq ,~ \chi^{\dag} H u^c e^c ,~  \chi H^{\dag} q \ell, & \\ 
& \chi H^{\dag} u^c d^c ,~ \chi H u^c u^c , & \\ 
&  \chi^{\dag} H^{\dag} d^c e^c ,~ D^\mu \chi q^\dag \overline{\sigma}_\mu u^c, & \\ 
&  D^\mu \chi^\dag q^\dag \overline{\sigma}_\mu e^c 
,~ D^\mu \chi d^{c\dag} \overline{\sigma}_\mu \ell &  
\end{array}$
& 5 &$\gg m_{\rm Pl}$ ($g_1$)  \\ 
\hline
\rowcolor{LightCyan}
0 & $(\overline{3},2,11/6)$ & ${4}/{3},{7}/{3}$ & $\chi H^{\dag} u^c u^c ,~ \chi^{\dag} H d^c e^c$ & 5 &$5.5 \times 10^{19}$ ($g_1$) \\ 
\hline
\rowcolor{LightCyan}
0 & $(3,3,2/3)$ & -$ {1}/{3}, {2}/{3}, {5}/{3} $ & 
$
\begin{array} {lcl}
& \chi^{\dag} H^{\dag} q e^c ,~ \chi H u^c \ell, & \\
& \chi H^{\dag} d^c \ell ,~ D^\mu \chi q^\dag \overline{\sigma}_\mu \ell &
\end{array}
$ 
& 5 &$\gg m_{\rm Pl}$ ($g_1$)  \\
\hline
\rowcolor{LightCyan}
0 & $(3,3,5/3)$ & $ {2}/{3}, {5}/{3}, {8}/{3}$ & $\chi^{\dag} H q e^c ,~ \chi H^{\dag} u^c \ell$ & 5 &$3.2 \times 10^{17}$ ($g_1$) \\  
\hline
\rowcolor{LightCyan}
0 & $(3,4,1/6)$ & $ \begin{array} {lcl} & -{4}/{3},-{1}/{3}, & \\  & {2}/{3}, {5}/{3} & \end{array}$   & $\chi H^{\dag} qq ,~ \chi^{\dag} H q \ell$ & 5 &$\gg m_{\rm Pl}$ ($g_2$) \\
\hline
\rowcolor{LightCyan}
0 & $(\overline{3},4,5/6)$ & $ \begin{array} {lcl} &  -{2}/{3}, {1}/{3}, & \\ & {4}/{3}, {7}/{3} & \end{array}$  & $\chi^{\dag} H qq,~ \chi H^{\dag} q \ell$  & 5 & $\gg m_{\rm Pl}$ ($g_2$) \\
\hline
\rowcolor{LightCyan}
0 & $(\overline{6},2,1/6)$ &  -${1}/{3}, {2}/{3}$ 
& $
\begin{array} {lcl}
& \chi H^{\dag} qq ,~\chi^{\dag} H u^c d^c , & \\
& \chi^{\dag} H^{\dag} d^c d^c ,~ D^\mu \chi^\dag q^\dag \overline{\sigma}_\mu d^c &
\end{array}
$ & 5 & $\gg m_{\rm Pl}$ ($g_1$) \\  
\hline
\rowcolor{LightCyan}
0 & $(6,2,5/6)$ & $ {1}/{3}, {4}/{3}$ & 
$
\begin{array} {lcl}
& \chi^{\dag} H q q ,~\chi H u^c u^c , & \\
& \chi H^{\dag} u^c d^c ,~ D^\mu \chi q^\dag \overline{\sigma}_\mu u^c &
\end{array}
$ & 5 & $\gg m_{\rm Pl}$ ($g_1$)  \\    
\hline
\rowcolor{LightCyan}
0 & $(\overline{6},2,7/6)$ &  $ {2}/{3}, {5}/{3}$  & $\chi^{\dag} H d^c d^c$ & 5 & $\gg m_{\rm Pl}$ ($g_1$) \\  
\hline
\rowcolor{LightCyan}
0 & $(8,1,0)$ &  0 
& $\begin{array} {lcl} 
& \chi H q u^c ,~\chi H^{\dag} q d^c, & \\
& D^\mu \chi D^\nu G_{\mu\nu} ,~ D^\mu \chi q^\dag \overline{\sigma}_\mu q, & \\
& D^\mu \chi u^{c\dag} \overline{\sigma}_\mu u^c 
,~ D^\mu \chi d^{c\dag} \overline{\sigma}_\mu d^c, & \\ 
& \chi G^{\mu \nu} G_{\mu \nu} ,~ \chi G^{\mu \nu} B_{\mu \nu}, & \\
& \chi \chi \chi H^{\dagger} H &
\end{array}$ 
& 5 & $\gg m_{\rm Pl}$ ($g_1$)  \\
\hline
\rowcolor{LightCyan}
0 & $(8,1,1)$ &  1 & 
$\begin{array} {lcl} 
& \chi H^{\dag} q u^c ,~\chi^{\dag} H q d^c , & \\
& D^\mu \chi^\dag u^{c\dag} \overline{\sigma}_\mu d^c 
,~ \chi \chi \chi^{\dagger} H^{\dagger} H^{\dagger} \\
\end{array}$ 
& 5 & $\gg m_{\rm Pl}$ ($g_1$) \\ 
\hline
\rowcolor{LightCyan}
0 & $(8,3,0)$ &  0,1 & 
$
\begin{array} {lcl} 
& \chi H q u^c ,~\chi H^{\dag} q d^c , & \\
& \chi G^{\mu \nu} W_{\mu \nu}  ,~ D^\mu \chi q^\dag \overline{\sigma}_\mu q, & \\
& \chi \chi \chi H^{\dagger} H &
\end{array}
$ 
& 5 & $\gg m_{\rm Pl}$ ($g_1$)  \\ 
\hline
\rowcolor{LightCyan}
0 & $(8,3,1)$ & 0,1,2 & $\chi H^{\dag} q u^c ,~\chi^{\dag} H q d^c ,~ \chi \chi \chi^{\dagger} H^{\dagger} H^{\dagger}  $ & 5 & $1.0 \times 10^{17}$ ($g_1$)  \\ 
\hline
\rowcolor{piggypink}
1/2 & $(6,1,1/3)$ & ${1}/{3}$ &$\chi^c \sigma^{\mu \nu} d^c G_{\mu \nu} $ & 5 & $\gg m_{\rm Pl}$ ($g_1$)  \\
%\hline
\rowcolor{piggypink}
1/2 & $(\overline{6},1,2/3)$ & $ {2}/{3}$ & $\chi \sigma^{\mu \nu} u^c G_{\mu \nu} $ & 5 & $\gg m_{\rm Pl}$ ($g_1$)  \\
%\hline
\rowcolor{piggypink}
1/2 & $(8,1,1)$ & 1 &  $\chi^c \sigma^{\mu \nu} e^c G_{\mu \nu} $  & 5 & $4.0 \times 10^{16}$ ($g_1$)  \\ 
  \hline
  \end{tabular}
  \caption{\label{summary2} Same as in \Table{summary1} but for colored states.
    }
\end{table}

\setlength{\tabcolsep}{5pt}

Interestingly, it turns out that such SM extensions generically possess extended accidental symmetries which ensure the stability of the lightest particles (LPs) in the new multiplets at the renormalizable level. If these are charge- and color-neutral, they can form viable dark matter candidates, a possibility, which has been throughly investigated in the literature~\cite{Cirelli:2005uq,Cirelli:2007xd,Cirelli:2009uv,Cirelli:2014dsa}. On the other hand, scenarios where the lightest component of the new multiplet is charged and/or colored are in general constrained by cosmological observations as well as by searches for exotic forms of matter on  Earth and in the Universe. Taking also these constraints into account, the final list of viable uncolored and colored weak representations are given in Tables~\ref{summary1} and~\ref{summary2}, respectively, which summarize the main results of our investigation. 

The details of the above sketched construction and analysis are contained in the rest of the paper which is organized as follows. In Sect.~\ref{safeext} we exploit accidental symmetries beyond the SM to construct SM extensions with new degrees of freedom at the weak scale, which are completely transparent to indirect low-energy probes. 
The way the set of all possible extra states is made finite is discussed in this section as well. 
In Sect.~\ref{sec:decays} we estimate the new particles' lifetimes. In turn in Sect.~\ref{cosmo} we consider bounds on possibly long lived states coming from early Universe cosmology. In particular, the effects on big bang nucleosynthesis (BBN) turn out to be the most important ones. Sect.~\ref{lhc} explores the collider phenomenology of the viable weak-scale SM extensions and estimates the current lower bounds on the new particles' masses coming from existing LHC searches. We conclude in Sect.~\ref{conclusions} while a more detailed technical discussion of the renormalization group (RG) evolution of the gauge couplings and the SU$(2)_L$ decomposition of the effective operators are relegated to the Appendices.

\section{Accidentally safe extensions of the SM}
\label{safeext}

Our starting point is the classification of SM extensions which automatically preserve 
the accidental  and approximate symmetry structure of the SM without 
imposing additional protective mechanisms (only SM gauge and Lorentz symmetries are required). 
For simplicity, we will limit our discussion to the 
case where a single extra representation $\chi$ is added to the SM field content.
While simultaneously adding more than one representation from our set is in principle possible, two additional restrictions need to be considered in that case: (i) adding more matter representations will in general lower the scale of the EFT validity (cf.~\sect{landau}), (ii) additional SM gauge invariants may be constructed, potentially breaking $\mathcal G_F$ and/or the new accidental symmetry associated with $\chi$ stability at the renormalizable level.

We start by listing all the $d \leq 3$ operators made of SM fields. 
If $\chi$ is a fermion, we require that the new state does not couple to SM fermions 
at the renormalizable level. 
In this way, $\mathcal G_F$ is automatically preserved and an extra accidental symmetry guarantees the stability of the new particle 
at the renormalizable level. 
On the other hand, the case of extra scalars is more involved since they can always couple to the Higgs field at the renormalizable level 
without breaking $\mathcal G_F$ and their stability depends on the allowed interactions with the Higgs field.

\begin{table}[htbp]
  \centering
  \begin{tabular}{@{} |c|c|c|c|c| @{}}
  \hline
    Spin & SM field & SU$(3)_c$ & SU$(2)_L$ & U$(1)_Y$ \\ 
% \hline
 \hline
    0 & $H$ &  1  & 2 & $+1/2$ \\
    1/2 & $q$ &  3  & 2 & $+1/6$ \\ 
    1/2 & $u^c$ &  $\overline{3}$  & 1 & $-2/3$ \\ 
    1/2 & $d^c$ &  $\overline{3}$  & 1 & $+1/3$ \\ 
    1/2 & $\ell$ &  1  & 2 & $-1/2$ \\ 
    1/2 & $e^c$ &  1  & 1 & $+1$ \\
    \hline
  \end{tabular}
  \caption{\label{notationtable} SM field content and quantum numbers.} 
\end{table}
\begin{table}[htbp]
  \centering
  \begin{tabular}{@{} |c|c|c|c|c| @{}}
  \hline
   & $\mathcal{O}_{\rm{SM}}$ & SU$(3)_c$ & SU$(2)_L$ & U$(1)_Y$ \\ 
 \hline
% \hline
   & $q H (H^\dag)$ &  $3$ & $1 \oplus 3$ & $+2/3 (-1/3)$ \\ 
   & $u^c H (H^\dag)$ &  $\overline{3}$ & $2$ & $-1/6 (-7/6)$ \\ 
   $\psi_{\text{SM}} H (H^\dag)$ & $d^c H (H^\dag)$ &  $\overline{3}$ & $2$ & $+5/6 (-1/6)$ \\ 
   & $\ell H (H^\dag)$ &  $1$ & $1 \oplus 3$ & $0 (-1)$ \\ 
   & $e^c H (H^\dag)$ &  $1$ & $2$ & $+3/2 (+1/2)$ \\ 
    \hline   
    & $q q$ &  $\overline{3} \oplus 6$ & $1 \oplus 3$ & $+1/3$ \\ 
    & $q u^c$ & $1 \oplus 8$ &  $2$ & $-1/2$ \\
    & $q d^c$ & $1 \oplus 8$ &  $2$ & $+1/2$ \\
    & $q\ell$ &  $3$ & $1 \oplus 3$ & $-1/3$  \\
    & $q e^c$ & $3$  & $2$  & $+7/6$   \\
    & $u^c u^c$ & $3 \oplus \overline{6}$ & $1$ & $-4/3$ \\
    & $u^c d^c$ & $3 \oplus \overline{6}$ & $1$ & $-1/3$ \\
    $\psi_{\text{SM}} \psi_{\text{SM}}$ & $u^c \ell$ & $\overline{3}$ & $2$ & $-7/6$ \\
    & $u^c e^c$ & $\overline{3}$ & $1$ & $+1/3$ \\
    & $d^c d^c$ & $3 \oplus \overline{6}$ & $1$ & $+2/3$ \\
    & $d^c \ell$ & $\overline{3}$ & $2$ & $-1/6$ \\
    & $d^c e^c$ & $\overline{3}$ & $1$ & $+4/3$ \\
    & $\ell \ell$ & $1$ & $1 \oplus 3$ &  $-1$ \\
    & $\ell e^c$ & $1$ & $2$ & $+1/2$ \\
    & $e^c e^c$ & $1$ & $1$ & $+2$ \\
 \hline
    & $H H$ &  $1$ & $3$ & $+1$ \\ 
    $H$ & $H H^\dag$ &  $1$ & $1 \oplus 3$ & $0$ \\
    combinations  & $H H H$ &  $1$ & $4$ &  $+3/2$ \\ 
    & $H H H^\dag$ &  $1$ & $2 \oplus 4$ &  $+1/2$ \\ 
    \hline
  \end{tabular}
  \caption{\label{SMdm3dec} List of all possible $d \leq 3$ operators made of SM fields. 
  Operators of the type $\psi_{\text{SM}}^\dag \psi_{\text{SM}}$ are not displayed since they couple to Lorentz vectors, 
  which are not considered in our analysis.}
\end{table}

A brief comment regarding larger Lorentz group representations is in order at this point.
The presence of extra Lorentz vectors (i.e.~spin 1 bosons) requires either the extension of the 
SM gauge group or new strong dynamics. In the former case, accidental preservation of $\mathcal G_F$ requires the extended gauge symmetry to be a direct product of the SM gauge group and possible new factors under which the SM fermions need to transform trivially.  Such setups have been thoroughly studied in the literature (c.f.~\cite{Jaeckel:2013ija} for a recent review) and we have nothing to add.
On the other hand, 
new vectors due to some strong dynamics at the TeV scale are incompatible with a large mass gap $\Lambda_{\rm{eff}} \gg$~TeV. The same argument applies to composite particles of higher spins.
Finally, extra fundamental particles with spins $3/2$ and $2$ can appear in theories of extended and gauged space-time symmetry (c.f.~\cite{VanNieuwenhuizen:1981ae,Nilles:1983ge}), but such constructions necessarily go beyond our EFT framework. 
We will hence limit our discussion to the inclusion of either spin $0$ or $1/2$ extra representations.

In the following, we adopt a two-component notation where all the fermion 
fields are Weyl spinors belonging to the same irreducible representation of the Lorentz group.
The accidental matter multiplets are collectively denoted by $\chi$. We use the subscripts $S$ and $F$ to denote the bosonic (spin 0) and fermionic (spin 1/2) SM gauge representations, respectively, where appropriate to avoid ambiguity. 
 The SM fermions are collectively denoted by $\psi_{\text{SM}}$ and their quantum numbers 
are fixed according to \Table{notationtable}. The list of all possible $d \leq 3$ operators made of SM fields is provided in \Table{SMdm3dec}.

\subsection{New fermions}
\label{newfermions}

If a fermionic $\chi$ transforms under a complex or pseudoreal representation of the gauge group (so that a Majorana mass term is forbidden), we introduce another field $\chi^c$ with conjugate quantum numbers. In this way, the new state is vector-like and a mass term can always be added.

According to our previous discussion, we want to forbid the interactions 
$\chi \psi_{\text{SM}}$, $\chi \psi_{\text{SM}} H$ and 
$\chi \psi_{\text{SM}} H^\dag$.\footnote{Notice that terms of the form $\chi \chi H$ or $\chi \chi H^\dag$ are forbidden by SU$(2)_L$ invariance.}
By inspecting \Table{SMdm3dec} we conclude that $\chi$ cannot have the following quantum numbers:
\begin{equation}
\label{chineqto}
\chi \neq \psi_{\text{SM}}, 
(1,1,0), (1,3,0), (1,3,1), (1,2,3/2), 
 (\overline{3},2,5/6), (3,2,7/6),
(\overline{3},3,1/3), (3,3,2/3) \, .   
\end{equation}
If $\chi$ transforms under a real representations of the SM group, then we can also add a Majorana mass term 
and the most general Langragian reads (see e.g.~\cite{Willenbrock:2004hu} for two-component notation)
\begin{equation}
\mathcal{L}= \mathcal{L}_{\text{SM}} + i \chi^{\dagger} \overline{\sigma}^{\mu} D_{\mu} \chi + \frac{1}{2} M (\chi^T \epsilon \chi + \text{h.c.}) \, ,
\end{equation}
which is invariant under a $Z_2$ transformation $\chi \to -\chi$. 
On the other hand, if $\chi$ transforms under a complex or pseudoreal representations of the SM group, we introduce an extra Weyl fermion $\chi^c$ 
with conjugate gauge quantum numbers with respect to $\chi$, so that a Dirac mass term is allowed, and get
\begin{equation}
\mathcal{L}=\mathcal{L}_{\text{SM}} + i \chi^{\dagger} \overline{\sigma}^{\mu} D_{\mu} \chi +  i \chi^{c \dagger} \overline{\sigma}^{\mu} D_{\mu} \chi^c 
+ M (\chi^T \epsilon \chi^c + \text{h.c.}) \, ,
\end{equation}
which is invariant under a U(1) transformation $\chi \to e^{i \theta} \chi$ and $\chi^c \to e^{- i \theta} \chi^c$.
In both cases an accidental symmetry implies stability of  the new particles 
at the renormalizable level and also requires that they are pair produced in high-energy particle colliders.

\subsection{New scalars}
\label{newscalars}

For scalar $\chi$, in order to preserve $\mathcal G_F$ we have to avoid all couplings of the form $\chi \psi_{\text{SM}} \psi_{\text{SM}}$. 
By inspecting \Table{SMdm3dec} we conclude that $\chi$ cannot have the following quantum numbers:
\begin{align}
\label{Xneqto}
\chi \neq \ & 
(1,1,1), (1,3,1), (1,1,2), 
(1,2,1/2),
(\overline{3},1,1/3), 
(3,1,2/3), 
(\overline{3},1,4/3),
(3,2,1/6),
(3,2,7/6), \nonumber \\
& (\overline{3},3,1/3), 
(6,1,1/3),  
(\overline{6},1,2/3),
(6,1,4/3),  
(6,3,1/3), 
(8,2,1/2)
\, .
\end{align}

\begin{table}[thbp]
\centering
\begin{tabular}{@{} |c|c|c|c|c| @{}}
\hline
Spin  & $\chi$ &  $\mathcal{O}_{\rm decay}$ & dim$(\mathcal{O}_{\rm decay})$ & Stability \\ 
\hline
%\hline
0 & $(1,1,0)$ &  $\chi H H^{\dag}$  & 3 & $\times$   \\  
0 & $(1,3,0)$ &  $\chi H H^{\dag}$  & 3 & $\times$  \\  
0 & $(1,4,1/2)$ &  $\chi H H^{\dag} H^{\dag}$  & 4 & $\times$  \\  
0 & $(1,4,3/2)$ &  $\chi H^{\dag} H^{\dag} H^{\dag}$  & 4 & $\times$  \\  
0 & $(R,2k,1/2)$ &  $\chi \chi H^{\dag} H^{\dag}$  & 4 & $Z_2$  \\  
0 & $(R,n,0)$ &  $\chi \chi H H^{\dag}$  & 4 & $Z_2$  \\  
0 & $(C,n,Y)$ &  $\chi \chi^{\dag} H H^{\dag}$  & 4 & U(1)  \\  
0 & $(C,2k,1/6)$ &  $\chi \chi \chi H^{\dag}$  & 4 & $Z_3$  \\  
0 & $(R,2k,1/2)$ &  $\chi \chi \chi^{\dag} H^{\dag}$  & 4 & $\times$  \\  
\hline
  \end{tabular}
  \caption{\label{renchiH} 
  Extra scalar representations which can couple to the Higgs at the 
  renormalizable level without breaking $\mathcal G_F$. 
  $(C,n,Y)$ denote generic quantum numbers under the SM gauge group 
  which are not already contained in the list of \eq{Xneqto}. 
  $R$ stands for a real SU$(3)_c$ representation (i.e.~$R = 1, 8, 27, \ldots$) 
  and $2 k$ for an even SU$(2)_L$ representation. 
  In the last column, we provide (when appropriate)
  the symmetry responsible for the stability of $\chi$. 
  The cases denoted by a ``$\times$'' lead instead to the decay of $\chi$ at the renormalizable level.}
\end{table}

Analogously to the case of extra fermions in \sect{newfermions}, 
gauge interactions alone cannot lead to the decay of $\chi$ at the renormalizable level, 
since the kinetic terms again exhibit a $Z_2$ or a U(1) 
invariance for the case of an extra real or complex scalar, respectively. 
The decay of the new particle is 
however possible (depending on the quantum numbers of $\chi$) 
due to the presence of extra renormalizable interactions between $\chi$ and $H$, 
which are listed in \Table{renchiH}.

\subsubsection{Scalar potential, CP and custodial symmetry}
\label{scalpot}

In the presence of any new scalar multiplet $\chi$ the scalar potential can be written as (see e.g.~\cite{AbdusSalam:2013eya})
\begin{multline}
\label{potentialXH}
V(H,\chi) = V_{\rm{SM}} + \eta \left( 
m_\chi^2 \abs{\chi}^2 + \alpha \abs{\chi}^2 \abs{H}^2
+ \beta (\chi^{\dagger} T^a_\chi \chi) (H^{\dagger} T^a_H H) \right) \\ 
+ \left[ \gamma (\chi^{\dagger} C_\chi T^a_\chi \chi^*) (H^{T} C_H T^a_H H) + \rm{h.c.} \right]
+ \ldots \, , 
\end{multline}
where $\eta$ is equal to $1(1/2)$ for a complex (real) representation, $T^a_R$ and $C_R$ 
denote respectively the SU$(2)_L$ generators and conjugation matrices in the representation $R$ 
(so, for instance, $T_H^a=\sigma^a/2$ and $C_H=i \sigma^2$ where $\sigma^a$ for $a=1,2,3$ 
are the Pauli matrices). We take $\vev{H}^T = (0, v/\sqrt{2})$ with $v = 246 \ \rm{GeV}$. The ellipses in \eq{potentialXH} 
stand for extra terms, 
like e.g.~$(\chi^\dag T^a_\chi \chi)^2$, which do not sizeably affect the mass splitting of $\chi$ (see below). In addition, $\chi$ in specific weak representations might allow for additional renormalizable operators listed in \Table{renchiH}. 

\begin{figure}[ht]
\centering
\includegraphics[angle=0,width=7cm]{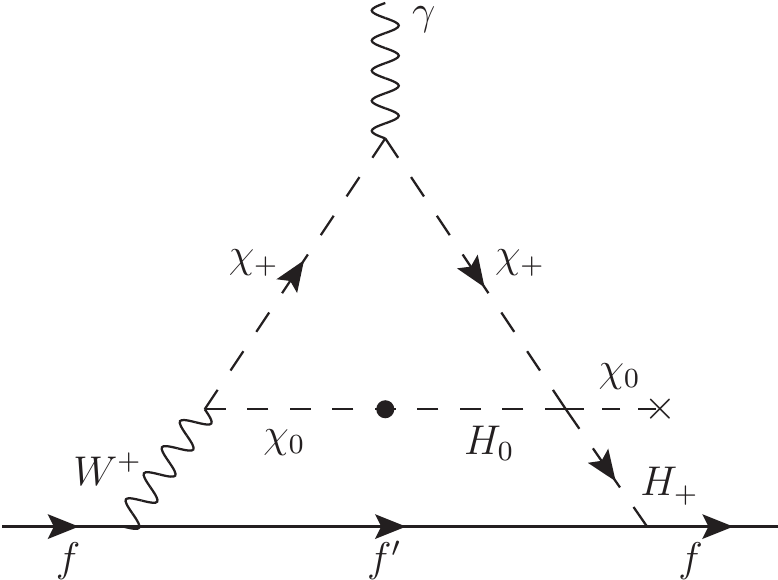}
\caption{\label{fig:EDM} 
Scalar loop contribution to the electron EDM.}
\end{figure}

The first accidental symmetry of the scalar potential that we wish to discuss is CP. 
Generic sources of CP violation are severely constrained by the measurement of 
electric dipole moments (EDMs) \cite{Pospelov:2005pr}. 
Among the accidental scalar matter extensions 
of \Tables{summary1}{summary2}, it turns out that only $(1,4,1/2)_S$ explicitly violates CP. 
This can be seen by noticing that for such a multiplet one can construct 
three non-hermitian invariants in the scalar potential (cf.~the third, fifth and last row in \Table{renchiH}) 
and that only one out of the three phases associated with the corresponding complex couplings 
can be rotated away by a re-phasing of $\chi$ and $H$. 
In this case the most significant experimental constraint comes from the searches for an electron EDM ($d_e$), defined through the effective operator $\mathcal L \ni -i (d_e/2)  \bar e (\sigma \cdot F) \gamma_5 e$\,. The $(1,4,1/2)_S$ contributes at two loops through the diagram in \fig{fig:EDM}, 
which corresponds to the diagram in Fig.~12 of Ref.~\cite{Inoue:2014nva} 
after replacing $H_1 \rightarrow H$ and $H_2 \rightarrow \chi$. 

Taking into account the extra $\vev{\chi} / v \lesssim 1 \%$ suppression due to EW precision constraints (see below), 
 assuming $\mathcal{O}(1)$ scalar couplings and mixing angles, and $m_\chi \sim v$ 
for the sake of a very conservative estimate (see also \cite{Inoue:2014nva,Abe:2013qla}), we obtain 
$|d_e| \lesssim 7 \times 10^{-29} e$\,cm. This has to be compared with the recent experimental bound from the ACME collaboration \cite{Baron:2013eja}  of  $|d_e^{\rm exp}| <  8.7 \times 10^{-29} e$\,cm at 90\% C.L.. While not constraining at the moment, interestingly,  future experimental improvements 
on the electron EDM  might start to probe CP violation in generic weak-scale scalar extensions 
of the SM involving the $(1,4,1/2)_S$ multiplet.

Another accidental symmetry of the SM scalar potential is the so-called custodial symmetry.
In the $g' \rightarrow 0$ limit the massive gauge bosons transform as a triplet of an unbroken 
global SU$(2)_C$, which is also responsible for the tree-level relation $\rho_{\rm{tree}} \equiv m_W^2 / m_Z^2 \cos^2\theta_W =1$. 
New sources of SU$(2)_C$ breaking which cannot be accounted in the SM 
are described by the $\rho_0 \equiv \rho / \rho_{\rm{SM}}$ parameter \cite{Beringer:1900zz}. 
Experimentally, $\rho_0^{\rm{exp}} = 1.0004 \substack{+0.0003 \\ -0.0004}$ \cite{Beringer:1900zz}, which is compatible 
with the SM prediction $\rho_0 = 1$. Thus the experimental value of $\rho_0$ can be used to constrain 
new sources of SU$(2)_C$ breaking due to the extra scalar $\chi$. 

If $\chi$ gets a vacuum expectation value (VEV), there is a tree-level contribution \cite{Gunion:1989we}
\begin{equation}
\rho^{\rm{tree}}_0 - 1 =  \left\{ \eta \left[ j (j+1)  -  Y^2 \right] - 2 Y^2 \right\}\left[4 \frac{ \vev{\chi}^2}{v^2} 
+\mathcal{O} \left( \frac{ \vev{\chi}^4}{v^4} \right)\right] \, ,
\end{equation}
where $j$ is the total weak-isospin quantum number of $\chi$ and $Y$ its 
hypercharge in the $Q = T^3 + Y$ normalization.  Apart for the safe representations yielding $\rho_0 = 1$ for any value of $\vev{\chi}$: 
$(1,1,0)$, $(1,2,1/2)$, $(1,7,2)$ \cite{Hisano:2013sn}, $(1,26,15/2)$, \etc., 
the $2\sigma$-level saturated bound is at the level of $\vev{\chi} / v \lesssim 1\%$. 

In general, whether a scalar field can develop a VEV depends on the choice of the 
parameters in the scalar potential.  
However, ``tadpole'' couplings of $\chi$ to some $H$'s always imply 
an induced VEV for $\chi$. 
From \Table{renchiH} we see that this is indeed the case for the states: 
$(1,1,0)$, $(1,3,0)$, $(1,4,1/2)$, $(1,4,3/2)$. 
While the VEV of the former does not contribute to $\rho_0$, the remaining ones 
can be in principle dangerous. 
By looking at the generic shape of the potential and its stationary equations, 
we estimate on dimensional grounds (for $\mathcal{O}(1)$ couplings and barring fine-tunings),
$\vev{\chi} \sim v^2/m_\chi$ (triplet case) and 
$\vev{\chi} \sim v^3/m_\chi^2$ (quadruplet cases). 
Hence, $\vev{\chi} / v \lesssim 1\%$ corresponds to 
$m_\chi \gtrsim 100 \ v \approx 20 \ \text{TeV}$ (triplet case) and 
$m_\chi \gtrsim 10 \ v \approx 2 \ \text{TeV}$ (quadruplet cases), 
which limits the visibility of these states at the LHC, unless 
a moderate fine-tuning is allowed in the scalar potential. 

Custodial symmetry also helps us to understand the properties of the theory beyond the tree level. 
Indeed, a tree-level splitting within the components of $\chi$ originating from the scalar potential in \eq{potentialXH} gives a radiative contribution 
to $\rho_0$.  In the following we assume $\vev{\chi} \ll v$ to suppress the tree-level contribution to $\rho_0$. 
Consequently $\vev{\chi}$ itself cannot sizably contribute to the mass splitting. 
Notice, also, that among the scalar states selected in \Table{summary1}, 
the coupling $\gamma$ is relevant 
only for $(1,4,1/2)$. 
However, since this state decays through a renormalizable operator, 
the details of its mass spectrum are not of particular interest.\footnote{
The coupling $\gamma$ induces mixing between 
the conjugate components of $\chi$ with the same $|Q| \neq 0$ and, 
for the $Q=0$ component, it splits its  
real and imaginary part. 
The contribution of $\gamma$ to $\rho^{\rm{1-loop}}_0$ 
has been considered for instance in \cite{AbdusSalam:2013eya}.}
We are hence left with the contribution of $\beta$ to the mass splitting, which yields
\begin{equation}
\label{m2treeI}
m^2_I = m_\chi^2 + \frac{1}{2} \alpha v^2 - \frac{1}{4} \beta v^2 I \equiv M^2 - \delta^2 I \, ,  
\end{equation}
where $-j \leq I \leq j$ denotes the $T^3$ eigenvalue of the $(2j+1)$-dimensional representation $\chi$ 
and we defined the parameter $M^2 \equiv m_\chi^2 + \frac{1}{2} \alpha v^2$ 
and $\delta^2 \equiv \frac{1}{4} \beta v^2$. 
Using the general formula for the one-loop correction 
in \cite{Lavoura:1993nq} and expanding the loop function for $\delta < M$ 
we find 
\begin{equation}
\label{deltarho1lsimply}
\rho^{\rm{1-loop}}_0 - 1= \frac{\eta N_C \alpha_{\rm{em}}}{16 \pi \sin^2\theta_W m_W^2} 
\left[ \frac{2}{9} \frac{\delta^4}{M^2}  j (j+1) (2j+1) +  \mathcal{O} \left( \frac{\delta^8}{M^6} \right)  \right] \, ,
\end{equation}
where $N_C$ is the dimensionality of $\chi$ under the color factor. 
Neglecting the higher-order $\delta/M$ terms, we finally obtain
\begin{equation}
\label{boundM}
M \gtrsim 72.5 \ \text{GeV} \left( \frac{0.001}{\rho^{\rm{exp}}_0 - 1} \right)^{1/2} \beta 
\sqrt{\eta N_C} \sqrt{j (j+1) (2j+1)} \, ,
\end{equation}
which is valid for $M > \delta \approx \sqrt{\beta} \ 123 \ \text{GeV}$. 
For $\mathcal{O}(1)$ values of the coupling $\beta$ the typical bounds on $M$ range in the 
few hundred GeV region, depending on the dimensionality of the representation. 
We hence conclude that the mass bounds coming from loop-level contributions to $\rho_0$ are less general (they depend on the value of $\beta$) and not particularly constraining when compared to existing direct searches limits (see~\sect{lhc}). 
This is, however, not necessarily true for higher dimensional representations.

Alternatively, the information from $\rho_0$ can be used to give an upper bound on the 
mass splitting $\Delta m = m_{I+1} - m_I \approx - \frac{\delta^2}{2 M}$.
As an example, let us mention that for the case $(1,5,2)_S$ we get 
$\Delta m \lesssim 20$ GeV.  
This information is exploited in \sect{neutralLP} when inferring collider bounds 
on the neutral state of such a multiplet by looking at the charged component production and decays.   

\subsubsection {Bounds on Higgs portal coupling}
\label{sec:Higgs}

%The recent measurement of the Higgs boson invisible decay width \cite{Aad:2014iia, Chatrchyan:2014tja} can constrain new particles' masses up to $m_H/2$ if the particles have a large enough coupling to the Higgs boson. This can also cover cases for which the invisible $Z$ boson width cannot give any constraints, namely if the particle is neutral and has a zero hypercharge. The CMS bound on the Higgs branching ratio into invisible particles amounts to BR$_{\rm inv}<0.58$ \cite{Chatrchyan:2014tja}. Only scalars are restricted by  the Higgs invisible width, since our fermionic states do not couple at the renormalizable level with the Higgs boson. 
The Higgs boson can couple to the new scalars via the portal coupling $\alpha$ of \eq{potentialXH}. This leads to two kinds of effects: (1) If some components of $\chi$ lie below half of the Higgs mass, they can contribute to the Higgs total decay width. Taking into account other existing collider constraints (see Table~\ref{summarybounds}) this is only possible for the neutral component $\chi_0$. In particular, it contributes to the Higgs invisible decay branching fraction. The partial decay width of the Higgs boson via the $\alpha$ coupling into a pair of $\chi_0$ states (for $\beta=0$) is found to be 
\begin{equation}
 \Gamma_{\rm inv}=\frac{\eta \alpha^2 v^2}{16 \pi} \frac{1}{m_H} \sqrt{1-\frac{4 m_{\chi_0}^2}{m_H^2}}\, .\label{eq:higgsinv}
\end{equation}
In addition, (2) all charged components of $\chi$ will contribute at 1-loop level to the $H\to \gamma\gamma$ (and $H\to \gamma Z$) decays, while colored $\chi$ will affect Higgs boson production through gluon fusion (GF) and also its decays to two gluons $H\to gg$\,. Using the results of~\cite{Djouadi:2005gj, Chang:2012ta,Dorsner:2012pp} we find
\begin{subequations}
\begin{align}
\mu_{\gamma\gamma} \equiv\frac{\Gamma_{\gamma\gamma}}{\Gamma^{\rm SM}_{\gamma\gamma}} & = \frac{|\mathcal A_{1}(x_W) + (4/3) \mathcal A_{1/2}(x_t) +  \eta \alpha d(R_\chi) \sum_i Q_i^2 (v/m_{\chi_i})^2 \mathcal A_0(x_{\chi_i})|^2 }{|\mathcal A_{1}(x_W) + (4/3) \mathcal A_{1/2}(x_t) |^2}\,,\\
\mu_{gg} \equiv \frac{\Gamma_{gg}}{\Gamma^{\rm SM}_{gg}} & = \frac{\sigma_{\rm GF}}{\sigma_{\rm GF}^{\rm SM}} = \frac{|(1/2) \mathcal A_{1/2}(x_t) + \eta \alpha C(R_\chi) \sum_i (v/m_{\chi_i})^2 \mathcal A_0(x_{\chi_i}) |^2}{|(1/2) \mathcal A_{1/2}(x_t)|^2}\,,
\end{align}
\end{subequations}
where $x_i\equiv m_H^2/4m_i^2$, the sums $\sum_i$ run over all $\chi$ weak multiplet components $\chi_i$, $d(R_\chi)$ is the dimension of the color representation of $\chi$ and $C(R_\chi)$ is the corresponding index ($C(3)=1/2$, $C(6) = 5/2$ and $C(8)=3$). The relevant loop functions $\mathcal A_{1}(x_W) \simeq -8.32$, $\mathcal A_{1/2}(x_t) \simeq 1.38$ and $A_{0}(x)$ with limits $A_0({x\to 0}) = 1/3$, $A_0{(x\to \infty)} = -1/x + \mathcal O(x^{-2})$ can be found e.g.~in~\cite{Dorsner:2012pp}. The total decay width of the Higgs can thus be written as
\begin{equation}
\Gamma_H = \Gamma^{\rm SM}_H \left[ 1+BR_{\gamma\gamma}^{\rm SM} (\mu_{\gamma\gamma}-1) + BR_{gg}^{\rm SM} (\mu_{gg}-1)    \right] + \Gamma_{\rm inv}\,,
\end{equation}
where $\Gamma^{\rm SM}_H = 4.07$ MeV, BR$_{\gamma\gamma}^{\rm SM} = 2.28 \times 10^{-3}$ and  BR$_{gg}^{\rm SM} = 8.57 \times 10^{-2}$~\cite{LHCcxn}.  The invisible branching ratio is then finally given by  BR$_{\rm inv} =  \Gamma_{\rm inv}/\Gamma_{H}$.

\begin{figure}[t]
\centering
\includegraphics[angle=0,width=9cm]{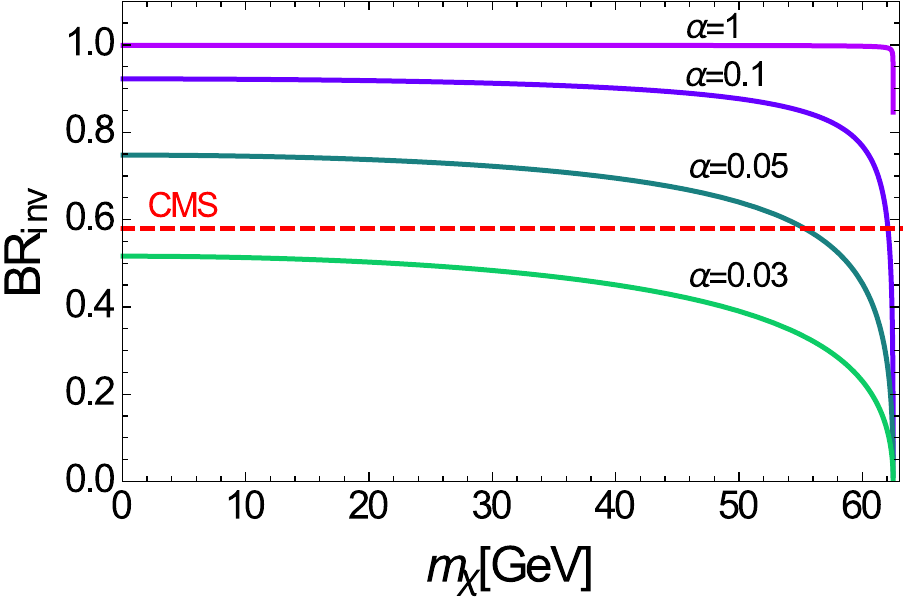}
\caption{\label{fig:HiggsinvBR} Invisible branching fraction of the Higgs boson as a function of the new scalars mass (real scalar) for different values of the portal coupling $\alpha$. The red dashed line shows the CMS exclusion limit of Ref.~\cite{Chatrchyan:2014tja}.}
\end{figure}

To analyze the resulting correlated effects in Higgs boson production and decays as measured at the LHC, we follow closely the procedure described in~\cite{Dorsner:2015mja} using also the same set of experimental results~\cite{TheATLAScollaboration:2013lia,Aad:2013wqa,ATLASttbb,Aad:2014iia,CMS:2013jda,CMS:2013sea,Chatrchyan:2013iaa,Chatrchyan:2013mxa,CMS:ril,Chatrchyan:2014tja,CMS:2014ala}. In particular, we find that in cases where the Higgs boson can decay to $\chi_0$, the  constraints on $\alpha$ are completely dominated by the bounds on the extra invisible decay rate. In Fig.~\ref{fig:HiggsinvBR} the invisible branching ratio as a function of the new scalar mass $m_{\chi}$ is shown for different values of the portal coupling $\alpha$. We used $\eta=1/2$ in Eq.~\eqref{eq:higgsinv}, assuming a real scalar. The red dashed line shows the CMS limit of Ref.~\cite{Chatrchyan:2014tja}.\footnote{Indirect bound on BR$_{\rm inv}$ coming from the global fit to all Higgs boson signal strenghts yields a slightly stronger bound of BR$_{\rm inv}\lesssim 0.2$.} It can be inferred from the plot that for a portal coupling $|\alpha|=\mathcal O(1)$, the new scalar states are excluded up to the kinematic limit for this decay. However for values of $|\alpha|\lesssim \mathcal O(0.01)$ currently no limit on $m_{\chi_0}$ can be given anymore.

\begin{figure}[t]
\centering
\includegraphics[angle=0,width=7cm]{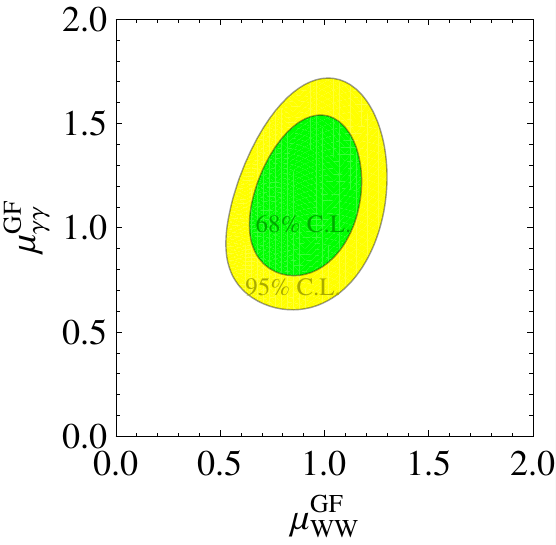}
\caption{\label{fig:HiggsFit} Exclusion bounds on the $\mu_{\gamma\gamma}^{\rm GF}$ and $\mu_{WW}^{\rm GF}$ LHC Higgs signal strengths allowing for arbitrary contributions to $\mu_{\gamma\gamma}$ and $\mu_{gg}$ but keeping $\Gamma_{\rm inv}=0$.}
\end{figure}

\begin{figure}[t]
\centering
\includegraphics[angle=0,width=5.5cm]{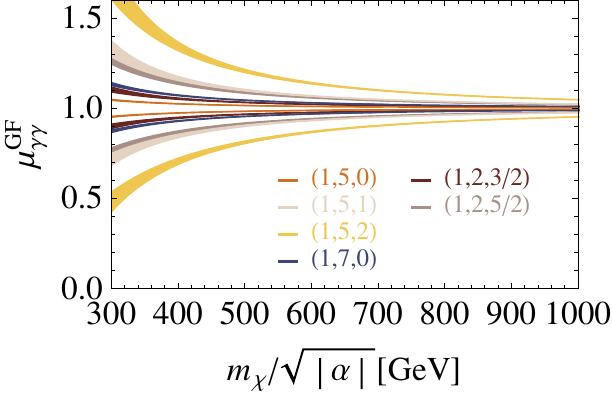}
\includegraphics[angle=0,width=5.5cm]{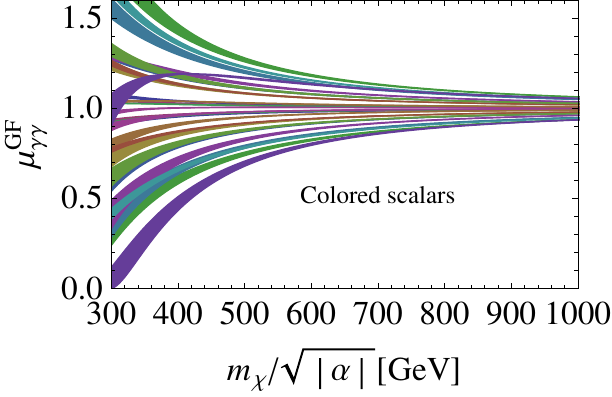}
\includegraphics[angle=0,width=5.5cm]{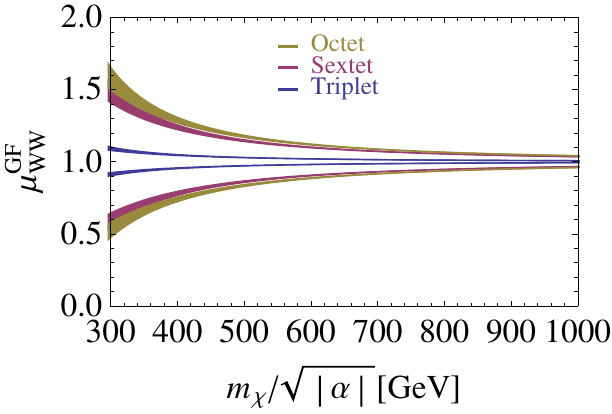}
\caption{\label{fig:HiggsFit1} Accidental scalar induced modifications to $\mu_{\gamma\gamma}^{\rm GF}$ and $\mu_{WW}^{\rm GF}$ LHC Higgs signal strengths as a function of $m_\chi/\sqrt{|\alpha|}$. The un-colored and colored scalar effects in $\mu_{\gamma\gamma}^{\rm GF}$ are shown in the left and middle panel, respectively. Single (complex) colored scalar effects in $\mu_{WW}^{\rm GF}$ are shown in the right panel. The shaded bands correspond to a scan $|\alpha| \in [0.1,1]$.}
\end{figure}

Even if $\chi_i$ are heavy ($m_{\chi_i}>m_H/2$), their contributions to $\mu_{\gamma\gamma}$ and $\mu_{gg}$ still lead to constraints on $\alpha$ from the measurements of the Higgs signal strengths at the LHC.  In particular, the most sensitive channels involve GF produced Higgs bosons decays to photons and W bosons, these being the two most precisely measured. Denoting the relevant signal strenghts as
\begin{equation}
\mu_{\gamma\gamma}^{\rm GF} \equiv \frac{\sigma_{\rm GF}}{\sigma_{\rm GF}^{\rm SM}} \frac{{\rm BR}_{\gamma\gamma}}{{\rm BR}_{\gamma\gamma}^{\rm SM}}\,, \quad \mu_{WW}^{\rm GF} \equiv \frac{\sigma_{\rm GF}}{\sigma_{\rm GF}^{\rm SM}} \frac{{\rm BR}_{WW}}{{\rm BR}_{WW}^{\rm SM}}\,, 
\end{equation}
the global fit of Higgs boson LHC data allowing for arbitrary contributions to $\mu_{\gamma\gamma}$ and $\mu_{gg}$ but keeping $\Gamma_{\rm inv}=0$ yields the 68\% and 95\% CL exclusion bounds shown in Fig.~\ref{fig:HiggsFit} . We observe that up to $50\%$ modifications in both observables are still allowed by the current data. These should be compared with $\alpha$ induced modifications shown in Fig.~\ref{fig:HiggsFit1} (assuming degenerate $\chi_i$). In particular, color-neutral $\chi$ predominantly affect $\mu^{\rm GF}_{\gamma\gamma}$ as shown in the left panel.  On the other hand, colored states can affect GF production and are thus constrained also from $\mu_{WW}^{\rm GF}$ as illustated in the right panel. The deviations in $\mu_{WW}^{\rm GF}$ are shown for single complex scalar in the given color representation. Finally, colored scalar effects in $\mu^{\rm GF}_{\gamma\gamma}$  are also shown in the middle panel.

Asymptotically, $\chi$ effects in both observables decouple as $\alpha/m_\chi^2$. The shaded bands in Fig.~\ref{fig:HiggsFit1} illustrate the amount of deviations from this limit as they correspond to a scan of $|\alpha| \in [0.1,1]$\,.  We observe that for $m_{\chi}\gtrsim 500$~GeV even $|\alpha|\lesssim \mathcal O(1)$ can be consistent with current Higgs data. Conversely $m_{\chi}\gtrsim 100$~GeV are perfectly allowed for small enough $|\alpha|\lesssim \mathcal O(0.1)$ Higgs portal couplings.

\subsection{Mass spectrum}
\label{massspectra}

The phenomenology of the new EW states is dictated by the mass spectrum. 
Typically, on top of a common mass term $m_\chi$, there is a radiative splitting within the SU$(2)_L$ multiplet and, 
for scalars only, a tree-level splitting due to the presence of non-trivial SU$(2)_L$ invariants in the scalar potential. 
In the $m_\chi \gg v$ limit the radiative contribution takes the form \cite{Cirelli:2005uq,DelNobile:2009st}
\begin{equation}
\label{deltaMrad}
\Delta m_{\rm{rad}} = m_{Q+1} - m_{Q} \approx 166 \ {\rm MeV} \left( 1 + 2\,Q + \frac{2\, Y}{\cos\theta_W} \right) \, , 
\end{equation}
which holds both for fermions and scalars. Notice that if $Y=0$ the LP in the multiplet is always the one with the smallest $|Q|$. This is 
not necessarily true when $Y \neq 0$.\footnote{E.g.~the LP of the fermion multiplet $(1,4,1/2)_F$ has $Q=-1$.} 

Similarly, the tree-level splitting in \eq{m2treeI} can be expanded 
in the $m_\chi \gg v$ limit, thus obtaining \cite{Cirelli:2005uq}
\begin{equation}
\label{deltaMtree}
\Delta m_{\rm{tree}} = m_{I+1} - m_I \approx \frac{\beta v^2 }{8 m_\chi} \approx \beta \times 7.6 \ {\rm GeV} \left( \frac{1 \ {\rm TeV}}{m_\chi} \right) \, .
\end{equation}
Notice that, while for fermions the mass spectrum is unambiguously fixed, 
for scalars it depends on the values of $\beta$ and $m_\chi$. Focussing on the $m_\chi <$ TeV region (relevant for LHC), 
if $\beta = \mathcal{O}(1)$ then the LP is always the one with the highest/lowest $I$, 
depending on the sign of $\beta$. 
However, for $\beta \in [10^{-3}, 1]$ the tree-level splitting can be comparable with the radiative one. 
In such cases it is possible to show (see below) that:
\begin{enumerate}
\item Any particle in the multiplet can be the LP for large domains of the model parameters, 
i.e.~without any fine-tuning. 
\item If the LP has charge $Q_{\rm{LP}}$, the next-to-LP has always charge $Q_{\rm{LP}} \pm 1$. 
\end{enumerate}
This latter fact turns out to be phenomenologically relevant, e.g.~when setting bounds on the neutral LP by looking at the decay of the next-to-LP. 

For completeness, we provide here a proof of the two statements above:  
by combining \eq{deltaMrad} and \eq{deltaMtree} one arrives at the expression 
$m_I=m_{-j}+a+bI+cI^2$, 
where $a$ and $b$ can have any sign (since they depend on $\Delta m_{\rm{tree}}$), and $c > 0$. 
The minimum of $m_I$ is obtained for $I_{\rm{min}}=-\tfrac{b}{2c}$. 
Hence, by an appropriate choice of the ratio $b/c$, the smallest $m_I$ can be anywhere in the range $I \in [-j,j]$. 
The fact that the next-to-LP has $Q_{\rm{LP}} \pm 1$ 
simply follows from the convexity of $m_I$ as a function of $I$. 
A similar argument holds as well in the $m_\chi \approx v$ regime, for which 
the full formula of the radiative splitting (see e.g.~Eq.~(6) in \cite{Cirelli:2005uq}) must be taken into account.

\subsection{Validity of the EFT}
\label{landau}

Our working hypothesis is that the SM$+\chi$ renormalizable 
theory is a low-energy effective description valid up to a cut-off scale $\Lambda_{\rm eff}$. 
In the spirit of a generic EFT with $\mathcal{O}(1)$ couplings 
and without any extra state beyond $\chi$ introduced at low energy,
$\Lambda_{\rm eff} \approx 10^{15}$ GeV is essentially fixed 
by neutrino masses through the $d=5$ Weinberg operator.  
Moreover, such a cut-off scale can automatically account for null results of all 
flavor, CP and B violating processes constraining $d=6$ operators made of SM fields. 
In particular, when the lowest-dimensional sources 
of breaking of the extra U(1) or $Z_2$ symmetry associated with the kinetic term of $\chi$ are the $d=5$ operators involving $\chi$ and SM fields, 
any $d=6$ operator involving only SM fields, generated by integrating out $\chi$, will have 
two insertions of such $d=5$ operators and hence at least a $1/\Lambda_{\rm eff}^2$ suppression. The situation changes only slightly if the extra U(1) or $Z_2$ is broken at the renormalizable level in the scalar potential, as in cases listed in  \Table{renchiH}. Namely, the only additional effect arises for $\chi \sim (1,4,1/2)_S$,  $(1,4,3/2)_S$ where integrating out $\chi$ induces a $\Delta L=2$ operator of the form $\ell \ell H H H H^\dagger$. Being suppressed by $1/\Lambda_{\rm eff} m_\chi^2$, it necessarily represents a subleading contribution to neutrino masses.

The  infinite set of states preserving $\mathcal{G}_F$  at the renormalizable level (see \eq{chineqto} and \eq{Xneqto})  can be reduced by requiring   that the EFT remains weakly coupled up to $\Lambda_{\rm eff} \approx 10^{15}$ GeV.  
The presence of extra matter multiplets drives the gauge couplings of the SM towards the non-perturbative 
regime.\footnote{We do not address here the question of the RG running of the 
scalar potential parameters, since it is a model dependent issue which also involves 
the analysis of the vacuum stability.}
Eventually, this might result in the presence of a Landau pole
below the cut-off scale
of the EFT. If the Landau pole is associated with 
a generic new dynamics, the accidental symmetries of the SM could be violated at that scale. 
Hence, for the self-consistency of the EFT approach, 
we require the absence of Landau poles below 
$\Lambda_{\rm{eff}} \approx 10^{15}$ GeV, which translates 
into an upper bound on the dimensionality of the extra representations. 

In light of stringent bounds on the inter-multiplet mass splittings (see \sect{massspectra}) we can safely integrate in all multiplet components at a single scale, which we choose to be the $Z$ mass in our numerical analysis. We note however, that for 
$m_\chi$ not much larger than the TeV scale the resulting Landau pole estimates scale linearly with $\chi$ masses. The analysis of the perturbativity bounds is detailed in \app{pertbounds} and the results are summarized in \Tables{LPSU2vsSU3}{summarydim5}.  
They provide a useful reference for the estimate of the Landau poles at two loops 
for the cases where the SM is extended with an extra multiplet charged under SU$(3)_c$ and/or SU$(2)_L$, 
and in particular for all the states considered in this work which can have a non-zero hypercharge as well. 

A crucial ingredient in order to make our list of 
extra states finite however, is given by cosmology. In fact, the only reason why we can disregard 
multiplets with an arbitrary hypercharge, e.g.~$Y=\pi$, is because these states 
feature an absolutely stable charged LP that cannot decay into SM particles 
because of electric charge conservation. The possibility of having 
an infinitesimal hypercharge is instead briefly 
discussed in \sect{cosmo}. 

A comment on the role of higher-order corrections in the RG equations is in order here. 
The determination of the Landau pole is often carried out at the one-loop level (see e.g.~\cite{Cirelli:2005uq}). 
However, for the non-abelian gauge factors there is an accidental cancellation 
in the one-loop beta function between matter and gauge contributions (cf.~\eq{oneloopbf}), 
so that two-loop effects may become  important.
Interestingly, among the cases that we found to be drastically affected by two-loop corrections 
there are the two minimal DM candidates: a real $(1,7,0)$ scalar and a Weyl $(1,5,0)$ 
fermion \cite{Cirelli:2005uq}.\footnote{Another situation where the two-loop RG analysis of the gauge couplings 
could change the qualitative UV behaviour of the theory 
is given by the Pati-Salam model presented in \cite{Giudice:2014tma}, 
where low-scale extensions of the SM providing total asymptotic freedom 
are investigated.}  
Following the results of \cite{Cirelli:2007xd} for the calculation of the relic density, 
we integrate in the scalar septuplet at $m_\chi = 25$ TeV and the fermionic 
quintuplet at $m_\chi = 10$ TeV. Hence we find, respectively
\begin{align}
\label{LP170}
&\Lambda^{\text{1-loop}}_{\text{Landau}} = 1.9 \times 10^{41} \ \text{GeV} 
\quad \longrightarrow \quad \Lambda^{\text{2-loop}}_{\text{Landau}} = 8.9 \times 10^{20} \ \text{GeV} \ \quad \text{$((1,7,0)_S$ case)} \, , \\
\label{LP150}
&\Lambda^{\text{1-loop}}_{\text{Landau}} = 9.0 \times 10^{28} \ \text{GeV} 
\quad \longrightarrow \quad \Lambda^{\text{2-loop}}_{\text{Landau}} = 4.0 \times 10^{21} \ \text{GeV} \ \quad \text{($(1,5,0)_F$ case)} \, .
\end{align}
If we associate the Landau pole with the cut-off of a generic EFT, this also sets 
the scale of the effective operator leading to the decay of the minimal dark matter candidate. 
Note however, that even for a cut-off of the order of the Planck mass, 
the framework of minimal DM is not endangered by $d \geq 6$ operators, 
since the lifetime of DM is still comfortably larger than the age of the 
Universe (and satisfies the indirect bounds on decaying DM). For a discussion of $d=5$ induced $(1,7,0)_S$ decays see \sect{loopdecays}.

In the selection of our states, the two-loop criterium proved to be important for several states. 
For instance, in the case of the real $(27, 1, 0)_S$ scalar multiplet we find that at one loop 
$\Lambda^{\text{1-loop}}_{\text{Landau}} = 1.9 \times 10^{41} \ \text{GeV}$, 
whereas at the two-loop level $\Lambda^{\text{2-loop}}_{\text{Landau}} = 1.3 \times 10^{7}\ \text{GeV}$,  
so that we can exclude this state from our list of accidental matter candidates.  

What about three-loop corrections then? As long as there are no accidental cancellations 
in the two-loop beta function (as it can be explicitly verified), 
they are not expected to drastically change the situation.\footnote{For instance, 
in the SM case where no strong cancellations are at play we find: 
$\Lambda^{\text{1-loop}}_{\text{Landau}} = 1.9 \times 10^{41}$ GeV, 
$\Lambda^{\text{2-loop}}_{\text{Landau}} = 5.2 \times 10^{40}$ GeV and 
$\Lambda^{\text{3-loop}}_{\text{Landau}} = 8.7 \times 10^{40}$ GeV. 
} It is then enough to rely on a two-loop estimate of the Landau pole 
in order to set an upper bound on the dimensionality of the extra representation.  

\section{Lifetimes}
\label{sec:decays}

The new extra states will eventually decay due to operators present in the EFT. 
There are essentially three classes of decays which we are going to consider in this section: 
\emph{i)} Inter-multiplet weak transitions where the heavier components within the 
SU$(2)_L$ multiplet decay via cascades involving the emission of (virtual) $W$ gauge bosons into the LP, 
\emph{ii)} Decays through renormalizable interactions (only for a specific class of new scalar states)
and \emph{iii)} Decays through non-renormalizable $d \geq5$ operators. 
We analyze each class of decays in turn below.

\subsection{Inter-multiplet weak transitions}
\label{imwt}

Heavier components within the SU$(2)_L$ multiplet can decay via EW transitions into lighter ones, 
with rates suppressed by a small phase space factor. 
Denoting the component of a total $j$-isospin representation with $T^3$-eigenvalue $I$ as $\chi^j_I$, 
for $\Delta m > m_{\pi^+}$, we have the decay width
(generalizing the expression in Ref.~\cite{DelNobile:2009st})
\begin{equation}
\label{imwtonepion}
\Gamma(\chi^j_{I+1} \rightarrow \chi^j_I \, \pi^+) 
= \frac{T_+^2 G_F^2 V_{ud}^2 \Delta m^3 f_{\pi^+}^2}{\pi} \sqrt{1-\frac{m_{\pi^+}^{2}}{\Delta m^2}}
\approx 
\frac{T_+^2}{7.5 \times 10^{-12} \ \rm{s}} \left( \frac{\Delta m}{500 \ \rm{MeV}} \right)^3 \, , 
\end{equation}
where $T_+ = \sqrt{j(j+1)-I(I+1)}$ and the approximation in the r.h.s.~of \eq{imwtonepion} is valid 
for $\Delta m \gg m_{\pi^+}$. 

Formula (\ref{imwtonepion}) is a reasonable approximation of the total width in the 
range $m_{\pi^+} \lesssim \Delta m \lesssim 1 \ \text{GeV}$.
For mass splittings close to the kinematical threshold of the decay into a pion, 
3-body decays involving leptons become important as well, while for 
$\Delta m \gtrsim 1$ GeV new hadronic channels open up 
(e.g.~involving kaons and other heavier hadrons) and the decay can be eventually computed at the 
partonic level, once quark-hadron duality sets in. 

\begin{figure}[t]
\centering
\includegraphics[angle=0,width=10cm]{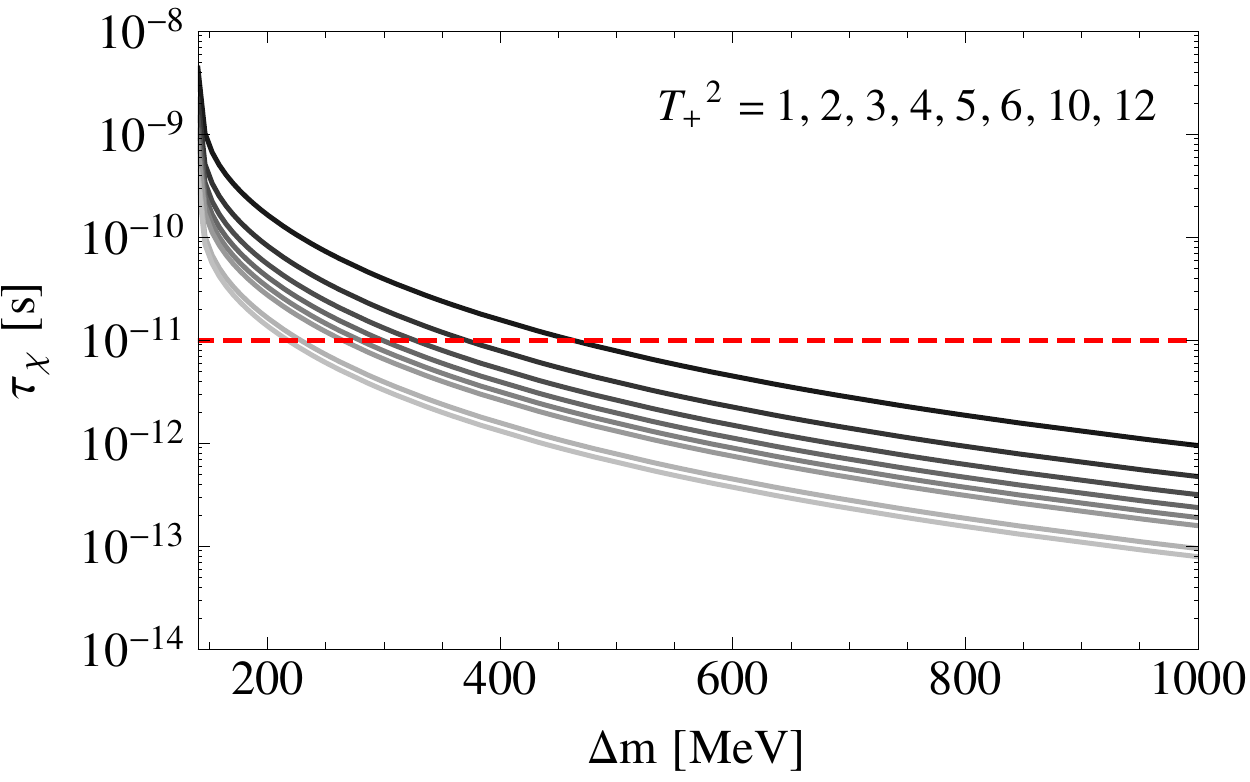}
\caption{\label{fig:IMWD_LifetimevsDeltam} 
Lifetimes associated with inter-multiplet weak transitions 
as a function of the mass splitting $\Delta m$. 
The grading of the curves (from black to gray) corresponds to different values of $T_+^2$ (from 1 to 12), 
as it can be found in representations up to $j=3$ (septuplet). The red dashed line corresponds to 
the typical freeze-out time for colorless 1~TeV-mass particles with weak interactions in the early Universe (cf.~\sect{cosmo}).}
\end{figure}

The typical lifetime of an SU$(2)_L$ multiplet component 
decaying via inter-multiplet weak transitions
is displayed in \fig{fig:IMWD_LifetimevsDeltam} 
as a function of the 
mass splitting and for different values of the ladder operator $T_+$, up to the $j=3$ (septuplet) case.

Within high energy collider experiments, the inter-multiplet decays are essentially prompt. On the  other hand, the LP at the end of these inter-multiplet cascades is stable on the detector scale, 
barring few exception which are discussed in the next subsection.

\subsection{Decays through renormalizable interactions}
\label{decren}

There exists the possibility that the new extra scalars 
retain renormalizable interactions with the SM Higgs which can induce their decay. 
These states are classified in \Table{renchiH} and correspond to the cases (labelled with the symbol ``$\times$")  where no accidental symmetry (e.g.~U(1), $Z_2$ or $Z_3$) forbids $\chi$ to decay.  
Let us comment in turn on the various possibilities. 

The case of the gauge singlet $(1,1,0)_S$ has been extensively studied in the literature 
(see e.g.~\cite{Barger:2007im}) 
and we do not have much to add here.
In the cases $(1,3,0)_S$, $(1,4,1/2)_S$, and $(1,4,3/2)_S$, $\chi$
can couple linearly  to Higgs operators. However, such ``tadpole" couplings also induce non-zero VEVs for $\chi$, 
which are severely constrained by EW precision observables. 
As already pointed out in \sect{scalpot}, unless a moderate fine-tuning is allowed in the scalar potential, 
the bounds on such dimensionally estimated VEVs  push the masses of these states beyond the kinematic reach of the LHC. 
Other multiplets which can possibly decay at the renormalizable level (those labelled with a ``$\times$" in \Table{renchiH}) are either not considered here because they break $\mathcal{G_F}$ by coupling 
to SM fermions (e.g.~$(8,2,1/2)_S$) or they generate 
a Landau pole below $\Lambda_{\text{eff}} \approx 10^{15}$ GeV 
(e.g.~$(1,6,1/2)_S$).

\subsection{Decays through $d \geq 5$ effective operators}
\label{effopdec}

Let us consider now the case where the decay of the new state $\chi$ is due to effective operators. 
Given an effective operator $\mathcal{O}_{\rm decay}$, we always absorb the Wilson coefficient 
in the definition of the effective cut-off scale $\Lambda_{\rm{eff}}$, 
e.g.
\begin{equation}
\mathcal{L} \ni
\frac{1}{\Lambda_{\rm{eff}}} \mathcal{O}_{\rm decay} + \text{h.c.} \, .
\end{equation}   
The differential decay rate of an unstable particle $\chi$ into $n_f$ final states reads 
\begin{equation}
\label{dGamma}
d\Gamma= \frac{1}{2 m_{\chi}} \left( \prod_{f} \frac{d^3 p_f}{2 \pi^3} \frac{1}{2 E_f } \right) \left| \mathcal{M} (m_{\chi} \to \{ p_f \} ) \right|^2 (2 \pi)^4 \delta^{(4)} \left( p_{\chi} - \sum_f p_f \right) \, .
\end{equation}
By assuming a constant matrix element and massless final states, the phase space factor can be 
integrated in the rest frame of the decaying particle, yielding
\begin{equation}
\label{phasespace}
{\rm PS}_{n_f} \equiv \int \left( \prod_{f} \frac{d^3 p_f}{2 \pi^3} \frac{1}{2 E_f } \right) (2 \pi)^4 \delta^{(4)} \left( p_{\chi} - \sum_f p_f \right) 
= \frac{1}{2(4 \pi)^{2 n_f -3}} \frac{m_{\chi}^{2 n_f-4}}{(n_f-1)!(n_f-2)!} \, . 
\end{equation}
So, for example, the phase space factors up to $n_f = 4$ are: 
${\rm PS}_2 = \frac{1}{8 \pi}$, 
${\rm PS}_3 = \frac{m^2_{\chi}}{256 \pi^3}$, 
${\rm PS}_4 = \frac{m^4_{\chi}}{24576 \pi^5}$.

In the case of a dimension $d$ effective operator, 
the amplitude squared for $n_f$ particles in the final 
state can be estimated by naive dimensional analysis (NDA) as
\begin{eqnarray}
\label{M2nda}
\left| \mathcal{M} (m_{\chi} \to \{ p_f \} ) \right|^2_{\rm{NDA}} = \frac{(\frac{v}{\sqrt{2}})^{2 n_c}}{\Lambda_{\text{eff}}^{2 (d-4)}} m^{2d -2 n_f -2 n_c -6}_{\chi} \, ,
\end{eqnarray}
where we also included the possibility of $n_c$ condensations of the Higgs boson. 
Hence, by putting \eqs{dGamma}{M2nda} together, we get the following expression for the total width for $m_\chi \gg v$ 
\begin{equation}
\label{GammaNDA}
\Gamma_{\rm{NDA}} = \frac{1}{4(4 \pi)^{2 n_f -3}} \frac{m_{\chi}^{2d -2n_c-7}}{(n_f-1)!(n_f-2)!} \frac{ (\frac{v}{\sqrt{2}})^{2 n_c}}{\Lambda_{\text{eff}}^{2 (d-4)}} \, .  
\end{equation}
Unless differently specified, we compute the lifetimes of the states decaying through the non-renormalizable operators in \Tables{summary1}{summary2}  using \eq{GammaNDA}. 
Whenever multiple operators can be responsible for the decay of $\chi$, we  sum over the several widths assuming the operators contribute with the same Wilson coefficient. 
What is missing in \eq{GammaNDA} with respect to the full decay width are the relevant SU$(2)_L$ Clebsch-Gordan coefficients, symmetry, color and flavor factors, the kinematical dependence of the matrix element, the masses of the final states and finally mixing effects induced when scalar $\chi$ obtain VEVs. 
In the region $m_\chi \gg v$ all of these are expected to give $\mathcal{O}(1)$ corrections. 
When more accuracy is required, for example when setting BBN bounds, 
we take all these factors into account, computing the relevant decay widths explicitly.

\subsubsection{Cascade decays}
\label{offshelldecays}

Whenever a Higgs doublet is contained in a SM-invariant operator, 
it can happen that not all the SU$(2)_L$ components of the multiplet $\chi$ can directly decay 
through the effective operator. This is easily understood by going to the unitary gauge, 
where some of the SU$(2)_L$ contractions end up into the goldstone directions of the Higgs doublet. 
See \app{SUtwodecomp} for a description of the SU$(2)_L$ decompositions of the relevant operators. 
Depending on the mass spectrum, the cases where the LP 
cannot directly decay through the effective operator are displayed in \Table{cascadedimgeq5}. 
\begin{table}[htbp]
\centering
\begin{tabular}{@{} |c|c|c|c| @{}}
\hline
Spin  & $\chi$ &  $Q_{\rm{LP}}$ & $\mathcal{O}_{\rm decay}$ \\ 
\hline
0 & $(1, 2, 5/2)$ &  $3$  & $\chi^{\dag} H e^c e^c$   \\  
0 & $(1, 5, 1)$ &  $-1,1,2,3$  & $ \chi^{\dag} H H H H^{\dag}$   \\  
0 & $(1, 5, 2)$ &  $1,2,3,4$  & $\chi^{\dag} H H H H$   \\  
0 & $(\overline{3},2,11/6)$ &  $7/3$  & $\chi H^{\dag} u^c u^c + \chi^{\dag} H d^c e^c$   \\  
0 & $(3,3,5/3)$ &  $8/3$  & $\chi^{\dag} H q e^c + \chi H^{\dag} u^c \ell$   \\  
0 & $(3,4,1/6)$ &  $5/3$  & $\chi H^{\dag} qq + \chi^{\dag} H q \ell$   \\  
0 & $(\overline{3},4,5/6)$ &  $7/3$  & $\chi^{\dag} H qq+ \chi H^{\dag} q \ell$   \\  
0 & $(\overline{6},2,7/6)$ &  $5/3$  & $\chi^{\dag} H d^c d^c$   \\  
0 & $(8,3,1)$ &  $2$  & $\chi H^{\dag} q u^c +\chi^{\dag} H q d^c $   \\  
\hline
  \end{tabular}
  \caption{\label{cascadedimgeq5} 
  Extra states decaying through off-shell cascades.}
\end{table}

It is possible, however, for the LP to cascade decay via off-shell heavier components (which  eventually 
decay through the effective operator) and $W$ bosons, thus resulting in lifetimes 
which are typically larger than in the case of the direct decay. 
Moreover, these decay rates must be evaluated numerically since the 
NDA formula in Eq.~\eqref{GammaNDA} cannot be straightforwardly applied 
due to the strong momentum dependence of the matrix element. 

\begin{figure}[t]
\centering
\includegraphics[angle=0,width=14cm]{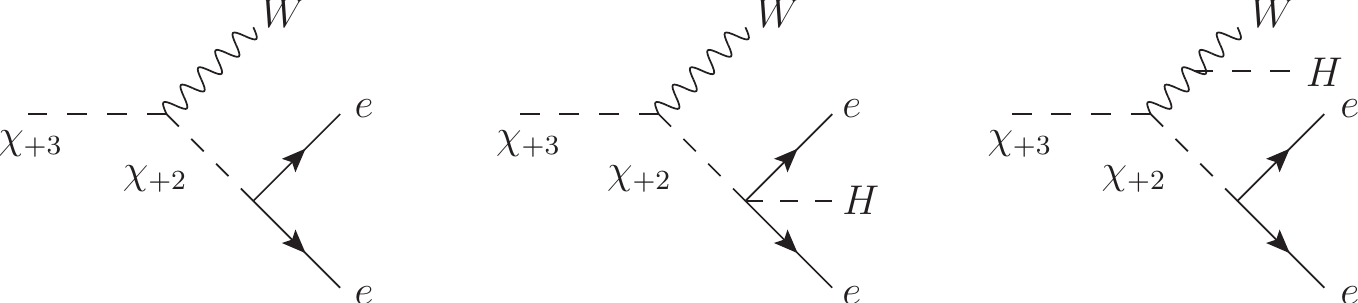}
\vspace*{0.2cm}
\caption{Feynman diagrams for the cascade decay of the $\chi_{+3}$ component of the $(1, 2, 5/2)_S$ state for $m_{\chi_{+3}}< m_{\chi_{+2}}$.\label{fig:cascadedecay}}
\end{figure}

For the computation of the decay width of the $(1, 2, 5/2)_S$ multiplet component with $Q=3$, $\chi_{+3}$, under the assumption that $\chi_{+3}$ is lighter than (or degenerate with) $\chi_{+2}$, we take into account the decays into two leptons and into two leptons together with a Higgs boson.  The relevant Feynman diagrams are shown in Fig.~\ref{fig:cascadedecay}. The numerical phase space integration is performed with the help of {\tt RAMBO} \cite{Kleiss:1985gy} and we neglect the effects of lepton masses.

For the $(1, 5, 1)_S$ and $(1, 5, 2)_S$ multiplets, longer decay chains are possible for the multiple charged components of the multiplet. 
In addition, for the neutral states within these multiplets the effective operator induces a mixing with the Higgs boson which in turn generates direct couplings to SM vector bosons. These contributions to the lifetimes do not decouple for large $m_{\chi}$ and hence need to be taken into account over the whole considered mass range. For all multiplet components we thus consider decays with final states comprising of $2-4$ SM gauge or Higgs bosons. The numerical results have been obtained with {\tt Madgraph 5}~\cite{Alwall:2011uj} using {\tt FeynRules}~\cite{Alloul:2013bka}  generated model files. 
\par
Finally, the cascade decays of the colored cases can be estimated from the one of the $\chi_{+3}$ component of the $(1,2,5/2)_S$ multiplet by appropriate replacements of Clebsch-Gordan coefficients and by multiplying with the respective color factors. Note that an accurate evaluation of the 
decay rates for the colored cases is not necessary since their relic abundance turns out to be very suppressed resulting in no relevant BBN constraints. More details can be found in \sect{cosmo}.
\par 
The SU$(2)_L$ factors needed in the evaluation of the cascade decays are exemplified in \app{SUtwodecomp}. For all the cases the cut-off scale $\Lambda_{\text{eff}}$ was set to $10^{15}\text{ GeV}$. We do not include 
off-shell effects of the $W$ bosons in the computation of the lifetimes.\footnote{In Ref.~\cite{Grober:2015fia} it was shown that in the case of stop decays these off-shell effects can be numerically relevant for mass differences between the decaying particle and the decay products up to 35 GeV.} 
\par
 \begin{figure}[t]
\centering
\includegraphics[angle=0,width=13cm]{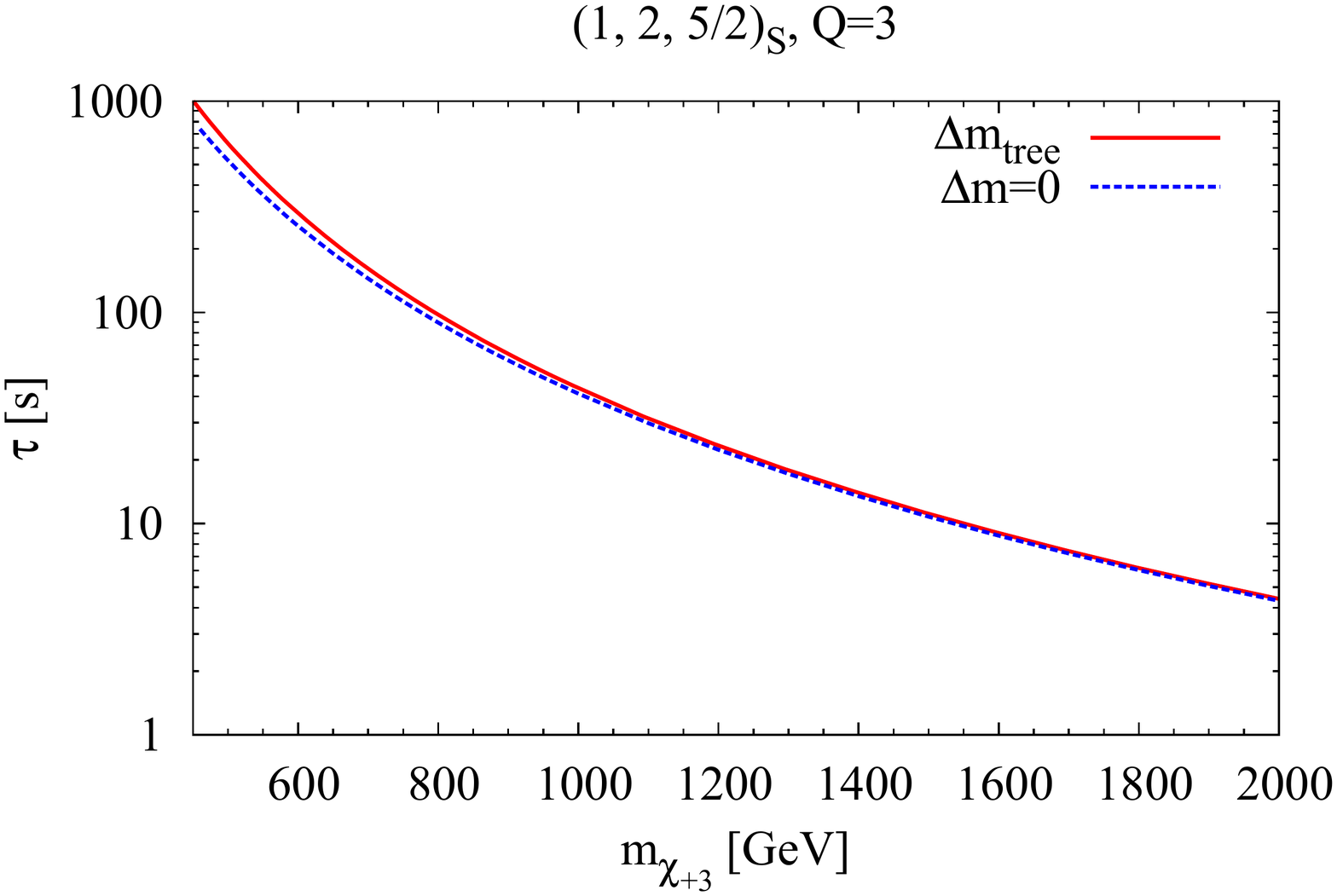}
\vspace*{-1.0cm}
\caption{Lifetime as a function of the mass of the $Q=3$  component
of the multiplet $(1, 2, 5/2)$, if it is the lightest. 
The red solid curve shows the lifetime for a NDA estimated tree-level mass splitting between $\chi_{+3}$ and $\chi_{+2}$ components, while the blue dashed
curve represents the zero mass splitting limit.\label{fig:lifetime_1252}}
\end{figure}
In Fig.~\ref{fig:lifetime_1252} the lifetime of the
 $Q=3$ component of $(1, 2, 5/2)_S$ is shown as a function of its mass,
 assuming that it is lighter than the $Q=2$ component and hence decays via an off-shell $\chi_{+2}$. 
The blue dashed curve shows the lifetime in the zero mass splitting approximation, while
 the red solid one stands for a NDA estimated tree-level mass splitting as given in Eq.~\eqref{deltaMtree}.
 In the plot we assume that the $d=5$ operator involves only one lepton flavor. If $\chi$ couples in the same way to all three flavors the corresponding lifetimes are reduced by a factor of three.  
From Fig.~\ref{fig:lifetime_1252} it can be inferred
that the presence of tree-level mass splitting only affects the lifetimes for low masses of $\chi_{+3}$, of the order $\mathcal O(v)$.
For larger masses it quickly becomes completely irrelevant and we henceforth work in the zero mass splitting limit whenever computing cascade decays.
\begin{figure}[h]
\centering
\vspace*{-1cm}
\hspace*{-1.7cm}
\includegraphics[angle=0,width=10cm]{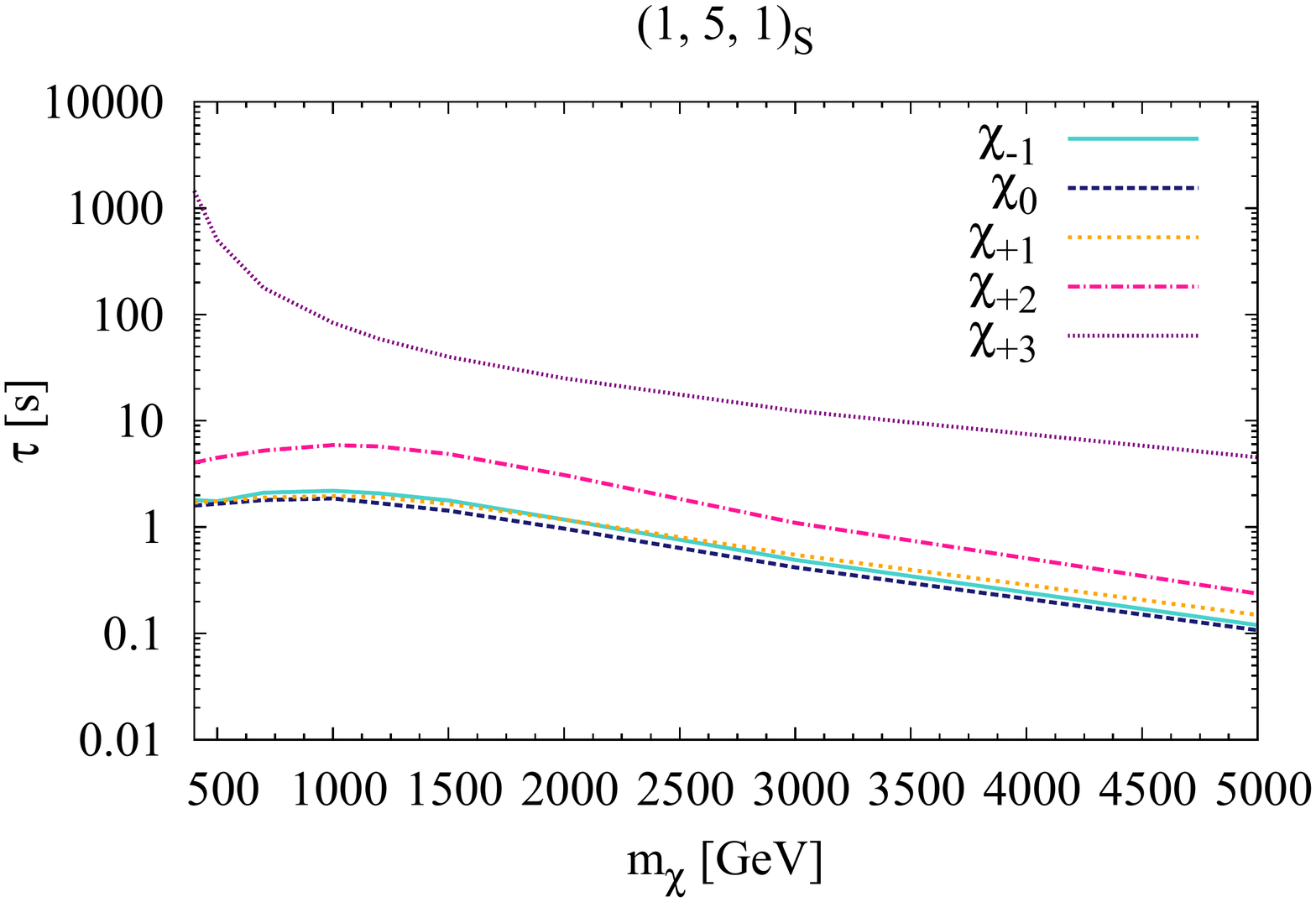}
\hspace*{-1.6cm}
\includegraphics[angle=0,width=10cm]{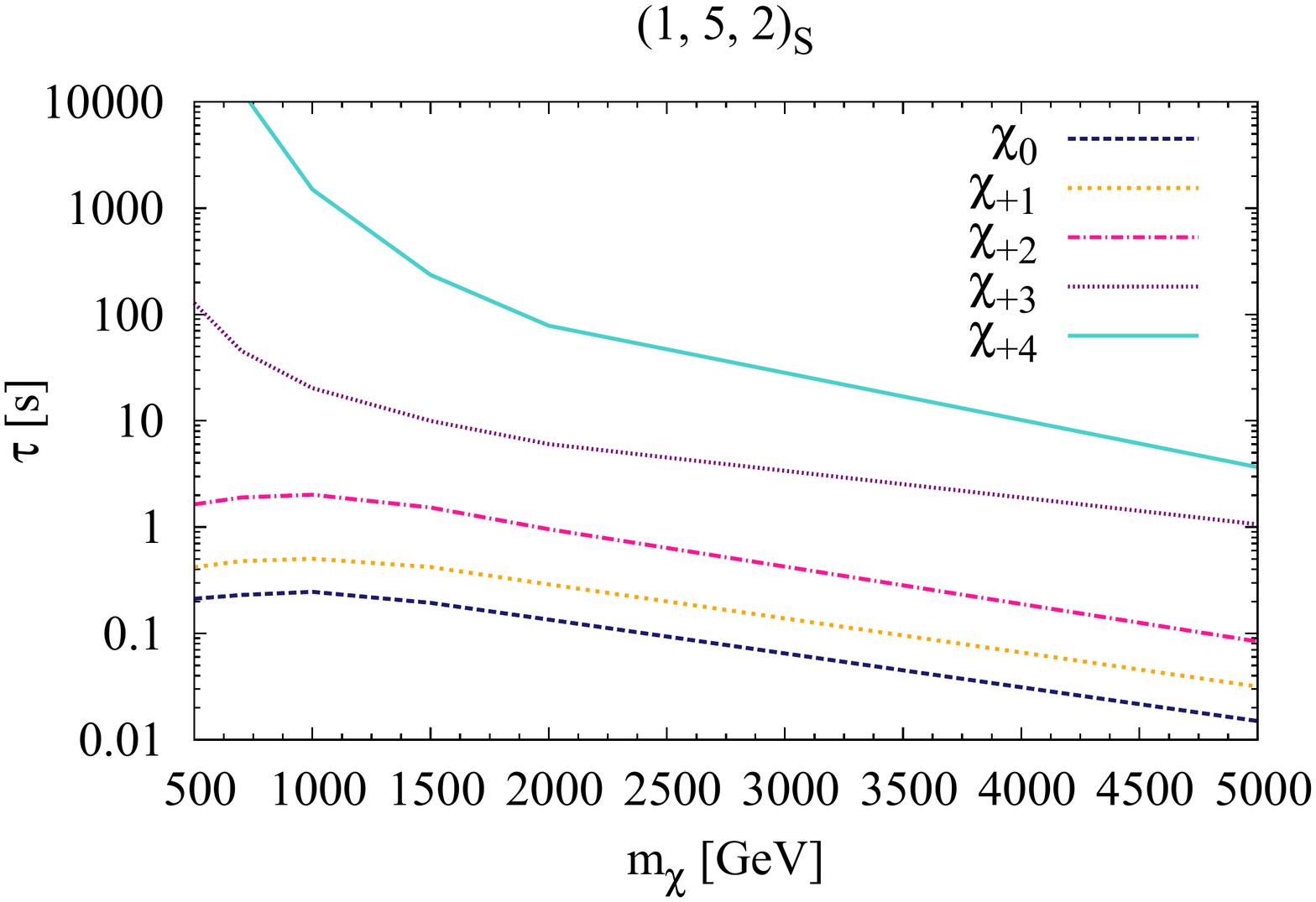} 
\hspace*{-1.8cm}
\vspace*{-1.0cm}
\caption{{\it Left:} Lifetimes of the $Q=0$ (dark blue dashed), $Q=1$ (yellow large dots), $Q=-1$ (turquoise solid), $Q=2$ (pink dash-dotted) and $Q=3$ (violet small dots)
states of the $(1, 5, 1)_S$ multiplet, assuming for each curve that the respective
component is the lightest one. {\it Right:} Same as for the left-hand side but for $(1,5,2)_S$. The turquoise  solid line corresponds to the $Q=4$ lightest state. \label{fig:lifetime_compdiffq} 
} 
\end{figure}
\par
In Fig.~\ref{fig:lifetime_compdiffq}, we show the lifetimes of all the components of the $(1, 5, 1)_S$ 
(left panel) and $(1,5,2)_S$ (right panel) 
multiplet assuming inter-multiplet mass degeneracy.
As it can be inferred from the plot, the same scaling behavior of all the components for large $m_{\chi}$ is found as expected in the SU$(2)_L$ limit. 
The lifetimes of $\chi_{+2}$,  $\chi_{+3}$ (and  $\chi_{+4}$ in case of $(1,5,2)_S$) are larger due to the fact that a smaller number of final states is available, especially at lower masses, and hence the decay widths are suppressed. 
For such long lifetimes there are potential issues with cosmology (see~\sect{bbn}). 

\subsubsection{Loop-induced decays}
\label{loopdecays}
In all the SM extensions considered in \Tables{summary1}{summary2} 
there is always an operator responsible for the decay of the new multiplet that 
is linear in $\chi$, except in the case of the
(real) scalar multiplet with SM gauge quantum numbers $(1,7,0)$. In this case 
the operator responsible for  $\chi$ decay is 
$\chi \chi \chi H^{\dagger} H.$\footnote{Different SU$(2)_L$ contractions give rise to different independent operators. In this section we consider the case where two fields $\chi$ are contracted in a $j=4$ weak isospin multiplet.} This can be understood 
by simply noticing that the SM extended with a real $(1,7,0)$ scalar has an accidental 
$Z_2$ symmetry, $\chi \rightarrow - \chi$, at the renormalizable level 
and the presence of an operator trilinear in $\chi$ clearly breaks such a symmetry.
We note that in the context of minimal DM~\cite{Cirelli:2005uq} this $d=5$ operator and its effect on the scalar septuplet lifetime have been previously overlooked. The decay only proceeds at one-loop level and, depending on the nature of the lightest particle in the multiplet, 
can result in the following final states with EW gauge 
bosons\footnote{For very large $\chi$ masses, final states containing Higgs bosons might be important as well.} 
\begin{itemize}
\item $\chi_0$:  
the possible two-body final states are $\gamma\gamma, \gamma Z, ZZ$ and $W^+W^-$. The relevant Feynman diagrams are shown in Fig.~\ref{fig:loopdecays}.
By neglecting the gauge boson masses in the final state we get
\begin{equation}
 \Gamma_{\chi_0} = \frac{857 C^2_0}{441548 \pi^5 } \frac{g^4 v^4}{\Lambda_{\rm eff}^2 m_{\chi}}=5.9 \times 10^{-8} \ {\rm s}^{-1} \left(\frac{10^{15}\, {\rm GeV}}{\Lambda_{\rm eff}}\right)^2 \left(\frac{1\, {\rm TeV}}{m_{\chi}}\right)  \, , 
\end{equation}
where $C_0 \approx -0.0966$ is a numerical factor coming from the evaluation of the relevant Passarino-Veltman functions. We observe that even for an EFT cut-off at the Planck scale, the fast decay of the neutral component of the 
septuplet effectively rules out this particular minimal scalar DM candidate~\cite{DNPN}.

\begin{figure}[t]
\centering
\includegraphics[angle=0,width=14cm]{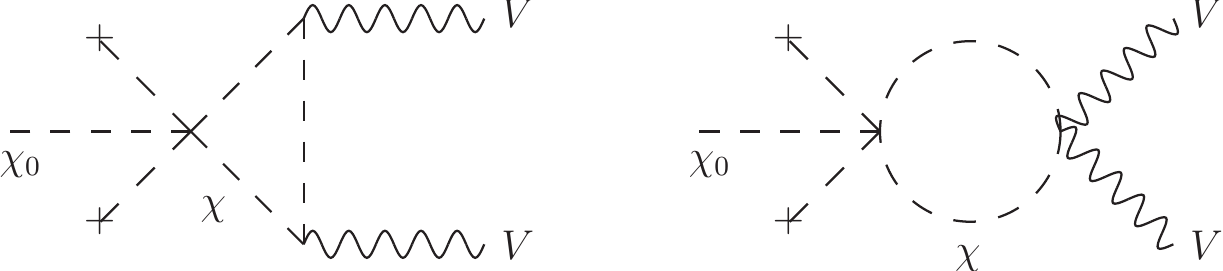}
\vspace*{0.2cm}
\caption{Feynman diagrams for the loop decay of the $\chi_{0}$ component of the $(1, 7, 0)_S$ multiplet with $VV=\gamma\gamma, \; \gamma Z, \; Z Z, \; W^+ W^-$. 
Electroweak VEV insertions are denoted by a cross. \label{fig:loopdecays}}
\end{figure}

\item $\chi_{+1}$:  
the two-body final states are $W\gamma$ or $WZ$, with a decay rate given by
 \begin{equation}
 \Gamma_{\chi_{+1}} = \frac{9 C^2_0}{34496 \pi^5 } \frac{g^4 v^4}{\Lambda_{\rm eff}^2 m_{\chi}}=7.9\times10^{-9} \ {\rm s}^{-1} \left(\frac{10^{15}\, {\rm GeV}}{\Lambda_{\rm eff}}\right)^2 \left(\frac{1\, {\rm TeV}}{m_{\chi}}\right) \, . 
\end{equation}
 \item $\chi_{+2}$: there is only a two-body decay into $WW$, yielding 
  \begin{equation}
 \Gamma_{\chi_{+2}} = \frac{9245 C^2_0}{2207744 \pi^5 } \frac{g^4 v^4}{\Lambda_{\rm eff}^2 m_{\chi}}=1.3\times10^{-7} \ {\rm s}^{-1} \left(\frac{10^{15}\, {\rm GeV}}{\Lambda_{\rm eff}}\right)^2 \left(\frac{1\, {\rm TeV}}{m_{\chi}}\right) \, .
\end{equation} 
\item $\chi_{+3}$: 
in this case there are no two-body decay channels into gauge bosons, 
while it is possible to show that if we ignore the effect of SM fermions 
$\chi_{+3}$ cannot decay into an odd number of gauge bosons.\footnote{At the one-loop level there are no contributions from SM fermions and the charge conjugation $C$ transformation is a symmetry of the gauge and scalar sectors. The selection rules for these decay channels follow from the presence of this symmetry.} 
Hence, we do expect that the leading contribution to this decay 
will be given by a final state containing four gauge bosons. 
Though we did not explicitly compute this decay rate, 
we can quote (and use in the numerical analysis) an NDA estimate given by
\begin{equation}
 \Gamma_{\chi_{+3}} = \frac{1}{3145728 \pi^7} \frac{g^8 v^4}{\Lambda_{\rm eff}^2 m_{\chi}} =1.9\times10^{-11} \ {\rm s}^{-1} \left(\frac{10^{15}\, {\rm GeV}}{\Lambda_{\rm eff}}\right)^2 \left(\frac{1\, {\rm TeV}}{m_{\chi}}\right) \, .
\end{equation}  
\end{itemize}
We end this section by noting that for $\Lambda_{\rm eff} \gtrsim 10^{15}$~GeV the loop-induced lifetimes when combined with cosmological considerations preclude the scalar septuplet to be within the kinematical reach of the LHC (see~\sect{bbn} for details).

\section{Cosmology}
\label{cosmo}

Most of the accidentally safe SM extensions are characterized by color- and weak multiplets of scalars or fermions, with weak-scale masses and no renormalizable interactions beyond their couplings to the SM gauge bosons (and the Higgs portal operators in the case of scalars). Thus they will be produced and thermalized in the early Universe, eventually freezing-out once their thermalizing interactions become slower than the Hubble expansion rate $H(T)$.  The details depend somewhat on the mass hierarchy within the $\chi$ multiplets but the decay rates of the lightest $\chi_i$ components (through higher dimensional operators) are typically much smaller than $H(T)$ at freeze-out for both weakly and strongly interacting $\chi$. We thus have effectively a two step process and we can treat freeze-out and decay separately. 

In case $\chi$ is a color singlet, the cosmological relic abundance will generically be determined by its (co)annihilations into EW gauge bosons resulting in a cosmological density of $\Omega_\chi h^2 \sim 0.01$\,. On the other hand, the final relic abundance of a colored multiplet is determined in two stages. At temperatures $T\sim m_\chi/30$ the relic abundance is determined by perturbative QCD annihilations resulting in $\Omega_\chi h^2 \sim 10^{-3} $. Then, $\chi$ undergoes a second stage of annihilation after the QCD phase transition, further reducing its relic abundance to a value roughly three orders of magnitude smaller~\cite{Kang:2006yd}. 

The $\chi$ lifetimes determine at which cosmological epoch they will decay. Such decays will involve the creation of energetic SM particles, which can produce a variety of observable effects. First, the decays of heavier multiplet components into the lightest $\chi_i$ state always happen well before nucleo-synthesis and give a negligible entropy release. On the other hand, if the lightest $\chi_i$ states can decay through $d=5$ operators, their lifetimes are at least of the order $(0.1-10^5)$~s, and  may thus affect the primordial generation of light nuclear elements~\cite{Fields:2006ga}. For longer lifetimes of the order $(10^{12}-10^{13})$~s, $\chi_i$ decays would create distortions in the thermalization of the cosmic microwave background (CMB) before recombination. Such distortions  of the spectrum by the injection of high-energy photons into the plasma lead to strong constraints~\cite{Hu:1993gc}.  Decays of $\chi_i$ after recombination can give rise to photons that free-stream to us, and are visible in the diffuse gamma ray background~\cite{Kribs:1996ac}. Observations by Fermi LAT~\cite{Ackermann:2012qk} limit the flux of these gamma rays and thus constrain such scenarios.  In general these observations of the diffuse gamma ray background rule out $\chi_i$ with lifetimes between $(10^{13}-10^{26})$~s. If $\chi_i$ only decay through $d \geq 6$ operators, they will survive to the present day. In case they are integer charged, they would act as heavy positively charged nucleons, producing anomalously heavy isotopes. A combination of  measurements places severe limits on the abundance of terrestrial heavy elements today~\cite{Burdin:2014xma}, effectively excluding such scenarios.\footnote{We note however that in principle these bounds can be evaded for $m_\chi \gg $TeV~\cite{Chuzhoy:2008zy}.} Alternatively, if their charge is a non-integer fraction of that of the electron, they are excluded by the null results of searches for fractionally charged
particles in bulk matter on Earth or meteoritic material~\cite{Langacker:2011db,Perl:2009zz}. Finally, sufficiently stable neutral $\chi_i$ can form (a fraction of) dark matter, a possibility, which has been thoroughly covered in the literature~\cite{Cirelli:2005uq,Cirelli:2007xd,Cirelli:2009uv,Cirelli:2014dsa}.
In principle, one could think about introducing an infinitesimal hypercharge ($\epsilon_Y$) which would make $\chi$ absolutely stable but still pass all the cosmological bounds. This would open up additional DM candidate scenarios like 
the complex scalars $(1,1,\epsilon_Y)$, $(1,3,\epsilon_Y)$, $(1,5,\epsilon_Y)$ or the Dirac fermions $(1,1,\epsilon_Y)$, $(1,3,\epsilon_Y)$. Representations having $Y \neq 0$ for $\epsilon_Y \rightarrow 0$ are excluded by 
direct DM searches \cite{Cirelli:2005uq}. On the other hand, higher-dimensional SU$(2)_L$ representations have a Landau pole below $10^{15}$ GeV (cf.~\Table{LPSU2vsSU3}).  We will not entertain such a possibility any further since it is a rather simple distortion of the minimal DM setup (see for instance~\cite{Cirelli:2014dsa}). We also refer the reader to existing literature for more details on experimental bounds on $\epsilon_Y$ (e.g.~\cite{Davidson:2000hf}).

We close this section with a few general comments about the possible interpretation of cosmological DM within our framework.
First of all, we note that the microscopic nature of the DM is still uncertain. For example Massive Astrophysical Compact Halo Object (MACHO) made of ordinary baryons (like black holes or neutron stars) could in principle be a viable option. 
It is known, however, that in such cases departures from the standard Big Bang theory are needed. 
The present cosmological data and various theoretical considerations 
favor the hypothesis of particle DM.
Besides the minimal DM cases, requiring DM of this type in our framework means departing from minimality. The easiest possibility then is to assume, on top of the (non DM) accidental matter state, the presence of the fermionic minimal DM multiplet at $\sim 10$\,TeV. However, this works only for some accidental matter states. In other cases extra $d=5$ operators can trigger too fast decay of the minimal DM candidate. 
Additional possibilities include the presence of extra gauge interactions where the stability of DM is again guaranteed by an accidental symmetry of the new gauge sector (for a recent work along these lines see \cite{Antipin:2015xia}) or axion DM with PQ symmetry breaking above $\Lambda_{\rm eff}$~\cite{Abbott:1982af,Dine:1982ah,Preskill:1982cy}.

\subsection{Relic abundance}

We first consider scenarios with uncolored $\chi$, where its lightest component is electrically charged. In these cases, direct searches already limit $m_\chi \gg m_Z$ (see Sect.~\ref{colorlessandcharged}) and we can compute the relevant thermally-averaged cross-sections in the SU$(2)_L$-symmetric limit. This approach is valid as long as all SU$(2)_L$ multiplet components are present in the thermal plasma. 

For inter-multiplet splittings of typical radiative size all heavier $\chi_i$ components decay into the lightest one with 
lifetimes (cf.~\fig{fig:IMWD_LifetimevsDeltam}) 
which can be comparable or even shorter than the inverse Hubble rate at freeze-out
(typically $\mathcal O(10^{-11}\,\rm s)$).
Thus the abundance of the lightest $\chi_i$ component (before itself starts decaying) is actually described by the sum of the densities of all $\chi_i$ states. And as long as $\chi_i \leftrightarrow \chi_j$ conversion rates are in equilibrium at freeze-out (which is always the case for color singlet weakly interacting $\chi$), the actual $\chi_i \to \chi_j X$ rates do not affect the total relic abundance~\cite{Edsjo:1997bg}, and the SU$(2)_L$ symmetric approximation can be justified. 

Finally, we also ignore thermal corrections. They mainly induce thermal mass splittings of the order $\Delta m_\chi \sim (g^2T)^2/m_\chi$, which can be neglected at the level of precision we are considering here~\cite{Cirelli:2007xd}. 

Due to the above approximations we can write a single Boltzmann equation that describes 
the evolution of the total abundance of all components $\chi_i$ of the multiplet as a whole. In particular, it includes all co-annihilations in the form of $\sum_{ij} \sigma_A (\chi_i\chi_j \to$ SM particles). The final $\chi$  abundance can be well approximated as~\cite{Cirelli:2005uq}
\beq
Y_\chi \equiv \frac{n_\chi(T)}{s(T)} \approx \sqrt{\frac{180}{\pi g_{\rm SM}}} \frac{1}{m_{\rm Pl} T_f \langle \sigma v\rangle}\,, ~~~ \frac{m_\chi}{T_f} \approx \ln \frac{g_\chi m_\chi m_{\rm Pl} \langle \sigma v \rangle}{240 \sqrt g_{\rm SM}}\,,
\label{eq:Ychi}
\eeq
where $g_\chi$ is the number of degrees of freedom of a whole $\chi$ multiplet including anti-particles in case of complex representations, $g_{\rm SM}$ is the number of SM degrees-of-freedom in thermal equilibrium at the freeze-out temperature $T_f$ (c.f.~\cite{Kolb:1990vq}), and $s$ is their total entropy. 
The typical freeze-out temperature is $T_f \sim m_\chi/26\ll m_\chi$, such that we can keep only the dominant s-wave (co)annihilation processes.  The relevant formulae for the corresponding thermally averaged annihilation cross-sections  $\langle \sigma v \rangle$ into $\text{SU(2)}_L \otimes \text{U(1)}_Y$ vector bosons for both scalar and fermionic $\chi$ with generic $\text{SU(2)}_L \otimes \text{U(1)}_Y$ quantum numbers can be found in~\cite{Cirelli:2005uq}. The resulting $Y_\chi $ estimates are within $10\%$ of the more complete treatment including $p-$wave annihilations and renormalization of the SM gauge couplings~\cite{Cirelli:2007xd}.  However, for $m_\chi \gtrsim 1$~TeV, the relic abundance is expected to be further reduced by $\mathcal O(1)$ non-perturbative (Sommerfeld) corrections due to the electrostatic Coulomb force effects~\cite{Hisano:2006nn}. In case of scalars, additional renormalizable Higgs portal interactions can also contribute to the annihilation cross section deferring freeze-out. In light of this our estimates of $Y_\chi$ using Eq.~\eqref{eq:Ychi} with dominant EW gauge boson contributions to $\langle \sigma v\rangle$ can be considered as upper bounds on the actual relic abundances of $\chi$.

\begin{figure}
  \begin{center}
    \includegraphics[width=.60\textwidth]{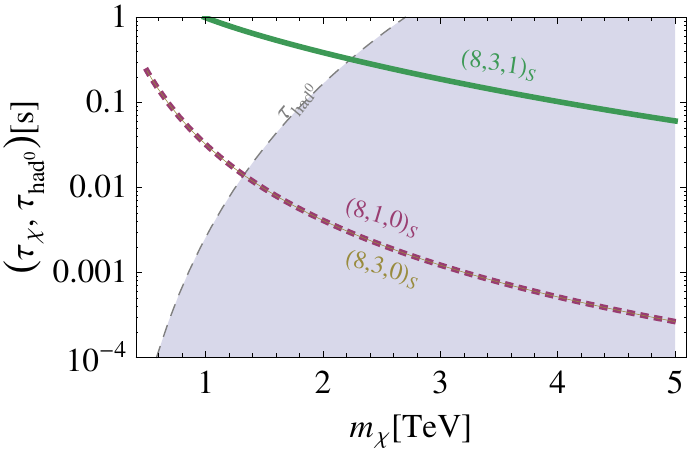}
  \end{center}
  \caption{\label{fig:lifetime} Comparison of NDA estimates for the $\chi_0$ lifetimes ($\tau_\chi$) in case of colored multiplets $ (8,1,0)_S$ (drawn in thick dashed purple), $(8,3,0)_S $ (drawn in thin brown) and $(8,3,1)_S$ (drawn in thick green) with the annihilation lifetimes of the corresponding $\chi_0$ hadronic bound states ($\tau_{\rm had^{0}}$, drawn in thin dashed gray). In the shaded region, the $\chi_0$ decay before their hadronic bound states fully annihilate. See text for details.}
  \end{figure}

In the case of colored $\chi$, one needs to consider two separate regimes of annihilation. The first era is before the QCD phase transition when all $\chi$ components are freely propagating in the QCD plasma and the annihilation cross-section can be determined using perturbative QCD. The second era is after the QCD phase transition when the heavier multiplet components have decayed to the lightest $\chi$, which in turn have become confined in color neutral bound states. The annihilation cross section in this second period turns out to be much higher than in the first, thus leading to a second period of annihilation which completely determines the final $\chi$ relic abundance~\cite{Kang:2006yd}. In particular, heavy colored particles are confined within hadronic states of typical size $R_{\rm had}\sim {\rm GeV}^{-1}$ which annihilate with a geometrical cross section yielding $\langle \sigma v \rangle \sim \pi R_{\rm had}^2 \sqrt{T_B/m_\chi}$, where $T_B \sim 180$~MeV is the temperature at which QCD confines and hadronic bound states form. 
The final $\chi$ relic abundance can thus be approximated as
\beq
Y_\chi  \sim 10^{-17} \left(\frac{R_{\rm had}}{{\rm GeV}^{-1}}\right)^{-2} \left( \frac{T_B}{180~\rm MeV} \right)^{-2/3} \left( \frac{m_\chi}{\rm TeV}  \right)^{1/2} \,,
\label{eq:YQCD}
\eeq
where we have used Eq.~\eqref{eq:Ychi} with $T_f=T_B$ and $g_{SM} \sim 15$ just below the QCD phase transition. The annihilation proceeds through intermediate excited bound states which decay by radiating away photons before annihilating into quarks and gluons~\cite{Kang:2006yd}. These processes need to be considered carefully, since such late decays to photons and hadronic jets could affect nucleosynthesis~\cite{Kawasaki:2004qu}. In case of electrically charged $\chi_i$, this process is fast with a lifetime of
\beq
\tau_{\rm had^+} \sim \frac{[\alpha_{s}(m_\chi)]^{1/2} m_{\chi}^2}{\alpha_{\rm EM} \Lambda_{\rm had}^3} \sim 3 \times 10^{-17}~{\rm s}~\left( \frac{\alpha_{s}(m_\chi)}{0.1} \right)^{1/2} \left( \frac{\Lambda_{\rm had}}{\rm GeV} \right)^{-3} \left( \frac{m_\chi}{\rm TeV} \right)^{2} \,.
\eeq 
where $\Lambda_{\rm had} \sim 1$~GeV is related to the QCD string tension $\sigma$ via $\sigma \sim \Lambda_{\rm had}^2$. On the other hand, for electrically neutral $\chi_i$, radiation of photons is loop suppressed, leading to a much longer annihilation process
\beq
\tau_{\rm had^0} \lesssim \frac{4\pi m_{\chi}^6}{\alpha^2_{\rm EM} \Lambda_{\rm had}^7} \left( \frac{T_B}{\Lambda_{\rm had}} \right)^{7/3} \sim 1~{\rm s}~ \left(\frac{m_{\chi}}{{2.7 \ \rm TeV}^{}}\right)^{6} \left(\frac{\Lambda_{\rm had}}{{ \rm GeV}^{}}\right)^{-28/3} \left(\frac{T_B}{{180 \ \rm MeV}^{}}\right)^{7/3} \,,
\eeq 
where the inequality is due to neglected non-local contributions to the decay rate. This scenario however, only applies to our cases $\chi \sim (8,1,0)_S,(8,3,0)_S $ and $(8,3,1)_S$  when the lightest component is neutral.  In Fig.~\ref{fig:lifetime} we plot the relevant lifetimes $\tau_\chi$ and $\tau_{\rm had^0}$ as a function of $\chi$ mass. 
We observe that in the low mass regime, $\tau_{\rm had^0} < \tau_\chi$ and we can use the non-perturbative result in Eq.~\eqref{eq:YQCD} to estimate the final $\chi$ abundance. However in the heavy $\chi$ limit, $\tau_{\rm had^0} > \tau_\chi$ and so $\chi$ decays before the second stage annihilation process is completed. In this case the relevant abundancies are those after the first stage of annihilation above the QCD phase transition. We can estimate them using the general SU$(N)$ annihilation cross-sections computed in~\cite{Berger:2008ti} after exchanging the relevant SU$(N)$ group invariants and correcting for the different number of degrees of freedom. In particular for the adjoint representation of QCD $\sum_{i,j,a,b} |\{ T^a, T^b \}_{ji}|^2 = 216$ and $\sum_{i,j,a,b} |[ T^a, T^b ]_{ji}|^2 = 72$. Velocity expanding the resulting $\chi\chi \to g g$ cross section we obtain (c.f.~\cite{Gondolo:1990dk})\,,
\beq
\langle \sigma v \rangle = \frac{27 \pi \alpha_s^2}{g_\chi m^2}\,,
\label{eq:sigmaQCD}
\eeq
in the conventions of~\cite{Cirelli:2005uq}. In estimating the resulting relic abundance we can safely assume that weak interactions keep $\chi_i \leftrightarrow \chi_j$ processes in equilibrium until decoupling~\cite{Edsjo:1997bg}. We have also checked that employing Eqs.~\eqref{eq:Ychi} and~\eqref{eq:sigmaQCD} and using the running $\alpha_s(2 m_\chi)$ to estimate the relic abundance reproduces the results of the full leading order perturbative QCD calculation and integration of the Boltzmann equation to within $20\%$ (in agreement with similar results for the fundamental QCD representation in~\cite{Berger:2008ti}). Non-perturbative Sommerfeld corrections are expected to lead to an  $\mathcal O(1)$ reduction in the final result and so our estimates can again be taken as upper bounds on the actual relic abundance of color octet scalar $\chi$ above the QCD phase transition. In the intermediate $m_\chi$ regime when $\tau_{\rm had^0} \sim \tau_\chi$ the actual evolution of the $\chi$ number density in the primordial plasma depends on the detailed dynamics of the neutral $\chi$ hadron annihilation and decays, the evaluation of which is beyond the scope of our study. As we show in the next section however, these details are never relevant, since they do not lead to observational constraints.

\subsection{Implications for Big Bang Nucleosynthesis}
\label{bbn}
In general, nucleosynthesis of primordial elements in the early Universe represents a sensitive probe of any metastable relic with lifetime of about $1$~s or longer~\cite{Fields:2006ga,Iocco:2008va}. The constraints come from two classes of processes: injection of very energetic photons or hadrons from decays during or after BBN adds an additional non-thermal component to the plasma and can modify the abundances of the light elements~\cite{Lindley:1984bg, Reno:1987qw, Dimopoulos:1987fz, Scherrer:1987rr, Ellis:1990nb}; in addition, if the relic particle is electromagnetically charged, bound states with nuclei may arise that strongly enhance some of the nuclear rates and allow for catalysed production of e.g.~${}^6$Li, ${}^7$Li~\cite{Pospelov:2006sc, Kohri:2006cn, Kaplinghat:2006qr}. The Standard BBN prediction for the ${}^6$Li abundance is actually significantly smaller than the observed one, so that the presence of a charged relic with appropriate lifetime can help reconciling BBN with the measured abundances of ${}^6$Li and ${}^7$Li~\cite{Bird:2007ge,Jittoh:2007fr,Jedamzik:2007cp,Jittoh:2008eq, Cumberbatch:2007me, Kusakabe:2007fu}.

In general, the decay can produce very energetic SM particles that can initiate either hadronic or electromagnetic showers in the plasma. The most stringent bounds are obtained for a relic that produces mostly hadronic showers, since electromagnetic particles like photons or electrons can thermalize very quickly by interacting with the tail of the CMB distribution until times of about $10^6$~s. In the following we will consider the constraints for relics producing a small number of energetic hadronic jets with a branching ratio $\mathcal B_{\rm had}= 1$ and $E_{\rm had} \sim m_\chi$, where $E_{\rm had}$ is the decay energy released in the form of hadronic showers  to obtain conservative upper bounds on $\chi$ number densities. This assumption is mostly valid if $\chi$ is colored  (in particular in this case always $\mathcal B_{\rm had}= 1$ and  $E_{\rm had} \gtrsim m_\chi/2$), while $\mathcal B_{\rm had} < 1$ is expected for non-colored $\chi$. Then the hadronic BBN bounds are relaxed accordingly by a factor $1/\mathcal B_{\rm had}$. Finally, for lifetimes $\tau_\chi \gtrsim 10^4$~s, electromagnetic interactions start having a significant effect and the bounds above $\tau_\chi \gtrsim 10^7$~s become effectively independent of $\mathcal B_{\rm had}$. 

In practice there are three regions of  lifetimes as discussed in~\cite{Kawasaki:2004qu}: for lifetimes $0.1~{\rm s} \lesssim \tau_\chi \lesssim 100$~s the dominant effect is the interconversion between protons and neutrons, that changes the ${}^4$He abundance by overproducing it; at longer lifetimes $100~{\rm s} \lesssim \tau_\chi \lesssim 10^7$~s hadrodissociation is the most efficient process and the bounds come from the non-thermal production of lithium and deuterium; finally at late times $10^7~{\rm s} \lesssim \tau_\chi \lesssim 10^{12}$~s photodissociation caused both by direct electromagnetic showers and by those generated by the daughter hadrons starts to dominate and result mainly in the overproduction of ${}^3$He.

In the following we use the results from the general analysis of~\cite{Kawasaki:2004qu} for the second and the third lifetime regions. In particular the bound coming from the abundance of ${}^3$He scales as $1/m_\chi $ and does not depend on $\mathcal B_{\rm had}$ (we neglect the decay energy released into neutrinos, which is always expected to be a small fraction of $m_\chi$). On the other hand due to the Li anomaly, in the second region we only consider bounds on $Y_\chi$ coming from the deuterium to hydrogen abundance ratio (D/H), which scale roughly as $E_{\rm had} ^{-1/2}$. Finally, we note that a charged thermal relic with $\tau_\chi \sim 10^{2}-10^{3}$~s and abundance just  below the D/H bound may (partly) ameliorate the standard BBN Lithium problems (c.f.~\cite{Fields:2011zzb}). 

The value of the observed ${}^4$He abundance $Y_p \equiv 4 (n_{\rm He}/n_{\rm H})/(1+4 n_{\rm He}/n_{\rm H}) \approx 2(n_n/n_p)/(1+n_n/n_p)$  which dominates the constraints in the first lifetime region has been updated since the analysis of~\cite{Kawasaki:2004qu} and currently reads $Y_p = 0.250(3)$~\cite{Iocco:2008va} to be compared with the prediction of standard BBN of $Y_p^{\rm SBBN} = 0.2483(5)$~\cite{Steigman:2007xt}. The bounds from~\cite{Kawasaki:2004qu} which assumed significantly smaller $Y_p$ thus need to be re-evaluated.  For this purpose we numerically solve the relevant Boltzmann equations~(c.f.~\cite{Kolb:1990vq})
\begin{align}
- H(T) T \frac{d y_\chi}{d T} & = - \frac{y_\chi}{\tau_\chi} \,, \nonumber\\
- H(T) T \frac{d y_n}{d T} & = - \lambda_{np} y_n + \lambda_{pn} y_p - \frac{y_n}{\tau_n} - \frac{\mathcal B_{\rm had}}{\tau_{\chi}} y_\chi K_\chi  \,, \nonumber\\
- H(T) T \frac{d y_p}{d T} & = - \lambda_{pn} y_p + \lambda_{np} y_n + \frac{y_n}{\tau_n} + \frac{\mathcal B_{\rm had}}{\tau_{\chi}} y_\chi K_\chi\,,
\end{align}
where $y_i \equiv n_i/n_b$ and $n_b$ is the baryon number density (we use $\eta \equiv n_b/n_\gamma = 6.1 \times 10^{-10}$), $H(T)=\pi (T^2/m_{\rm Pl}) \sqrt{g_{\rm SM} / 90}$ is the Hubble rate 
and $\tau_n = 880(1)$~s is the neutron lifetime. We have furthermore defined $K_\chi \equiv K_{n\to p} - K_{p\to n}$ and $\lambda_{np} \simeq \lambda_{pn} \exp(1/y) \simeq (1443/\tau_n) y^3 (y+0.25)^2$~\cite{Lopez:1998vk,Mukhanov:2003xs}, where $y=T/Q$ and $Q=1.293$~MeV is the neutron-proton mass difference. We have checked that this approximate form of $\lambda_{np}$ and $\lambda_{pn}$ reproduces the final results using exact numerical integration of the weak nucleon conversion rates (c.f.~\cite{Weinberg:2008zzc, Lopez:1998vk}) to better than $0.5\%$. Finally, for the catalyzed nuclear conversion rates we employ the formulae for $K_{N\to N'}$ including all the numerical inputs as defined in~\cite{Kawasaki:2004qu}. At temperatures much bigger than $Q$ we expect $y_n = (1-y_p) = 1/(1+\exp(Q/T))$ and $y_\chi$ given by its thermal relic abundance $\bar y_\chi$. Finally, the resulting ${}^4$He abundance is well determined by $y_{n,p}$ at ($T\simeq 8.5 \times 10^9$~K) when BBN begins~\cite{Steigman:2007xt,Pospelov:2010hj}. As a cross-check of our approach, we have determined $Y_p$ in absence of $y_\chi$ and obtained $Y_p(\bar y_\chi=0) = 0.243$, which is consistent with the expected precision in light of our approximations. In particular, neglected higher order effects would increase $Y_p$ by $2\%$~\cite{Lopez:1998vk}, reproducing the standard BBN result. In setting constraints we use our estimates only to compute the deviations of $Y_p$ from the standard BBN value $\Delta Y_p = Y_p(\bar y_\chi) - Y_p(\bar y_\chi=0) $ and compare $Y_p^{\rm SBBN} + \Delta Y_p$ to the $2\sigma$ region of the observed $Y_p$ value. As discussed above, the effects of $\chi$ decays on $Y_p$ scale as $1/\mathcal B_{\rm had}$, and for $E_{\rm had} \gtrsim 100$~GeV (in the form of a fixed number of hadronic jets) also approximately as $E_{\rm had}^{-1/3}$. 

\begin{figure}[!t]
  \begin{center}
    \includegraphics[width=.49\textwidth]{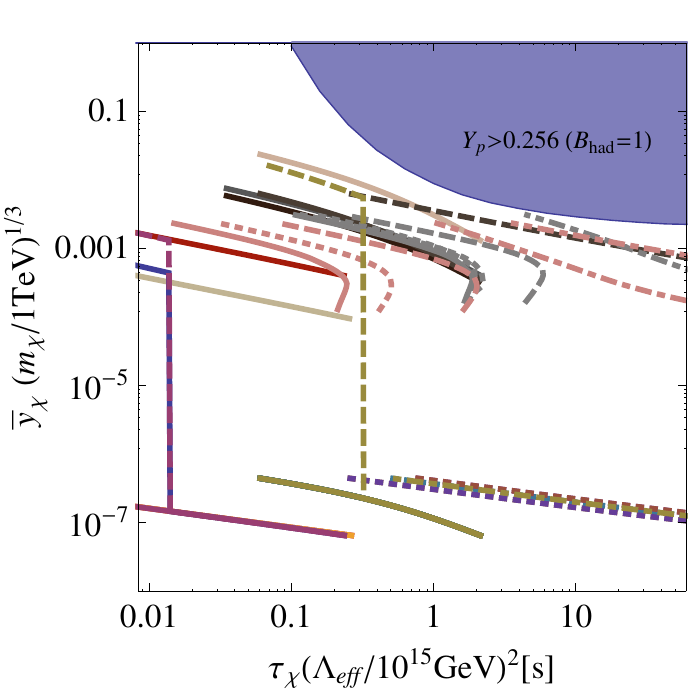}~~
    \includegraphics[width=.49\textwidth]{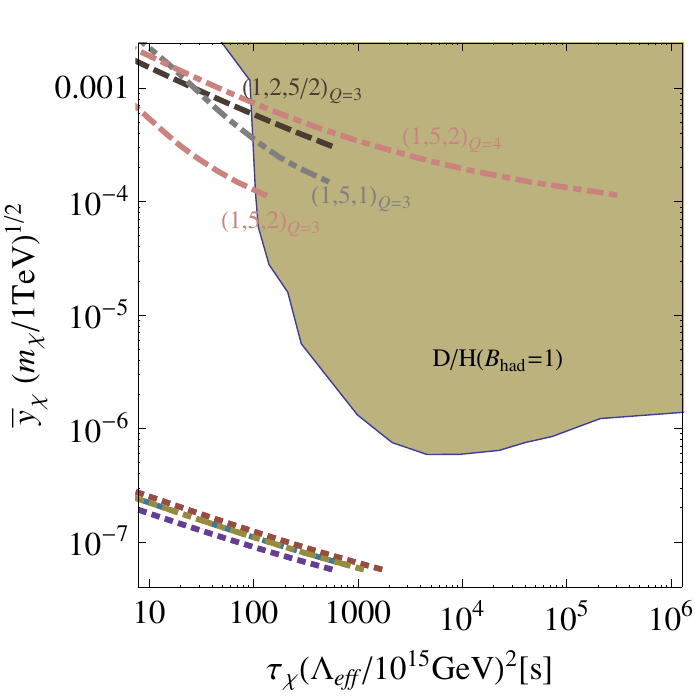}
  \end{center}
  \caption{\label{fig:BBN} Comparison of BBN $Y_p$ (left hand panel) and D/H (right hand panel) constraints on the abundances and lifetimes of metastable hadronically decaying particles with the corresponding estimates for the cases of viable $\chi$ multiplets. Each line, going from bottom to top, corresponds to the mass range $0.5~{\rm TeV} < m_\chi < 5$~TeV. The only explicitly labeled examples are $(1,5,2)_{S,Q_{\rm LP}=3}$, $(1,5,2)_{S,Q_{\rm LP}=4}$, $(1,5,1)_{S,Q_{\rm LP}=3}$ and $(1,2,5/2)_{S,Q_{\rm LP}=3}$, which are potentially constrained. The case $(1,7,0)$ is not shown as its lifetime is longer than $10^6$~s for $0.5~{\rm TeV} < m_\chi<5$~TeV and $\Lambda_{\rm eff} =10^{15}$~GeV.}
\end{figure}

\begin{figure}[!t]
  \begin{center}
    \includegraphics[width=.60\textwidth]{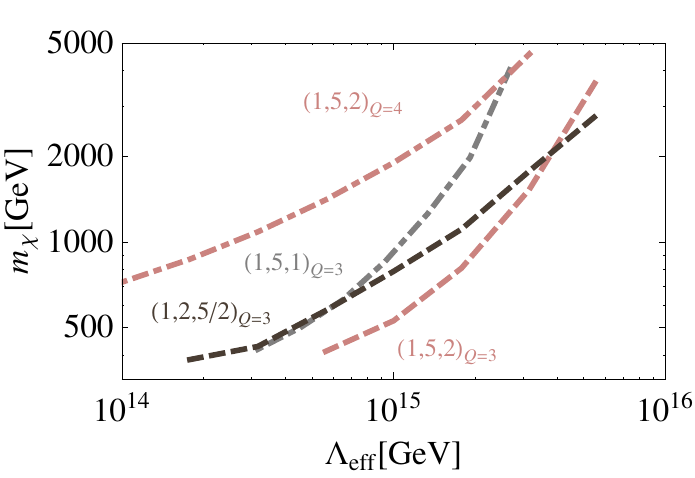}~~
  \end{center}
  \caption{\label{fig:BBNScale} Dependence of the D/H bound on the $\chi$ mass as a function of the EFT cut-off scale $\Lambda_{\rm eff}$ suppressing the relevant $d=5$ decay mediating operators. See text for details.}
\end{figure}

We finally determine the upper bound on the possible contributions of $\chi$ decays to $Y_p$ by fixing $\mathcal B_{\rm had}=1$ and assuming $\chi$ decay to two hadronic jets.  The resulting effects then scale as $m^{-1/3}_\chi$ and we can use a single reference value to constrain the abundances of $\chi$ at different $m_\chi$. 
As discussed above, this leads to a conservative $\mathcal O(1)$ overestimate of the actual $Y_p$ constraints for decays of $\chi$ involving also uncolored final states. The comparison of the lifetimes and relic abundances of all $\chi$ candidates from Tables~\ref{summary1} and~\ref{summary2} in the mass range of $0.5~{\rm TeV} < m_\chi < 5$~TeV with the $Y_p$ bound estimated in this way is shown in Fig.~\ref{fig:BBN} (left hand side). 
Notice the almost discontinuous drop of the abundance 
for some representations. This is due to the fact that 
for colored multiplets featuring a neutral component there 
is a qualitative change of behaviour when the 
lifetime of the particle becames smaller than the annihilation lifetime 
of the associated hadronic bound state (cf.~\fig{fig:lifetime}).
We observe that all the $\chi$ are consistent with the $Y_p$ constraint. Also, most of the candidates have lifetimes shorter than $\sim 10$~s (for $\Lambda_{\rm eff}\sim 10^{15}$~GeV), so that no further bounds from BBN processes at later times can be derived. The only exceptions are the cases in Table~\ref{cascadedimgeq5} where the lightest $\chi$ component can only decay through long cascades involving off-shell heavier components 
and $W$ bosons as well as $(1,7,0)_S$ decaying exclusively through loop-induced processes. For these cases the D/H bound applies as shown in Fig.~\ref{fig:BBN} (right hand side). In particular, while all the colored multiplets (including those decaying with cascades) are consistent with this constraint due to their low relic abundance after the second stage of strong annihilations, all the long-lived uncolored cases are in general constrained. In the relevant region of relic abundances and lifetimes, the bound turns out to be insensitive to the exact $\chi$ relic abundance or decay mode and so even our crude estimates suffice to extract fairly robust lower bounds on $\chi$ masses. They are shown in Table~\ref{table:BBNbound} for a fixed value of $\Lambda_{\rm eff}=10^{15}$~GeV, while the $\Lambda_{\rm eff}$ dependence is shown explicitly in Fig.~\ref{fig:BBNScale}.  

We finally note that it is close to these D/H exclusion bounds where the primordial Lithium problem might be addressed by the presence of $\chi$ listed in the first three rows of Table~\ref{table:BBNbound}. A detailed exploration of this possibility goes however well beyond the scope of the present analysis and we leave it for future study.

\begin{table}[htbp]
\centering
\begin{tabular}{@{} |c|c|c|c| @{}}
\hline
Spin  & $\chi$ &  $Q_{\rm{LP}}$ & Mass bound [GeV] \\ 
\hline
%\hline
0 & $(1, 2, 5/2)$ &  $3$  & 790   \\  
0 & $(1, 5, 1)$ &  $3$  & 920   \\  
0 & $(1, 5, 2)$ &  $3,4$  & 530, 1900   \\  
0 & $(1,7,0)$ &  $0,1,2,3$  & $\gg 5000$   \\  
\hline
  \end{tabular}
  \caption{\label{table:BBNbound} BBN bounds on the masses of long lived $\chi$ multiplet components assuming fixed $\Lambda_{\rm eff}=10^{15}$~GeV.   }
\end{table}

\section{Collider phenomenology}
\label{lhc}
In this section, we discuss the collider phenomenology of the new states 
and give bounds on the masses of the new particles.
The states of Table \ref{summary1} can be ordered into two classes: 
the ones which can decay by renormalizable interactions and the ones which 
decay via effective operators. In case the states decay via renormalizable interactions,
they can be detected via their decay products. We will shortly comment on the renormalizable 
cases in \sect{rencases}, mostly referring to the existing literature.
If they only decay via effective operators
they are rather long-lived and can eventually leave the detector before decaying.
 If the new particles are uncolored, the signature depends on whether the particle is charged or not.
 A summary of the different mass bounds for the uncolored cases can be found
 in Table \ref{summarybounds}.  
\par 
All the extra colored states,
given in Table \ref{summary2}, can only decay via $d=5$ operators and are hence long-lived.
 They hadronize and build exotic new mesons or baryons. We will discuss them in \sect{phenocolored}.
\par
The production of the new exotic fermions and scalars proceeds via Drell-Yan processes. Throughout this section, we
use the LO Drell-Yan production cross sections.
Formulae are given e.g.~in Ref.~\cite{DelNobile:2009st}. The cross section for scalars is in general more than one order of magnitude smaller than for fermions, which explains the lower exclusion bounds on the scalars.

\begin{table}[htbp]
\centering
\begin{tabular}{|c|c|c|c|}\hline
Spin & $\chi$ & $Q_{\textrm{LP}}$ & Mass bound [GeV]\\ \hline
0 & $(1, 2, 3/2)$ & 1, 2 & 430, 420\\ 
0 & $(1, 2, 5/2)$ & 2, 3 & 460, 460 \\ 
0 & (1, 5, 0) & 0, 1, 2 & 75, 500, 600 \\ 
0 & (1, 5, 1) & -1, 0, 1, 2, 3 & 640, $50^\star$ (85), 320, 490, 600\\ 
0 & (1, 5, 2) & 0, 1, 2, 3, 4 & 85, 530, 410, 500, 570\\ 
0 & (1, 7, 0) & 0, 1, 2, 3    & 75, 500, 600, 670 \\ 
1/2 & $(1, 4, 1/2)$ & -1 & 860 \\ 
1/2 & $(1, 4, 3/2)$ & 0 & 90 \\ 
1/2 & (1, 5, 0) & 0 & 95 \\ \hline
\end{tabular}
 \caption{\label{summarybounds} LHC-I/LEP summary bounds for uncolored accidental matter multiplets 
 decaying via $d \geq 5$ operators. Bounds on the neutral particles are given under the assumption of very 
 small mass splitting to the $|Q|=1$ component. The exclusion bound in braces corresponds to the case where the next-to-LP
has $Q=-1$ instead of $Q=1$. $^\star$A stronger exclusion bound, 
depending on the size of the portal coupling $\alpha$ (see \eq{potentialXH}), 
can be obtained from the Higgs data.}
\end{table}

\subsection{Renormalizable cases}
\label{rencases}

Among the extra multiplets which preserve the flavor group of the SM, compatibly with 
cosmology and a cut-off scale of $\Lambda_{\text{eff}} \approx 10^{15}$ GeV, we identified four 
states which decay via renormalizable interactions, namely  
$(1,1,0)_S$, $(1,3,0)_S$, $(1, 4, 1/2)_S$ and $(1, 4, 3/2)_S$. 
In all of these cases, the new state acquires a VEV. With the only exception of the SM singlet, 
these VEVs must be small in order to comply with EW precision measurements. 
For $\mathcal{O}(1)$ couplings in the scalar potential and barring fine-tunings 
this implies $m_\chi \gtrsim 2 - 20$ TeV, cf.~\sect{scalpot}. 
\par
The gauge singlet can sizeably mix with the Higgs boson. 
Such mixing is constrained by the current Higgs data, 
see e.g.~Ref.~\cite{Englert:2014uua}. 
The triplet and quadruplet scalar multiplets can only have a very small 
mixing with the Higgs boson due to its effects on EW precision observables. 
Nevertheless, their charged components can modify the Higgs to $\gamma\gamma$ and to $Z\gamma$ rates 
by their loop contributions. Whether these loop contributions suppress or enhance the diphoton rates depends on the sign of the couplings in the scalar potential \cite{Carena:2012xa, Picek:2012ei}. Finally, masses of the new neutral scalars below 62.5 GeV can be probed by the invisible Higgs boson width. For more details  see~\sect{sec:Higgs}. 
\par
Low masses of the triplets and quadruplets can be constrained by the $Z$ width (see~\sect{neutralLP} for more details). Their charged components can also be directly detected. They decay into vector bosons or via cascades into vector bosons and the (off-shell) neutral components of the multiplet. The coupling to two vector bosons is proportional to the VEV of the multiplet. Apart from searches for singly charged Higgs bosons, searches for multiple charged Higgs bosons can provide a distinctive
probe for large scalar multiplets.  
By now, searches for doubly charged Higgs bosons~\cite{ATLAS:2012hi, Chatrchyan:2012ya} have only been performed for decays of the charged scalars into fermions, as in e.g.~the case for models with $Y=1$ triplets 
\cite{Melfo:2011nx} and quadruplets with additional vector-like matter for seesaw mass generation of neutrinos~\cite{Ren:2011mh, Babu:2009aq}. Bounds on doubly charged Higgs bosons decaying to $W^{\pm}W^{\pm}$ can be obtained by reinterpreting SUSY searches for dileptons, missing energy and jets, and can exclude masses of the doubly charged scalars up to roughly 200 GeV at $\sqrt{s}=7\text{ TeV}$ for SU$(2)_L$ triplets~\cite{Englert:2013wga}.

\subsection{Colorless and charged LP}
\label{colorlessandcharged}
Charged stable particles will undergo charge exchange with the detector material.
For masses larger than 100 GeV the time of flight till to the outer detector is significantly
larger than for lighter objects such as muons. They can hence be distinguished 
by their longer time of flight and by their anomalous energy loss in the detector.
The energy loss in the detector is described by the Bethe-Bloch formula and depends on the speed
and the charge of the particle. Searches for such ionizing tracks have been performed in Refs.~\cite{Barate:1997dr, Abreu:2000tn,Achard:2001qw, Abbiendi:2003yd, Aktas:2004pq, Abazov:2008qu, Aaltonen:2009kea, Abazov:2012ina, Khachatryan:2011ts, Aad:2011mb, Aad:2011hz, Chatrchyan:2012sp, Aad:2013pqd,Chatrchyan:2013oca}.
The strongest bounds come from the CMS search of Ref.~\cite{Chatrchyan:2013oca}, where the exclusion limits on the production cross sections of fractionally, singly and 
multiply charged particles are presented assuming vanishing quantum numbers under SU$(2)_L$. Hence, 
in order to use the results of Ref.~\cite{Chatrchyan:2013oca} they need to be recast. Reference~\cite{CMS-PAS-EXO-13-006} gives tabulated efficiency values in terms of the transverse momentum, the pseudo rapidity and the velocity $\beta$ of the heavy charged particle. These can be used to reinterpret the results of Ref.~\cite{Chatrchyan:2013oca}  without running a full detector simulation. 
\par
In order to compute the efficiencies, the models were implemented into {\tt Madgraph 5} \cite{Alwall:2011uj} with the help of {\tt FeynRules} \cite{Alloul:2013bka}. The cross sections, computed at LO at the scale $Q=\sqrt{\hat{s}}$ using the MSTW2008 \cite{Martin:2009iq} parton distribution functions, were than rescaled by a factor accounting for the change in the efficiencies with respect to the cases considered in Ref.~\cite{Chatrchyan:2013oca}. 
\par
We find that the efficiencies for fermions with non-vanishing SU$(2)_L$ quantum numbers barely change compared to the case with $T_3=0$. The exclusion limits from Ref.~\cite{Chatrchyan:2013oca} on the cross section can hence be applied naively. For scalars, the efficiencies change slightly compared to the fermions. For masses of 300 GeV the efficiency is slightly smaller than the one for fermions, for 800 GeV it is roughly $15\%$ larger. For the cases we considered, the efficiency for scalars is always within $3\%$ of the case with vanishing quantum SU$(2)_L$ quantum numbers.
\par
In order to derive exclusion limits for the scalars, we adopt the following procedure. We compute the cross section and derive an approximate bound using the 95\% C.L. upper limits
given in Ref.~\cite{Chatrchyan:2013oca}. With this approximate bound at hand, we compute at the naive bound the efficiencies for the scalar and compare it to the efficiencies of a fermion with $T_3=0$. The results can then be recast by the appropriate factor. We note, however, that such refined  bounds  are  in good agreement with the results obtained naively. 
\par 
Even though the efficiency values in Ref.~\cite{CMS-PAS-EXO-13-006} are given for the singly charged analysis only, we apply the same procedure to the $|Q|>1$ case to check whether also here, the naive method gives sensible results, as the basic cuts in both analysis are the same. Indeed we find also here that within the precision of our results, the naive estimate is very good. The results can be found in Table~\ref{summarybounds}. The limits for charged fermions are stronger than for scalars due to the larger production cross section. The weakest exclusion limits are obtained for $T_3=0$.
 
Before concluding this subsection let us mention the recently approved LHC experiment MoEDAL \cite{Pinfold:2009oia,Acharya:2014nyr}, 
whose target is the study of new physics phenomena (e.g.~magnetic monopoles) 
which manifest themselves through the presence of 
highly-ionizing particles. 
In particular, the nuclear track detectors of MoEDAL are sensitive to particles with $|Q|/\beta \gtrsim 5$, 
where $Q$ is the charge and $\beta$ is the velocity of the particle in units of the speed of light.
For our framework, with $Q$ ranging from $1$ to $4$ (cf.~\Table{summarybounds}), 
the discovery potential of the MoEDAL experiment will be relevant at low values of the $\beta$ distribution.

\subsection{Colorless and neutral LP}
\label{neutralLP}
The search for stable (on detector scale) neutral and colorless particles is very challenging at the LHC. Limits can either be set directly on the mass of the neutral particle as e.g.~by mono-x searches or from constraints on the invisible $Z$ width 
or, indirectly, by giving bounds on the mass of the second lightest particle of the multiplet, such as in disappearing track signatures. Let us discuss in turn all these possibilities:
\begin{itemize}
\item {\bf Mono-x searches}
\newline
Neutral stable particles are searched for at the LHC in mono-x searches, in which large missing energy is accompanied by a radiation of an additional high-energetic particle x, where ``x'' can stand for a jet, a photon, a $W$ or $Z$ boson, a top quark or a Higgs bosons. 
Nevertheless, we find that the monojet searches of Ref.~\cite{Khachatryan:2014rra}, 
which potentially have the strongest reach \cite{Zhou:2013fla}, are not sensitive to our states yet. 
Similar results were found for instance in Ref.~\cite{Cirelli:2014dsa}, in the case of 
a fermionic $(1,3,0)$ multiplet.  
Monojet searches can, however, become sensitive at 14 TeV \cite{Cirelli:2014dsa}. 

\item {\bf Invisible Z width}
\\
At LEP, the $Z$ boson width was determined with high accuracy \cite{Alcaraz:2006mx, Agashe:2014kda}. This measurements set a tight bound on new physics contributions to the invisible $Z$ width at the level of 
$\Gamma^{\rm new}_{\rm inv}< 2 \text{ MeV}$. This hence excludes charged particles, or particles with non-trivial SU$(2)_L$ quantum numbers, up to the kinematic bound for the $Z\to \chi \chi^\dag$ decay, meaning that masses $m_{\chi} \lesssim 45\text{ GeV}$ are excluded.

\item {\bf Disappearing tracks}
\\
Disappearing tracks can be observed at the LHC if a rather long-lived charged particle decays within the sensitive volume into a neutral particle and a soft pion, which is not detected. 
The strongest limits on these searches \cite{Aad:2013yna}, are sensitive 
to lifetimes of the charged particle between $0.1$ ns and $10$ ns. 
We checked whether the typical lifetimes for our particles lie within this range. 
It turns out, however, that for all fermionic states of Table \ref{summary1} with a lightest neutral state, the mass splitting between the neutral component and the charged component is always so large, that the lifetime is smaller than 0.1 ns 
(cf.~\fig{fig:IMWD_LifetimevsDeltam}). This is due to the fact that the radiative mass splitting increases with the hypercharge and the SU$(2)_L$ quantum number. For the scalar states featuring a lightest neutral component and which do not decay through renormalizable interactions, the same argument holds. In addition, the mass splitting does not need to be purely radiative 
but a larger mass splitting can also stem from the potential term. 

\item {\bf LEP bounds on charginos}
\\
The LEP experiments set bounds on charginos that are nearly mass degenerate with the lightest neutralino. These bounds can be reinterpreted for our purposes in order to derive limits on the mass of the lightest neutral particles, 
since we showed in \sect{massspectra} that the next-to-LP has 
always charge $Q_{\rm LP} \pm1$. References~\cite{Heister:2002mn, Abreu:2000as, Acciarri:2000wy, Abbiendi:2002vz} cover a mass splitting $\Delta m$ between $200\text{ MeV}\lesssim \Delta m \lesssim 5\text{ GeV}$ and are based on soft events with an initial state radiated photon. In order to estimate the limits for the case where the LP of the multiplet is neutral, we took the OPAL results of Ref.~\cite{Abbiendi:2002vz}. There, the results were given in terms of a 95\% C.L. upper limit on the cross section. We implemented the models into {\tt MadGraph 5} with the help of {\tt FeynRules}~\cite{Alloul:2013bka}, computed the cross section values  for the $|Q|=1$ charged component of the multiplet, and compared them to the given limits in Ref.~\cite{Abbiendi:2002vz}. In order to verify this procedure, we computed the efficiencies for example points on parton level, using {\tt MadAnalysis}~\cite{Conte:2012fm}. 
For the fermionic states we found that the efficiencies are basically unchanged compared to the chargino case. For the scalars they turned out to be a bit reduced, which is however not relevant given the precision to which we estimate the bounds. The hence obtained limits on the charged components are given in 
\Table{summarybounds}. At the accuracy we are working this essentially corresponds 
to the bounds on the neutral components, which are obtained after subtracting the small mass splitting.
\\
For the case of the scalar $(1,5,1)$ multiplet, either the $+1$ charged or the $-1$ charged component can be the next-to-LP. These two cases lead to different exclusion bounds. The exclusion bound from the $Q=1$ state being the second-lightest component is much smaller due to the smaller production cross section, as this state corresponds to $T^3=0$. In such a case a stronger bound can come from the Higgs invisible width. 
\par
What about mass splittings larger than $5$ GeV not covered by the chargino search of Ref.~\cite{Abbiendi:2002vz}?
If the neutral LP of a scalar multiplet is the component with the smallest/largest isospin, 
then the mass splitting between the $|Q|=1$ next-to-LP and the neutral LP 
can also be larger than $5$ GeV. This is not true for a generic value of the isospin $-j < I < j$, 
as for the neutral state to be the lightest a cancellation between the tree-level and radiative mass splitting is 
required. In particular, the only case in which we have to consider a mass splitting larger than 5 GeV is 
for $(1, 5, 2)_S$. 
Even in such a case, however, mass splittings larger than about $20$ GeV 
are excluded by EW precision observables 
(cf.~\sect{scalpot}).
\par
Searches for charginos decaying into neutralinos and $W$ bosons (with the $W$s decaying hadronically, semileptonically or leptonically) and for $\Delta m > 5\text{ GeV}$ were performed also at 
LEP \cite{Abbiendi:2003sc}.\footnote{LHC searches for pair production of charginos, 
with the charginos decaying to $W$s and neutralinos, are not yet sensitive to such low mass splittings.} 
In the $(1, 5, 2)_S$ case, for mass splittings around 5 GeV the estimated bounds on the $|Q|=1$ particle decaying to the neutral state turn out to be weaker (by roughly $10$ GeV) than the ones given in Table \ref{summarybounds}. For mass splittings larger than 15 GeV the limits become tighter (by around 5 GeV). A more detailed analysis is however beyond the scope of this paper.

\end{itemize}

\subsection{Colored LP}
\label{phenocolored}
The description of long-lived colored particles is complicated by the effect of non-perturbative QCD interactions. 
In fact, in all the cases of \Table{summary2} the decay of the new states is induced by a $d=5$ operator and the lifetimes are long enough that heavy colored particles hadronize before decaying. The theoretical description of the hadron formation and of the nuclear interactions of such states with matter represents the main source of uncertainty. 
In this section, we will briefly describe the various steps for the path of our new states, from their production to their escape from the detector or, in case the initial velocity $\beta$ is small enough, to their eventual stopping and decay inside the detector. We finally conclude by looking at the recent LHC results that are tuned to the case of QCD bound states of SUSY particles with quarks and gluons.
We refer to \cite{Fairbairn:2006gg} for a review on various phenomenological aspects of stable massive particle at colliders.

\begin{description}
 \item[Production.] 
 The production mechanism of the new particles in our framework is determined by the color quantum number and by the value of the mass. 
 We are interested in the fundamental, the two indices symmetric and the adjoint representations of SU$(3)_c$ and we denote such cases respectively as $C_3$, $C_6$ and $C_8$. 
 At the renormalizable level the presence of a U(1) or a $Z_2$ accidental symmetry guarantees these states to be pair produced. 
 The fate of the produced state crucially depends on the velocity at the production time. Relativistic particles will lose energy throughout the detector but eventually escape it, while slow particles will be stopped in the detector and decay at a later time. Typical velocity distributions are displayed in \fig{distributions}. 
\begin{figure}[t]
\centering
\hspace*{-1.6cm}
\includegraphics[angle=0,width=9.9cm]{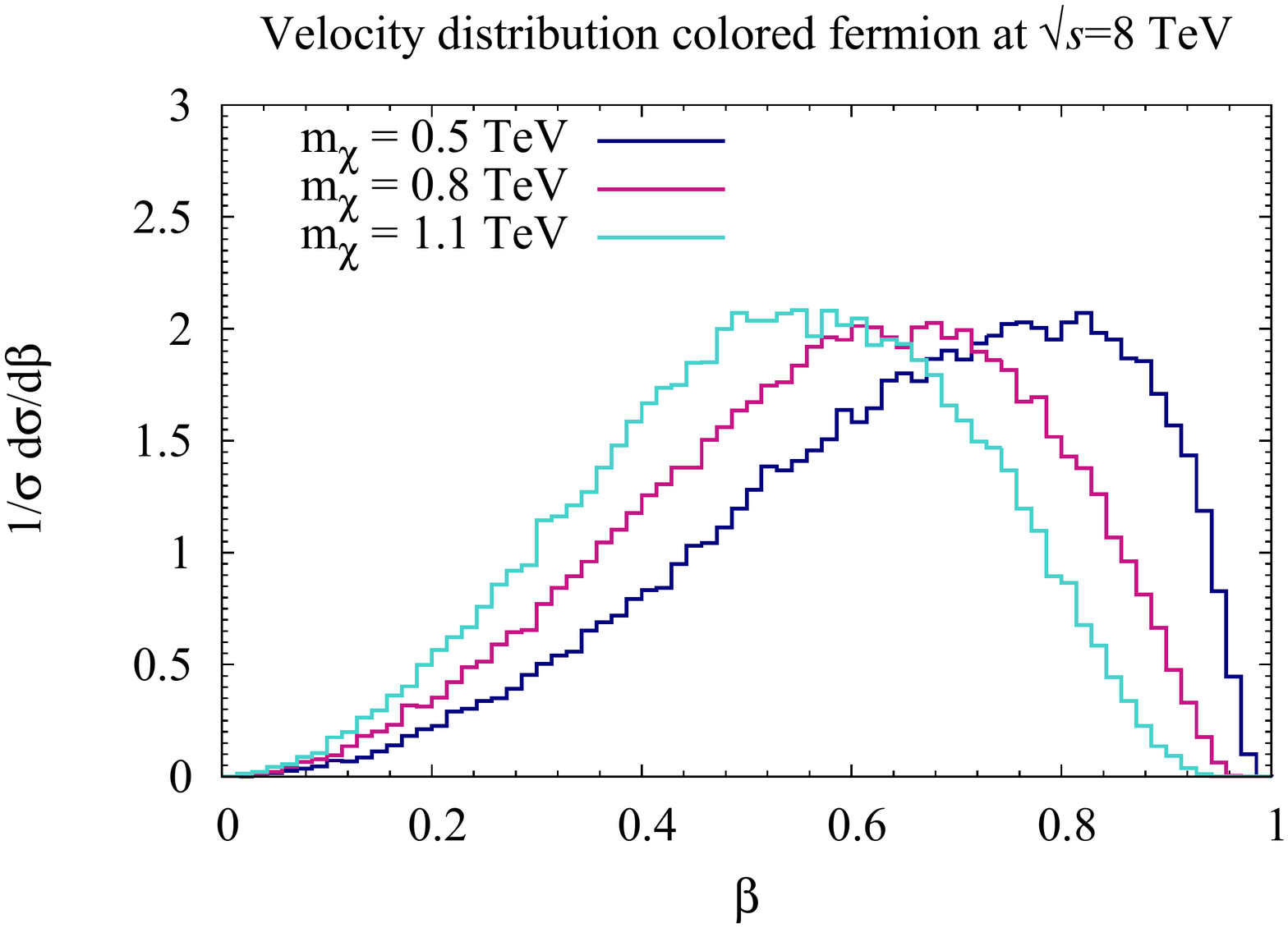}
\hspace*{-1.6cm}
\includegraphics[angle=0,width=9.9cm]{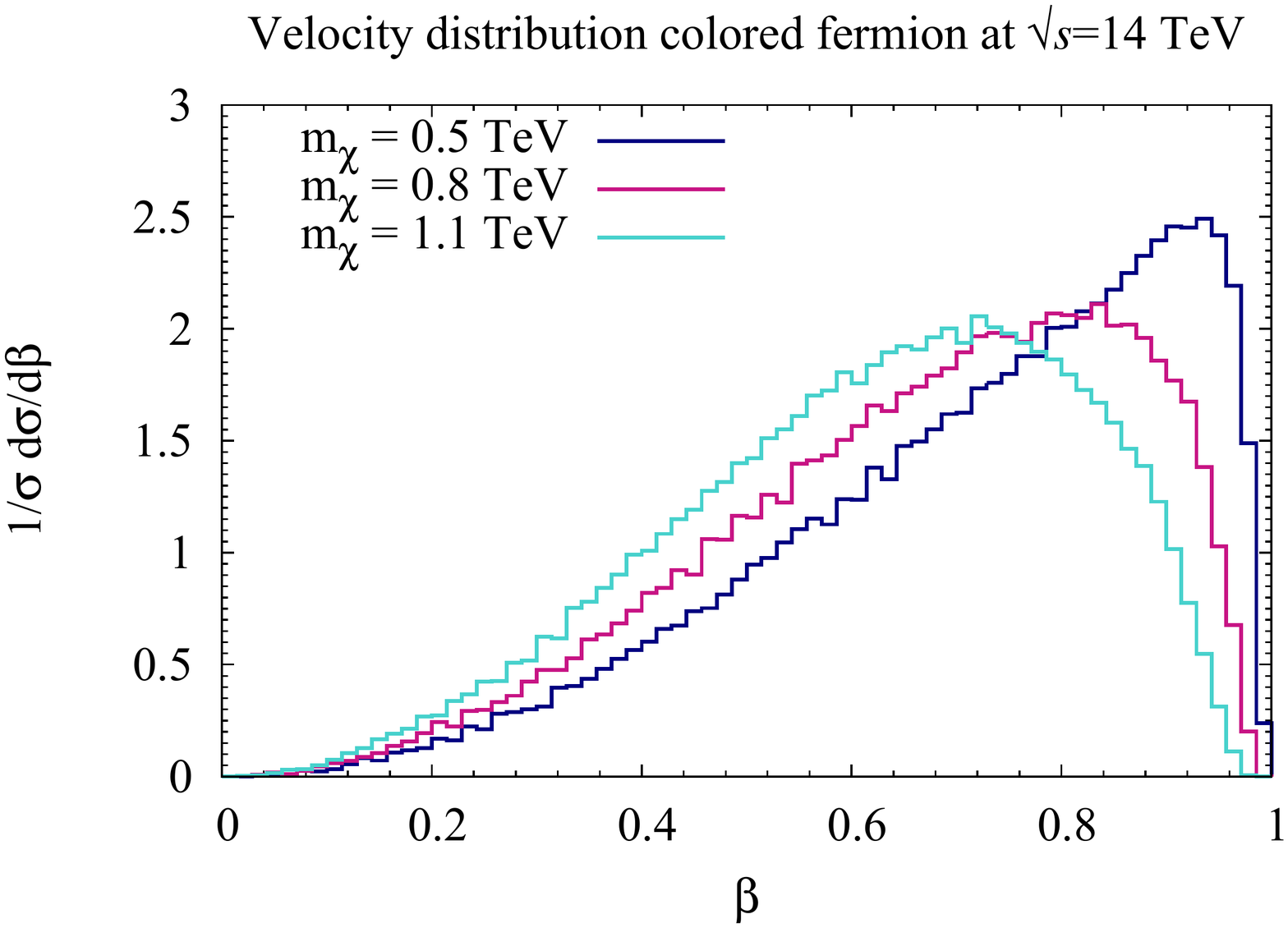} 
\hspace*{-1.7cm}
\\ \vspace*{-0.5cm}
\hspace*{-1.6cm}
\includegraphics[angle=0,width=9.9cm]{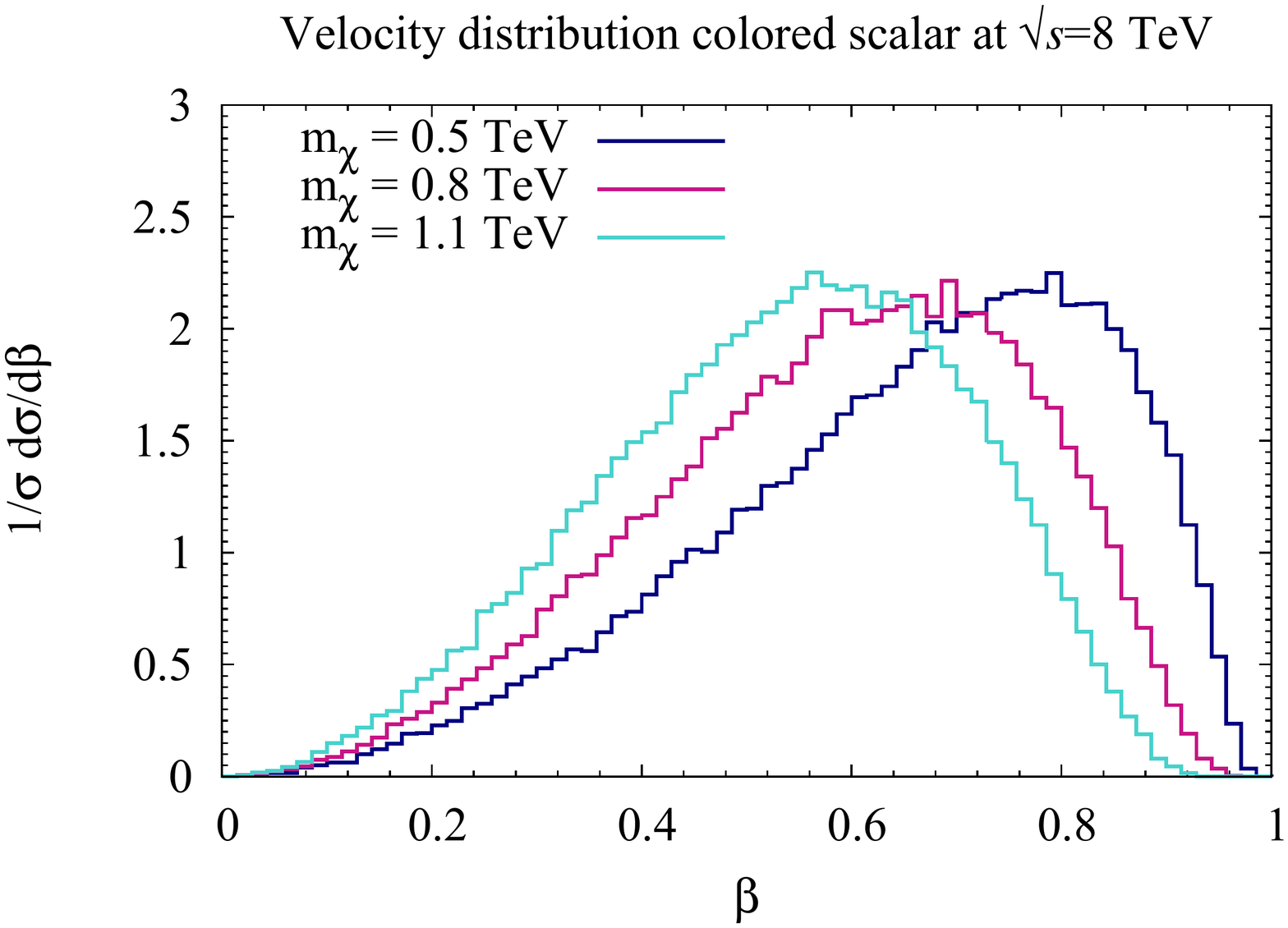}
\hspace*{-1.6cm}
\includegraphics[angle=0,width=9.9cm]{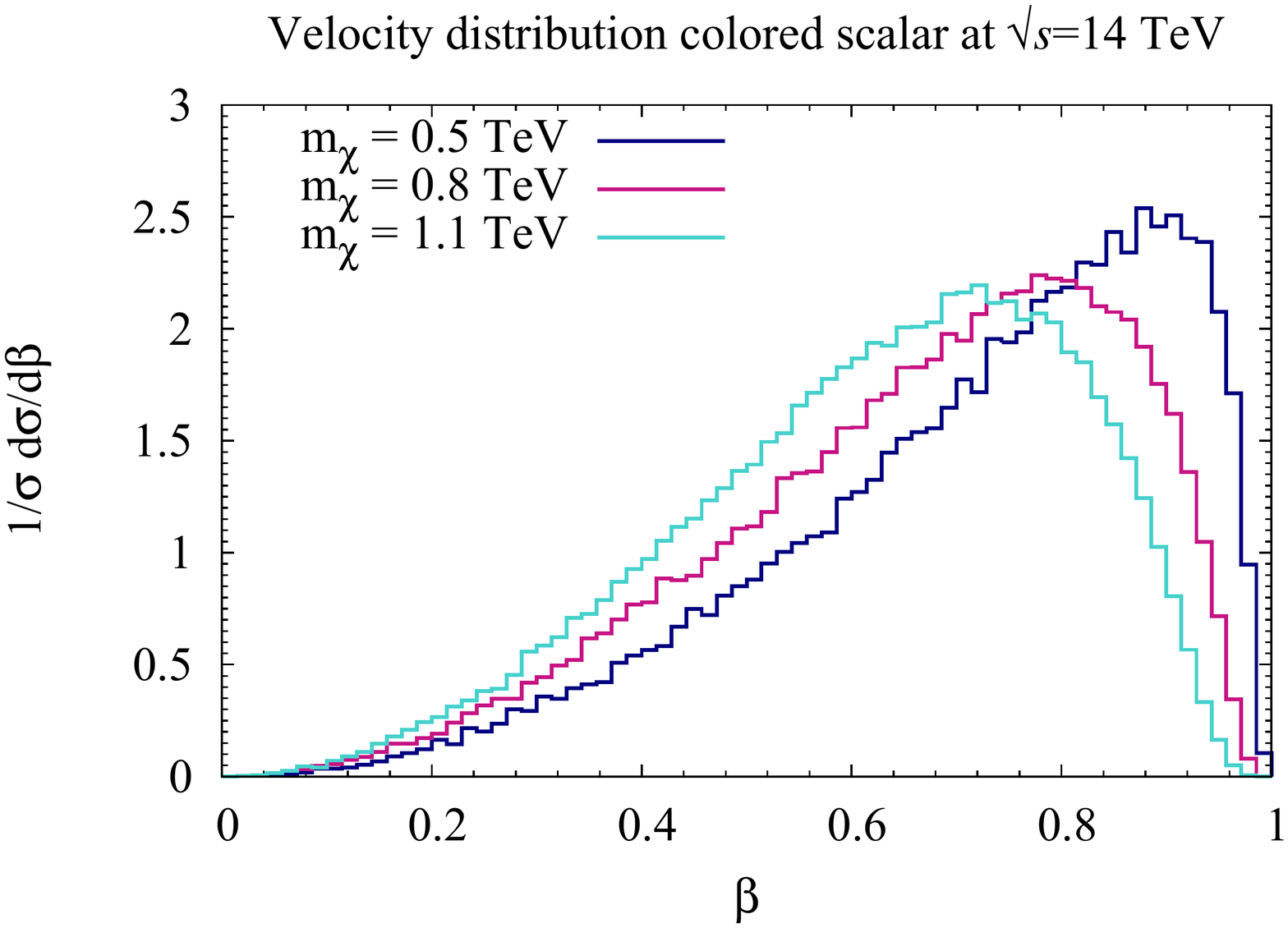}
\hspace*{-1.7cm}
\caption{Velocity distribution normalized to the total cross section for colored fermonic states (top) and colored scalars (bottom). Left (right) panels correspond to $\sqrt{s}=8\text{ TeV}$ ($\sqrt{s}=14\text{ TeV}$) for $m_{\chi}=500\text{ GeV}$ (blue),  $m_{\chi}=800\text{ GeV}$ (pink) and 
$m_{\chi}=1.1\text{ TeV}$ (light blue). \label{distributions}}
\end{figure} 
There is no significant difference between scalars and fermions. For higher center-of-mass energies and lower masses, 
$m_{\chi}$, higher velocities are more probable. Notice also that due to the normalization with respect to the total cross section there is no difference between the different color quantum numbers. Another issue that has to be considered for a complete description of the production mechanism is the Sommerfeld enhancement due to ladder exchange of gluons. This effect is relevant only for slowly produced states.
 \item[Hadronization.] 
 Once produced, a colored particle combines with quarks and gluons to form a colorless hadron state. For example, a color-triplet can form bound states such as $C_3  \overline{q}$ or $C_3 {q_1 q_2}$, an octet state can form invariants such as $C_8  \overline{q} q$,  $C_8  q_1 q_2 q_2$ or $C_8  g$, while the sextuplet can hadronize in states of the form $C_6 q g, C_6 q \overline{q} q$ and $C_6 \overline{q} \overline{q}$. The probability for $C_i$ of hadronizing in a given bound state are rather uncertain, different models  give quite different values.\footnote{See for example the comparison between the string model adopted by {\tt Pythia}~\cite{Sjostrand:2006za} and the cluster model used in {\tt HERWIG}~\cite{Moretti:2002eu,Bellm:2013lba} in Table 3 of \cite{Fairbairn:2006gg}.}  Bound states made of larger numbers of quarks and gluons are expected to be heavier \cite{DeRujula:1975ge} and, even if the hadronization in this channels could be non-negligible, the newly produced state could decay into a lighter one plus ordinary baryons and mesons, through QCD gauge interactions. Despite the fact that the hadronization processes are very uncertain, in some cases a detailed knowledge is not very important. As we are going to comment soon, nuclear conversions can wash out the information on the original hadron state at production. 
 \item[Propagation through matter.]
As soon as produced and during their propagation, the long lived colored particles interact with the electrons, protons and neutrons present in the detector.
The dominant interactions we consider here are the electromagnetic and the strong ones. 
\begin{itemize}
  \item {\it Electromagnetic interactions.}
  
A particle with electric charge can interact with atomic electrons as well as with protons and  neutrons in the nuclei. In the first case the net effect is the ionization of the atoms while interactions with atomic nuclei generate displacements of atoms from the lattice. In both cases the heavy long lived particle loses energy, however the energy loss $d E/d x$ from ionization is much larger than the one generated in the non-ionizing way. 
 
The main role of the heavy elementary parton is to contribute to the electric charge of the hadron. Indeed the SM gauge quantum number and the requirement to obtain a color-singlet hadron state has an important influence on the total charge of the resulting bound state. We notice that considering all the cases in \Table{summary2}, the resulting bound state has always integer charge. The most unfavourable situation, 
from the point of view of detection, happens when the resulting hadron is electromagnetically neutral.
  
\item {\it Strong interactions.}

Before discussing the various kind of interactions between the heavy hadrons and the matter in the detector, let us clarify the role of the parton $C_i$ in the nuclear reactions.
Due to their large mass, the wave-functions of the $C_i$'s are expected to be highly localized as compared to those of the light constituents (quarks and gluons) that are spread in space as in ordinary QCD. From this observation we can draw the conclusions that the probability for the heavy parton $C_i$ to interact with matter is very low, while the typical cross section of the hadron with matter, being due to the effect of the light partons, is expected to be of the same order of those for pion scatterings.

Our heavy long-lived hadron can have elastic as well as inelastic reactions with nucleons. Elastic scatterings are not particularly relevant, indeed the energy loss is small because the long-lived hadron scatters on a much lighter target nucleus. Inelastic processes are instead those responsible for the slowing-down of the hadron. In an inelastic reaction it is also possible to exchange baryon and electric charge. 

The importance of baryon exchange has been emphasized in \cite{Kraan:2004tz}, in these reactions a heavy hadron is transformed into another one with different baryon number. For the case of the gluino it has been argued that, bound states with null baryon charge ($R$-mesons) are very efficiently converted into baryonic states in reactions like 
$(C_8 d\bar{d}) + p \to \pi + (C_8 udd)$. The reverse reaction is suppressed mainly because of the mass split ordering of the various hadrons and by the low presence of pions as targetd in the detector material. As a consequence, early in the detector, mesons are converted into baryons. 

Processes with charge exchange are particularly relevant for detecting the presence of the heavy parton. Indeed, tracks generated by the passage of electric particles can be easily detected, so it is important to understand the value of the electric charge of the hadron through all its travel in the calorimeter material. Even in the most pessimistic case of a hadron generated as a neutral bound state, reactions with charge exchange can covert it in a charged state that could be detected.

We finally mention that there are a series of phenomenological approaches to describe the strong interactions between heavy hadrons and matter \cite{Drees:1990yw,Mafi:1999dg,Baer:1998pg}. Despite the fact they agree on several general qualitative aspects, they give rise to rather different quantitative results. 
  \end{itemize}
 \item[Stopping and decay.]
 Depending on the $C_i$ mass, a non-negligible number of particles could stop in the detector. 
 In our framework most of the states are supposed to decay within few seconds. 
This constitutes a really interesting possibility to understand the structure of the $d=5$ effective operators responsible for the decay. However, the detection of these processes represents a severe experimental challenge.
\end{description}

Having described the most important aspects of the phenomenology of our long lived particles, we now move to comment about direct searches.
At the LHC, searches for long-lived colored particles are performed
in the context of $R$-hadrons, which are bound states of gluinos/
squarks and quarks/gluons.
The $R$-hadrons can be detected by
the longer time-of-flight to the outer detectors and their anomalous energy loss.
Complementary to the searches relying simply on the longer time-of-flight and 
the anomalous energy loss, are searches for stopped $R$-hadrons. 
They are particularly suited for velocities $\beta\ll 1$.
\par
In the former case, the strongest limits come from the CMS search 
of Ref.~\cite{Chatrchyan:2013oca}, which excludes gluino masses up to 1276 GeV, 
if the fraction of gluinos hadronizing into $\tilde{g}-g$ bound states is 0.5. 
If the fraction is equal to one, gluino masses are excluded up to 1250 GeV.
In such a case the $R$-hadron is neutral in the inner tracker which leads to
 a smaller energy loss and hence a lower exclusion bound.
In Ref.~\cite{Chatrchyan:2013oca} stop masses were excluded up to 935 GeV (818 GeV).
The exact exclusion bound depends on the modeling of the interactions of the stop 
with the detector material. The exclusion bound is here given for the so-called cloud 
model of Ref.~\cite{Kraan:2004tz, Mackeprang:2006gx} (the charged-suppressed model 
of Ref.~\cite{Mackeprang:2009ad}). Very similar exclusion bounds were obtained in a recent ATLAS study~\cite{ATLAS:2014fka}
both for long-lived stops and gluinos.
\par
In the search for out-of-time decays of stopped gluinos or squarks of Ref.~\cite{Khachatryan:2015jha}, gluinos 
are excluded up to masses of 880 GeV, assuming BR($\tilde{g}\to\tilde{\chi}_1^0 g$)=100\%
 and a gluino lifetime between 1 $\mu$s and 1000 s.
Stop masses are excluded up to 470 GeV for 
BR($\tilde{t}\to \tilde{\chi}_1^0 t$)=100\% and stop lifetimes between 1 $\mu$s and 1000 s. 
The exclusion bounds require that the neutralino mass is kinematically consistent with the used requirements on
the energies of the gluon or respectively top decay products. The search furthermore assumes a cloud model for the $R$-hadron interactions.
Reference~\cite{Aad:2013gva}  
 excludes sbottom masses up to 344 GeV for BR($\tilde{b}\to \tilde{\chi}_1^0 b$)=100\% for lifetimes 
between 1 $\mu$s and 1000 s.
\par
The results on these searches of $R$-hadrons cannot straightforwardly be applied for our cases. 
In order to give exclusion bounds on our states a detailed study of the hadronization of the  
different states and a full detector simulation would be necessary. 
We expect that the exclusion limits not only depend strongly on the SU$(3)_c$ quantum number, but 
they depend also on the charge and SU$(2)_L$ quantum number. The charge influences the energy loss 
directly, whereas different SU$(2)_L$ quantum numbers lead to different time intervals, in which the 
$R$-hadron propagates as a neutral particle in the detector due to pion exchange between the 
different members of an isomultiplet \cite{Arvanitaki:2005nq}. 
For stopped $R$-hadrons, in addition, the searches depend on the BRs. For our cases if the 
new exotic particle decays into a missing energy and jet signature, there are always several
 other operators allowing the 
particle to decay (cf.~\Table{summary2}), such that the corresponding BR likely deviates from 1. 

\par
A further discussion on the bounds on long-lived colored states is  beyond the scope
of this paper, as, even in  the well-studied case of $R$-hadrons the exclusion bounds depend
significantly on the modelling of the hadronization and the nuclear scattering model.

\section{Conclusions}
\label{conclusions}

Low-energy tests of fundamental symmetries provide a powerful probe of new physics 
scales up to energies of about $10^{15}$ GeV.  
Given the accidental (B and L) and approximate 
(CP, flavor and custodial) symmetry structure of the SM, 
it is somewhat surprising that signals of physics 
beyond the SM (if it exists) have not been observed so far. 

This last statement hinges on the theoretical prejudice that new physics effects at low energies can be described by a generic EFT, where the Wilson coefficients of the effective operators are $\mathcal{O}(1)$. There are, of course, exceptions to this point of view. 
The simplest one is maybe to allow for ultraweak couplings in the theory, 
so that the generic EFT power counting fails -- an extreme example being the dissolution 
of the Weinberg operator when RH neutrinos 
are below the EW scale. 
On the other hand, the effective operators might not be there due to an exact symmetry of the 
Lagrangian in the full theory (as e.g.~${\rm B}-{\rm L}$ in 
left-right symmetric models \cite{Mohapatra:1979ia}) 
or they might be suppressed due to an approximate symmetry (as e.g.~in minimal flavor violation \cite{D'Ambrosio:2002ex}). 
In this paper, we explored yet another possibility: 
the quantum numbers of the new physics states below the EFT cut-off are such that by only 
requiring Lorentz and SM gauge invariance, the accidental and approximate 
symmetries of the SM are automatically preserved at the renormalizable level. 
The resulting new physics dynamics is practically invisible to low-energy indirect searches, and the 
only way of experimentally probing these scenarios is by direct production 
and detection of new particles at colliders. We hence focused on the phenomenological 
possibility that the new states lie within the kinematical reach of the LHC. 

Barring few exceptions, the new matter multiplets are subject to extra accidental 
$Z_2$ or U(1) symmetries which forbid their decays at the renormalizable level. 
Whenever the LP in the multiplet is color- and charge-neutral, it forms a 
DM candidate~\cite{Cirelli:2005uq,Cirelli:2007xd,Cirelli:2009uv,Cirelli:2014dsa}.  
Generally however, the extra multiplets will decay due to the presence of higher dimensional operators in the EFT. 
In the spirit of generic EFT we choose to work with a cut-off scale of $\Lambda_{\rm eff} \approx 10^{15}$ GeV which is large enough not to require any further protection mechanisms in the full theory and is moreover suggested by the observations of neutrino masses. 
 The infinite set of possible states which satisfy the accidental symmetry conditions 
can then be reduced thanks to cosmological considerations. In particular, 
since the new states are long-lived, 
scenarios where the lightest component of the new multiplet is charged and/or colored 
are constrained by cosmological observations as well as 
by searches for exotic forms of matter on the Earth and in the Universe. 
The latter practically exclude all the cases where the 
charged and/or colored LP decays via $d > 5$ operators. 

Another handle in order to further reduce the list of possible states is the 
requirement that the theory remains perturbative up to the cut-off scale of the EFT. 
In particular, we required that no Landau poles are generated below 
$\Lambda_{\rm eff} \approx 10^{15}$ GeV.
As a byproduct of the perturbativity analysis we noticed that, due to accidental 
cancellations in the coefficients of the one-loop beta functions for the 
non-abelian gauge factors, two-loop corrections can become  important 
and hence the one-loop determination of the Landau pole can be misleading. 
A typical example is given by the two minimal DM cases in \eqs{LP170}{LP150}. 
In this respect, we also pointed out the existence of a previously overlooked $d = 5$ 
operator which is responsible for a fast decay of $(1,7,0)_S$, thus ruling out the 
scalar minimal DM candidate \cite{DNPN}.  

The final set of states which satisfy all the above constraints is collected in \Tables{summary1}{summary2}. 
For these, we studied current bounds on their masses coming from their potential effects on BBN as well as from their production and detection at colliders. 
In particular, we found that for most of the states decaying through $d=5$ operators and being thermally produced in the early Universe, their abundances are sufficiently diluted not to affect standard BBN. The notable exceptions are those uncolored cases, where the decay rates are either loop suppressed or proceed through long cascades leading to high-multiplicity final states (they are listed in Table~\ref{table:BBNbound}).
At colliders, the lightest particle of the multiplets in \Tables{summary1}{summary2} are, barring few exceptions, stable 
on the detector scale.  
For the color singlets we found that the current mass bounds are of few hundred GeV if the lightest particle of the multiplet is charged, whereas for neutral states the detection is more difficult and hence the bounds lie below 100 GeV (cf.~\Table{summarybounds}). On the other hand, for colored multiplets the mass bounds strongly depend on the hadronization process and the nuclear interactions with the detector material. 

A this point, a natural question to ask is the following: 
What is accidental matter good for? Who ordered that? 
Besides the case of minimal DM, we note that the scalar multiplets in \Tables{summary1}{summary2} 
could easily improve the stability of the renormalizable Higgs 
potential.\footnote{On the other hand, we explicitly checked that none of the 
weak-scale accidental matter states improves on gauge coupling unification 
with respect to the SM.}  Close to the EFT cut-off, the potential can again be destabilized in cases where $d=5$ operators exist containing only scalars, namely $(1,5,0)_S$, $(1,5,1)_S$, $(1,5,2)_S$, $(1,7,0)_S$, $(8,1,0)_S$, $(8,1,1)_S$, $(8,3,0)_S$ and $(8,3,1)_S$. However, at such large field values the whole tower of operators should be considered and stabilization is expected to be recovered via $d\geq6$ operators.
Finally, a charged thermal relic with $\tau_\chi \sim (10^2 - 10^{3})$~s and abundance just below the 
D/H bound (which can be achieved for some of the accidental matter states listed in Table~\ref{table:BBNbound}) may also help to resolve the standard BBN Lithium problems~\cite{Bird:2007ge,Jittoh:2007fr,Jedamzik:2007cp,Jittoh:2008eq,Cumberbatch:2007me,Kusakabe:2007fu}. 
 More generally, accidental matter should be seen as a purely phenomenological possibility.  
New physics might manifest itself in a way we were not expecting and thus 
the direct search strategies should cover diverse scenarios. In particular, the typical signature 
of accidental matter is the presence of charged/colored particles which are 
stable on the scale of particle detectors and which have no chances to be detected through indirect searches. 
Consequently, high-energy colliders will be the only means of probing such scenarios. 
New experiments in the near future (LHC-II, MoEDAL, \etc) will have the capabilities to further explore their parameter space.

We end by noting that an improvement in $p$-decay bounds by an order of magnitude~\cite{Abe:2011ts} or failure to observe 
neutrinoless double beta decay with inverse neutrino mass hierarchy in the next generation of experiments~\cite{Garfagnini:2014nla} would put some pressure on this setup. In particular, (i) if neutrino oscillation experiments were to confirm the inverse neutrino mass hierarchy, then our EFT setup predicts an observable neutrinoless double beta decay signal. Failure to observe one in the next generation of experiments would imply the presence of NP degrees of freedom below the EFT cut-off which couple to SM fermions. Finally, (ii) there is already a mild tension between the proton decay bounds and 
neutrino mass measurements, if one assumes a common EFT scale for both phenomena. This tension would be strengthened by future improvements in $p$-decay bounds or by a positive indication of a quasi-degenerate light neutrino spectrum, requiring a significant scale separation between the relevant L and B violating operators.

\section*{Acknowledgments}

We thank Sergio Cecotti, Talal Ahmed Chowdhury, Ben Gripaios, Miha Nemev\v sek, Paolo Panci, Giovanni Marco Pruna and 
Christian Reuschle for useful discussions. 
This work was supported in part by the Slovenian Research Agency. 
The work of L.D.L.~is supported by the Marie Curie CIG program, project number PCIG13-GA-2013-618439. 
L.D.L., R.G.~and M.N.~are grateful to the theoretical physics group of the Jo\v{z}ef Stefan Institute for hospitality and support 
during the development of this project. 
L.D.L.~would like to thank the high-energy physics group of the University of Roma Tre 
for hospitality during the completion of this work. 
R.G.~would like to thank the theoretical physics group of the University of Genova 
for hospitality during the completion of this work. 

\appendix

\section{Two-loop Landau Poles}
\label{pertbounds}

In this Appendix we provide the RG evolution of the gauge couplings and 
study the emergence of the associated Landau poles. In this way one can set an upper bound on the 
dimensionality of the extra representation, by requiring that no Landau poles are generated below 
$\Lambda_{\text{eff}} \approx 10^{15}$ GeV (cf.~the discussion in \sect{landau}).

The two-loop RG equation for the three gauge couplings $g_i$ ($i=1,2,3$), 
read
\begin{equation}
\label{alpha2loops}
\frac{{\rm d}}{{\rm d}t}\alpha^{-1}_{i}=-a_{i}-\frac{b_{ij}}{4\pi}\alpha_{j} \, ,
\end{equation}
where $\alpha_i = \frac{g_i^2}{4\pi}$ and $t=\frac{1}{2\pi}\log \frac{\mu}{M_{Z}}$. 
The one- and two-loop beta function are \cite{Machacek:1983tz} (no summation over $i$) 
\begin{align}
\label{oneloopbf}
a_{i}&=- \frac{11}{3} C_2(G_i) + \frac{4}{3} \sum_F \kappa S_2(F_i) + \frac{1}{3} \sum_S \eta S_2(S_i)\,, \\
\label{twoloopbf}
b_{ij}&= 
\left[- \frac{34}{3} \left( C_2(G_i) \right)^2 
+  \sum_F \left( 4 C_2(F_i) + \frac{20}{3} C_2(G_i) \right) \kappa S_2(F_i) \right.  \\
& \left.  + \sum_S \left( 4 C_2(S_i) + \frac{2}{3} C_2(G_i) \right) \eta S_2(S_i) \right]\delta_{ij} 
+ 4 \Big[  \sum_F \kappa C_2(F_j) S_2(F_i) + \sum_S \eta C_2(S_j) S_2(S_i)  \Big] \, , \nonumber
\end{align}
where $G_i$ denotes the $i$-th gauge factor, $S_{2}$ and $C_{2}$ are the index (including multiplicity 
factors) and the quadratic Casimir of a given (fermionic ($F$) or scalar ($S$)) 
irreducible representation; $\kappa=1,\frac{1}{2}$ for Dirac and Weyl fermions and
$\eta=1, \frac{1}{2}$ for complex and real scalar fields, respectively. 
The Yukawa contribution in the two-loop beta function is 
neglected. In fact, the extra states we want to introduce do not couple with SM fermions, so that 
the Yukawa contribution does not grow with the dimensionality of the extra representation. 
Employing the GUT normalization for the abelian factor, we use the values 
$\alpha_1 (m_Z) = 0.016923$,
$\alpha_2 (m_Z) = 0.03374$,
and $\alpha_3 (m_Z) = 0.1173$ for the onset of the RG running \cite{Mihaila:2012pz,Beringer:1900zz}. 
For simplicity, the extra state $\chi$ is integrated in at $m_Z = 91.188$ GeV \cite{Beringer:1900zz}. 
The scaling of the Landau pole with $m_\chi$ is approximately linear. 

For the cases where the SM is extended with a representation charged under SU$(3)_c$ and/or SU$(2)_L$, 
the results are summarized in \Table{LPSU2vsSU3}, which provide  
a useful reference for estimating the bound on the dimensionality of the 
extra representations, by requiring that no Landau poles are generated below a given scale.  

The analysis has been repeated for all the states considered in this work, 
which can simultaneously transform under SU$(3)_c$ and SU$(2)_L$, and have a non-zero hypercharge as well.  
These include extra representations interacting with SM fields via $d=5$ operators
(but which cannot decay into SM states via renormalizable interactions), for which the results are reported 
in \Table{summarydim5}. 
Moreover, we investigated the Landau pole constraints for those extra scalar representations 
that can couple to the Higgs boson at the renormalizable level (cf.~\Table{renchiH}). Among them the only ones 
that do not couple to SM fermions at the renormalizable level 
and that survive the perturbativity criteria are the renormalizable cases of \Table{summary1}. 
In this respect, we mention a marginal case: $(1,6,1/2)_S$ 
for which $\Lambda_{\rm{Landau}}^{\rm{2-loop}} = 6.6 \times 10^{13}$ GeV. 
Finally, we also checked the possibility of having multiplets decaying via $d>5$ operators and whose neutral LP might be compatible with cosmological constraints. No cases beyond those of minimal DM (cf.~\Table{summary1}) 
and with a Landau pole above $10^{15}$ GeV are found.

\begin{landscape}
\begin{table}[htbp]
  \centering
  \begin{tabular}{@{} |c|c|c|c|c|c|c|c|c|c|c| @{}}
\hline 
RS & 1 & 2 & 3 & 4 & 5 & 6 & 7 & 8 & 9 & 10 \\ 
    \hline
    1 & $\gg m_{\rm Pl}$ & $\gg m_{\rm Pl}$ & $\gg m_{\rm Pl}$ & $\gg m_{\rm Pl}$ & $\gg m_{\rm Pl}$ & $\gg m_{\rm Pl}  $ &  $1.4 \times 10^{16} $ & \Red{$4.0 \times 10^{8}$} & \Red{$4.7 \times 10^{5}$} 
    & \Red{$1.7 \times 10^{4}$} \\
    \hline
    8 & $\gg m_{\rm Pl}$ & $\gg m_{\rm Pl}$ & $\gg m_{\rm Pl}$ &  \Red{$8.3 \times 10^{12}$} & \Red{$5.0 \times 10^{5} $} 
    & \Red{$5.2 \times 10^{3} $} & \Red{$< 10^3$} & \Red{$< 10^3$} & \Red{$< 10^3$} 
    & \Red{$< 10^3$}  \\
    \hline
    27 &  \Red{$1.3 \times 10^{7} $} & \Red{$1.4 \times 10^{3} $} & \Red{$< 10^3$}
    & \Red{$< 10^3$} & \Red{$< 10^3$} & \Red{$< 10^3$} & \Red{$< 10^3$} & \Red{$< 10^3$} & \Red{$< 10^3$} & \Red{$< 10^3$}  \\
    \hline
    \hline
CS & 1 & 2 & 3 & 4 & 5 & 6 & 7 & 8 & 9 & 10 \\ 
    \hline
    1 & $\gg m_{\rm Pl}$  & $\gg m_{\rm Pl}$  & $\gg m_{\rm Pl}$  & $\gg m_{\rm Pl}$  & $\gg m_{\rm Pl}$ & \Red{$7.2 \times 10^{13}$} 
    & \Red{$2.0 \times 10^{7}$} & \Red{$7.0 \times 10^{4}$} & \Red{$4.7 \times 10^{3}$} 
    & \Red{$1.1 \times 10^{3}$} \\
    \hline
    3 & $\gg m_{\rm Pl}$  & $\gg m_{\rm Pl}$  & $\gg m_{\rm Pl}$  & $> m_{\rm Pl}$ & \Red{$1.1 \times 10^{8}$}  & \Red{$4.7 \times 10^{4} $} & 
    \Red{$2.4 \times 10^{3}$} & \Red{$< 10^3$} & \Red{$< 10^3$} & \Red{$< 10^3$}  \\
    \hline
    6 & $\gg m_{\rm Pl}$  & $\gg m_{\rm Pl}$  & $\gg m_{\rm Pl} $ & \Red{$5.5 \times 10^{7} $} & \Red{$1.6 \times 10^{4} $} 
    & \Red{$1.1 \times 10^{3} $} & \Red{$< 10^3$} & \Red{$< 10^3$} & \Red{$< 10^3$} & \Red{$< 10^3$}  \\
    \hline
    8 & $\gg m_{\rm Pl}$  & $\gg m_{\rm Pl}$  & $> m_{\rm Pl} $ & \Red{$1.1 \times 10^{6} $} & \Red{$4.2 \times 10^{3} $} 
    & \Red{$< 10^3$} & \Red{$< 10^3$} & \Red{$< 10^3$} & \Red{$< 10^3$} & \Red{$< 10^3$}  \\
    \hline
    10 & $\gg m_{\rm Pl}$  & \Red{$3.8 \times 10^{7} $} & \Red{$8.5 \times 10^{3} $} & \Red{$1.3 \times 10^{3} $} 
    & \Red{$< 10^3$} & \Red{$< 10^3$} & \Red{$< 10^3$} & \Red{$< 10^3$} & \Red{$< 10^3$} & \Red{$< 10^3$}  \\
    \hline
    15 & $\gg m_{\rm Pl}$  & \Red{$7.3 \times 10^{4} $} & \Red{$2.0 \times 10^{3} $} 
    & \Red{$< 10^3$} & \Red{$< 10^3$} & \Red{$< 10^3$} & \Red{$< 10^3$} & \Red{$< 10^3$} & \Red{$< 10^3$} & \Red{$< 10^3$}  \\
    \hline
    $15'$ & \Red{$1.3 \times 10^{4} $} & \Red{$< 10^3$} & \Red{$< 10^3$} 
    & \Red{$< 10^3$} & \Red{$< 10^3$} & \Red{$< 10^3$} & \Red{$< 10^3$} & \Red{$< 10^3$} & \Red{$< 10^3$} & \Red{$< 10^3$}  \\
    \hline
    24 & \Red{$1.9 \times 10^{3} $} & \Red{$< 10^3$} & \Red{$< 10^3$} 
    & \Red{$< 10^3$} & \Red{$< 10^3$} & \Red{$< 10^3$} & \Red{$< 10^3$} & \Red{$< 10^3$} & \Red{$< 10^3$} & \Red{$< 10^3$}  \\
    \hline
    \hline 
WF & 1 & 2 & 3 & 4 & 5 & 6 & 7 & 8 & 9 & 10 \\ 
    \hline
    1 & $\gg m_{\rm Pl}$ & $\gg m_{\rm Pl}$ & $\gg m_{\rm Pl}$ & $\gg m_{\rm Pl}  $ & $8.3 \times 10^{17} $ & \Red{$5.0 \times 10^{8} $} 
    & \Red{$3.7 \times 10^{5} $} & \Red{$1.3 \times 10^{4} $} & \Red{$2.3 \times 10^{3} $} & \Red{$< 10^3$}  \\
    \hline
    8 & $\gg m_{\rm Pl}$ & $\gg m_{\rm Pl}$ & \Red{$7.3 \times 10^{9}$} & \Red{$3.2 \times 10^{4} $} 
    & \Red{$1.4 \times 10^{3} $} & \Red{$< 10^3$} & \Red{$< 10^3$} & \Red{$< 10^3$} & \Red{$< 10^3$} & \Red{$< 10^3$} \\
    \hline
    27 & \Red{$< 10^3$} & \Red{$< 10^3$} & \Red{$< 10^3$} & \Red{$< 10^3$} & \Red{$< 10^3$} & \Red{$< 10^3$} & \Red{$< 10^3$} & \Red{$< 10^3$} & \Red{$< 10^3$} & \Red{$< 10^3$} \\
    \hline
    \hline
DF & 1 & 2 & 3 & 4 & 5 & 6 & 7 & 8 & 9 & 10 \\ 
    \hline
    1 & $\gg m_{\rm Pl}$ & $\gg m_{\rm Pl}$ & $\gg m_{\rm Pl}$ & $8.5 \times 10^{18}$ & \Red{$9.4 \times 10^{7}$} & \Red{$8.4 \times 10^{4}$} 
    & \Red{$4.4 \times 10^{3}$} & \Red{$1.0 \times 10^{3}$} & \Red{$< 10^3$} & \Red{$< 10^3$} \\ 
    \hline
    3 & $\gg m_{\rm Pl}$ & $\gg m_{\rm Pl}$ & \Red{$8.1 \times 10^{14}$} & \Red{$5.4 \times 10^{5}$} & \Red{$4.3 \times 10^{3}$} & \Red{$< 10^3$}
    & \Red{$< 10^3$} & \Red{$< 10^3$} & \Red{$< 10^3$} & \Red{$< 10^3$} \\ 
    \hline
    6 & $\gg m_{\rm Pl}$ & \Red{$1.9 \times 10^{12}$} & \Red{$6.4 \times 10^{4}$} & \Red{$3.3 \times 10^{3}$} & \Red{$< 10^3$} & \Red{$< 10^3$}
    & \Red{$< 10^3$} & \Red{$< 10^3$} & \Red{$< 10^3$} & \Red{$< 10^3$} \\ 
    \hline
    8 & $\gg m_{\rm Pl}$ & \Red{$1.5 \times 10^{7}$} & \Red{$1.0 \times 10^{4}$} & \Red{$1.3 \times 10^{3}$} & \Red{$< 10^3$} &\Red{$< 10^3$}
    & \Red{$< 10^3$} & \Red{$< 10^3$} & \Red{$< 10^3$} & \Red{$< 10^3$} \\ 
    \hline
    10 & \Red{$2.6 \times 10^4$} & \Red{$< 10^3$} & \Red{$< 10^3$} & \Red{$< 10^3$} & \Red{$< 10^3$} & \Red{$< 10^3$}
    & \Red{$< 10^3$} & \Red{$< 10^3$} & \Red{$< 10^3$} & \Red{$< 10^3$}\\ 
        \hline
    15 & \Red{$3.2 \times 10^3$} & \Red{$< 10^3$} & \Red{$< 10^3$} & \Red{$< 10^3$} & \Red{$< 10^3$} & \Red{$< 10^3$}
    & \Red{$< 10^3$} & \Red{$< 10^3$} & \Red{$< 10^3$} & \Red{$< 10^3$} \\ 
    \hline    
\end{tabular}
  \caption{\label{LPSU2vsSU3} 
  Two-loop Landau poles (GeV) for the SM augmented with an extra multiplet 
  (integrated in at $m_Z$) which is charged under SU$(3)_c$ and/or SU$(2)_L$ ($Y=0$). 
  The rows and columns denote respectively the SU$(3)_c$ and SU$(2)_L$ representations,
  while the four subtables correspond to the cases of an extra real scalar (RS), complex scalar (CS), Weyl fermion (WF) and Dirac fermion (DF).  
  $m_{\rm Pl} \approx 10^{19}$ GeV is the Planck mass. The cases where the two-loop 
  Landau pole is below $10^{15}$ GeV are emphasized in red.}
\end{table}
\end{landscape}

\begin{table}[htbp]
  \centering
  \begin{tabular}{@{} |c|c|c| @{}}
  \hline
   Spin & $\chi$ & $\Lambda_{\rm{Landau}}^{\rm{2-loop}}$[GeV]  \\ 
 \hline
 \hline
\rowcolor{LightCyan}  $0$ & $(1,2,3/2)$ & $\gg m_{\rm Pl}$ ($g_1$)  \\ 
\rowcolor{LightCyan}  $0$ & $(1,2,5/2)$ & $\gg m_{\rm Pl}$ ($g_1$) \\ 
\rowcolor{LightCyan}  $0$ & $(1,5,0)$ & $\gg m_{\rm Pl}$ ($g_1$)  \\ 
\rowcolor{LightCyan}  $0$ & $(1,5,1)$ & $\gg m_{\rm Pl}$ ($g_1$)  \\ 
\rowcolor{LightCyan}  $0$ & $(1,5,2)$ & $3.5 \times 10^{18}$ ($g_1$)  \\ 
\rowcolor{LightCyan}  $0$ & $(1,7,0)$ & $1.4 \times 10^{16}$ ($g_2$)  \\ 
\rowcolor{LightCyan}  $0$ & $(3,1,5/3)$ &  $\gg m_{\rm Pl}$ ($g_1$)   \\ 
\rowcolor{LightCyan}  $0$ & $(\overline{3},2,5/6)$ &  $\gg m_{\rm Pl}$ ($g_1$)  \\ 
\rowcolor{LightCyan}  $0$ & $(\overline{3},2,11/6)$ & $5.5 \times 10^{19}$ ($g_1$) \\ 
\rowcolor{LightCyan}  $0$ & $(3,3,2/3)$ & $\gg m_{\rm Pl}$ ($g_1$)  \\
\rowcolor{LightCyan}  $0$ & $(3,3,5/3)$ & $3.2 \times 10^{17}$ ($g_1$)  \\  
\rowcolor{LightCyan}  $0$ & $(3,4,1/6)$ &  $\gg m_{\rm Pl}$ ($g_2$)  \\
\rowcolor{LightCyan}  $0$ & $(\overline{3},4,5/6)$ &  $\gg m_{\rm Pl}$ ($g_2$)  \\
\rowcolor{LightCyan}  $0$ & $(\overline{6},2,1/6)$ & $\gg m_{\rm Pl}$ ($g_1$)  \\  
\rowcolor{LightCyan}  $0$ & $(6,2,5/6)$ & $\gg m_{\rm Pl}$ ($g_1$)  \\    
\rowcolor{LightCyan}  $0$ & $(\overline{6},2,7/6)$ & $\gg m_{\rm Pl}$ ($g_1$)  \\  
$0$ & $(6,2,11/6)$ & $4.0 \times 10^{12}$ ($g_1$) \\
$0$ & $(\overline{6},4,1/6)$ &  $5.5 \times 10^{7}$ ($g_2$)  \\
$0$ & $(6,4,5/6)$ & $5.0 \times 10^{7}$ ($g_2$)  \\
\rowcolor{LightCyan}  $0$ & $(8,1,0)$ & $\gg m_{\rm Pl}$ ($g_1$)   \\  
\rowcolor{LightCyan}  $0$ & $(8,1,1)$ & $\gg m_{\rm Pl}$ ($g_1$)   \\ 
\rowcolor{LightCyan}  $0$ & $(8,3,0)$ & $\gg m_{\rm Pl}$ ($g_1$)  \\ 
\rowcolor{LightCyan}  $0$ & $(8,3,1)$ & $1.0 \times 10^{17}$ ($g_1$) \\ 
$0$ & $(27,1,0)$ & $1.3 \times 10^{7}$ ($g_3$) \\ 
$1/2$ & $(1,3,2)$ &  $1.4 \times 10^{13}$ ($g_1$)  \\ 
\rowcolor{piggypink}    $1/2$ & $(1,4,1/2)$ &  $8.1 \times 10^{18}$ ($g_2$)  \\ 
\rowcolor{piggypink}     $1/2$ & $(1,4,3/2)$ &  $2.7 \times 10^{15}$ ($g_1$)  \\ 
    $1/2$ & $(\overline{3},3,4/3)$ & $9.3 \times 10^{10}$ ($g_1$)  \\
    $1/2$ & $(\overline{3},3,5/3)$ & $1.6 \times 10^{8}$ ($g_1$)  \\
    $1/2$ & $(3,4,1/6)$ & $5.4 \times 10^{5}$ ($g_2$)  \\
    $1/2$ & $(\overline{3},4,5/6)$ & $5.3 \times 10^{5}$ ($g_2$) \\
    $1/2$ & $(3,4,7/6)$ & $5.2 \times 10^{5}$ ($g_2$)  \\
\rowcolor{piggypink}    $1/2$ & $(6,1,1/3)$ & $\gg m_{\rm Pl}$ ($g_1$)  \\
\rowcolor{piggypink}    $1/2$ & $(\overline{6},1,2/3)$ & $\gg m_{\rm Pl}$ ($g_1$)  \\
    $1/2$ & $(\overline{6},2,1/6)$ & $1.9 \times 10^{12}$ ($g_3$)  \\
\rowcolor{piggypink}     $1/2$ & $(8,1,1)$ & $4.0 \times 10^{16}$ ($g_1$)  \\ 
    $1/2$ & $(8,2,1/2)$ & $1.5 \times 10^{7}$ ($g_3$)  \\ 
    $1/2$ & $(\overline{15},1,1/3)$ & $3.2 \times 10^{3}$ ($g_3$)  \\
    $1/2$ & $(15,1,2/3)$ & $3.2 \times 10^{3}$ ($g_3$)  \\
    $1/2$ & $(15,2,1/6)$ & $3.6 \times 10^{2}$ ($g_3$)  \\
    \hline
    \end{tabular}
    \caption{\label{summarydim5} List of extra multiplets which can decay into SM particles 
    via $d=5$ operators (states decaying via $d=4$ operators 
    have been already subtracted) and corresponding two-loop Landau poles 
    evaluated by integrating in the new states at $m_Z$ (fields with zero hypercharge are 
    understood to be real). 
    In the third column, the symbol in the bracket 
    stands for the gauge coupling responsible for the emergence of the Landau pole. 
}
\end{table}

\clearpage

\section{SU$\mathbf{(2)_{\it L}}$ decompositions}
\label{SUtwodecomp}

By denoting the generators in the fundamental representation of SU$(2)_L$ as $T^a = \sigma^a/2$ 
(with $\sigma^a$ being the Pauli matrices and $a=1,2,3$), 
we define their action on the $(2j+1)$-dimensional completely symmetric tensor $\chi_{i_{1}i_{2} \ldots i_{2j}}$ 
($i_{1}, i_{2}, \ldots,  i_{2j} =1,2$) as
\begin{equation}
\delta^a (\chi_{i_{1}i_{2} \ldots i_{2j}}) = 
T^a_{i_1 k} \, \chi_{k i_{2} \ldots i_{2j}} 
+ T^a_{i_2 k} \, \chi_{i_{1} k \ldots i_{2j}} 
+ \ldots 
+ T^a_{i_{2j} k} \, \chi_{i_{1} i_{1} \ldots k} 
\, . 
\end{equation}
In general, we arrive at the following embedding of the properly normalized $T^3$ eigenstates: 
\begin{equation}
\begin{array}{l}
\chi_{11 \ldots 1} = \frac{1}{\sqrt{B_{2j,0}}} \chi^{j} \\
\chi_{11 \ldots 2} = \frac{1}{\sqrt{B_{2j,1}}} \chi^{j-1} \\ 
\vdots \\ 
\chi_{12 \ldots 2} = \frac{1}{\sqrt{B_{2j,2j-1}}} \chi^{-j+1} \\ 
\chi_{22 \ldots 2} = \frac{1}{\sqrt{B_{2j,2j}}} \chi^{-j} \, ,
\end{array}
\end{equation}
where the superscripts denote the $T^3$ eigenvalue, $B_{n,k}$ is the binomial factor 
$B_{n,k} = \frac{n!}{k! (n-k)!}$ and the normalization of the states is such that 
\begin{equation}
\chi^{*i_{1}i_{2} \ldots i_{2j}} \chi_{i_{1}i_{2} \ldots i_{2j}} = |\chi^{j}|^2 + |\chi^{j-1}|^2 + 
\ldots + |\chi^{-j+1}|^2 + |\chi^{-j}|^2 \, . 
\end{equation}
Let us consider, for instance, the case of the SU$(2)_L$ Higgs doublet: 
\begin{equation}
\begin{array}{l}
H_{1} = H_+ \\
H_{2} = H_0 \, ,
\end{array} 
\end{equation}
where the electric charge eigenstates are obtained through the formula $Q = T^3 + Y$.  
In particular, since in the unitary gauge: $H_{+} = 0$, $\Im H_{0} = 0$ and $\Re H_{0} = \tfrac{1}{\sqrt{2}} (v + h)$, 
whenever the Higgs doublet is contained in the effective operator responsible for the decay of $\chi$, 
it might happen that not all of the components of $\chi$ can directly decay through the effective operator. 
In the following, we provide the SU$(2)_L$ decomposition for the three uncoloured multiplets whose decay, 
depending on the mass spectrum, might proceed via off-shell cascades (cf.~\Table{cascadedimgeq5}): 
\paragraph{$\chi = (1,2,5/2)_S$}

\begin{itemize}
\item SU$(2)_L$ embedding: 
\begin{equation}
\begin{array}{c}
\chi_{1} = \chi_{+3} \\
\chi_{2} = \chi_{+2} \, .
\end{array} 
\end{equation}
\item Operator: 
\begin{equation}
\mathcal{O}_1 = \chi^{*i} e^c e^c H_i \, . 
\end{equation}
\item Decomposition in the unitary gauge:
\begin{equation}
O_1 = \tfrac{1}{\sqrt{2}} \, \chi^*_{-2} e^c e^c (v + h) \, . 
\end{equation}
Notice that $\chi_{+3}$ does not couple directly to SM particles. 
Hence, whenever it is the LP it will decay through an off-shell emission of $\chi_{+2}$ (cf.~\fig{fig:cascadedecay}).
\end{itemize}

\paragraph{$\chi = (1,5,1)_S$}

\begin{itemize}

\item SU$(2)_L$ embedding: 
\begin{align}
\chi_{1111} &= \chi_{+3} \nonumber \\
\chi_{1112} &= \tfrac{1}{\sqrt{4}} \chi_{+2} \nonumber \\
\chi_{1122} &= \tfrac{1}{\sqrt{6}} \chi_{+1}  \\
\chi_{1222} &= \tfrac{1}{\sqrt{4}} \chi_{0} \nonumber \\
\chi_{2222} &= \chi_{-1}  \nonumber \, .
\end{align}

\item Operator:
\begin{equation}
O_1 = \chi^{*ijkl} H_{i} H_{j} H_{k} H^{*l'} \epsilon_{l l'} 
\, . 
\end{equation}
\item Decomposition in the unitary gauge:
\begin{equation}
O_1 = \frac{1}{8} \chi^*_0 (v+h)^4  \, .
\end{equation}
Notice that only $\chi_0$ can directly decay through $O_1$. 
If $\chi_0$ is not the LP in the multiplet, 
the charged LP will cascade decay through off-shell components 
which end up into $\chi_0$.

\end{itemize}

\paragraph{$\chi = (1,5,2)_S$}

\begin{itemize}
\item SU$(2)_L$ embedding: 
\begin{align}
\chi_{1111} &= \chi_{+4} \nonumber \\
\chi_{1112} &= \tfrac{1}{\sqrt{4}} \chi_{+3} \nonumber \\
\chi_{1122} &= \tfrac{1}{\sqrt{6}} \chi_{+2}  \\
\chi_{1222} &= \tfrac{1}{\sqrt{4}} \chi_{+1} \nonumber \\
\chi_{2222} &= \chi_{0}  \nonumber \, .
\end{align}

\item Operator:
\begin{equation}
O_1 = \chi^{*ijkl} H_{i} H_{j} H_{k} H_{l}
\, . 
\end{equation}

\item Decomposition in the unitary gauge:
\begin{equation}
O_1 = \frac{1}{4} \chi_0^* (v+h)^4  \, .
\end{equation}
Similarly to the previous case, 
only $\chi_0$ can decay through $O_1$, while the charged components  
decay through off-shell cascades. 
\end{itemize}

\bibliographystyle{utphys.bst}
\bibliography{bibliography}

\providecommand{\href}[2]{#2}\begingroup\raggedright\begin{thebibliography}{100}

\bibitem{Isidori:2010kg}
G.~Isidori, Y.~Nir, and G.~Perez, ``{Flavor Physics Constraints for Physics
  Beyond the Standard Model},''
  \href{http://dx.doi.org/10.1146/annurev.nucl.012809.104534}{{\em
  Ann.Rev.Nucl.Part.Sci.} {\bfseries 60} (2010) 355},
\href{http://arxiv.org/abs/1002.0900}{{\ttfamily arXiv:1002.0900 [hep-ph]}}.
%%CITATION = ARXIV:1002.0900;%%.

\bibitem{Cirigliano:2013lpa}
V.~Cirigliano and M.~J. Ramsey-Musolf, ``{Low Energy Probes of Physics Beyond
  the Standard Model},''
  \href{http://dx.doi.org/10.1016/j.ppnp.2013.03.002}{{\em
  Prog.Part.Nucl.Phys.} {\bfseries 71} (2013) 2--20},
\href{http://arxiv.org/abs/1304.0017}{{\ttfamily arXiv:1304.0017 [hep-ph]}}.
%%CITATION = ARXIV:1304.0017;%%.

\bibitem{Kamenik:2014xya}
J.~F. Kamenik, ``{Flavor constraints on new physics},''
\href{http://dx.doi.org/10.1142/S0217732314300213}{{\em Mod.Phys.Lett.}
  {\bfseries A29} no.~22, (2014) 1430021}.
%%CITATION = MPLAE,A29,1430021;%%.

\bibitem{Cirelli:2005uq}
M.~Cirelli, N.~Fornengo, and A.~Strumia, ``{Minimal dark matter},''
  \href{http://dx.doi.org/10.1016/j.nuclphysb.2006.07.012}{{\em Nucl.Phys.}
  {\bfseries B753} (2006) 178--194},
\href{http://arxiv.org/abs/hep-ph/0512090}{{\ttfamily arXiv:hep-ph/0512090
  [hep-ph]}}.
%%CITATION = HEP-PH/0512090;%%.

\bibitem{Cirelli:2007xd}
M.~Cirelli, A.~Strumia, and M.~Tamburini, ``{Cosmology and Astrophysics of
  Minimal Dark Matter},''
  \href{http://dx.doi.org/10.1016/j.nuclphysb.2007.07.023}{{\em Nucl.Phys.}
  {\bfseries B787} (2007) 152--175},
\href{http://arxiv.org/abs/0706.4071}{{\ttfamily arXiv:0706.4071 [hep-ph]}}.
%%CITATION = ARXIV:0706.4071;%%.

\bibitem{Cirelli:2009uv}
M.~Cirelli and A.~Strumia, ``{Minimal Dark Matter: Model and results},''
  \href{http://dx.doi.org/10.1088/1367-2630/11/10/105005}{{\em New J.Phys.}
  {\bfseries 11} (2009) 105005},
\href{http://arxiv.org/abs/0903.3381}{{\ttfamily arXiv:0903.3381 [hep-ph]}}.
%%CITATION = ARXIV:0903.3381;%%.

\bibitem{Cirelli:2014dsa}
M.~Cirelli, F.~Sala, and M.~Taoso, ``{Wino-like Minimal Dark Matter and future
  colliders},'' \href{http://dx.doi.org/10.1007/JHEP01(2015)041,
  10.1007/JHEP10(2014)033}{{\em JHEP} {\bfseries 1410} (2014) 033},
\href{http://arxiv.org/abs/1407.7058}{{\ttfamily arXiv:1407.7058 [hep-ph]}}.
%%CITATION = ARXIV:1407.7058;%%.

\bibitem{Jaeckel:2013ija}
J.~Jaeckel, ``{A force beyond the Standard Model - Status of the quest for
  hidden photons},'' {\em Frascati Phys.Ser.} {\bfseries 56} (2012) 172--192,
\href{http://arxiv.org/abs/1303.1821}{{\ttfamily arXiv:1303.1821 [hep-ph]}}.
%%CITATION = ARXIV:1303.1821;%%.

\bibitem{VanNieuwenhuizen:1981ae}
P.~Van~Nieuwenhuizen, ``{Supergravity},''
\href{http://dx.doi.org/10.1016/0370-1573(81)90157-5}{{\em Phys.Rept.}
  {\bfseries 68} (1981) 189--398}.
%%CITATION = PRPLC,68,189;%%.

\bibitem{Nilles:1983ge}
H.~P. Nilles, ``{Supersymmetry, Supergravity and Particle Physics},''
\href{http://dx.doi.org/10.1016/0370-1573(84)90008-5}{{\em Phys.Rept.}
  {\bfseries 110} (1984) 1--162}.
%%CITATION = PRPLC,110,1;%%.

\bibitem{Willenbrock:2004hu}
S.~Willenbrock, ``{Symmetries of the standard model},''
\href{http://arxiv.org/abs/hep-ph/0410370}{{\ttfamily arXiv:hep-ph/0410370
  [hep-ph]}}.
%%CITATION = HEP-PH/0410370;%%.

\bibitem{AbdusSalam:2013eya}
S.~S. AbdusSalam and T.~A. Chowdhury, ``{Scalar Representations in the Light of
  Electroweak Phase Transition and Cold Dark Matter Phenomenology},''
  \href{http://dx.doi.org/10.1088/1475-7516/2014/05/026}{{\em JCAP} {\bfseries
  1405} (2014) 026},
\href{http://arxiv.org/abs/1310.8152}{{\ttfamily arXiv:1310.8152 [hep-ph]}}.
%%CITATION = ARXIV:1310.8152;%%.

\bibitem{Pospelov:2005pr}
M.~Pospelov and A.~Ritz, ``{Electric dipole moments as probes of new
  physics},'' \href{http://dx.doi.org/10.1016/j.aop.2005.04.002}{{\em Annals
  Phys.} {\bfseries 318} (2005) 119--169},
\href{http://arxiv.org/abs/hep-ph/0504231}{{\ttfamily arXiv:hep-ph/0504231
  [hep-ph]}}.
%%CITATION = HEP-PH/0504231;%%.

\bibitem{Inoue:2014nva}
S.~Inoue, M.~J. Ramsey-Musolf, and Y.~Zhang, ``{CP-violating phenomenology of
  flavor conserving two Higgs doublet models},''
  \href{http://dx.doi.org/10.1103/PhysRevD.89.115023}{{\em Phys.Rev.}
  {\bfseries D89} no.~11, (2014) 115023},
\href{http://arxiv.org/abs/1403.4257}{{\ttfamily arXiv:1403.4257 [hep-ph]}}.
%%CITATION = ARXIV:1403.4257;%%.

\bibitem{Abe:2013qla}
T.~Abe, J.~Hisano, T.~Kitahara, and K.~Tobioka, ``{Gauge invariant Barr-Zee
  type contributions to fermionic EDMs in the two-Higgs doublet models},''
  \href{http://dx.doi.org/10.1007/JHEP01(2014)106}{{\em JHEP} {\bfseries 1401}
  (2014) 106},
\href{http://arxiv.org/abs/1311.4704}{{\ttfamily arXiv:1311.4704 [hep-ph]}}.
%%CITATION = ARXIV:1311.4704;%%.

\bibitem{Baron:2013eja}
{\bfseries ACME} Collaboration, J.~Baron {\em et~al.}, ``{Order of Magnitude
  Smaller Limit on the Electric Dipole Moment of the Electron},''
  \href{http://dx.doi.org/10.1126/science.1248213}{{\em Science} {\bfseries
  343} (2014) 269--272},
\href{http://arxiv.org/abs/1310.7534}{{\ttfamily arXiv:1310.7534
  [physics.atom-ph]}}.
%%CITATION = ARXIV:1310.7534;%%.

\bibitem{Beringer:1900zz}
{\bfseries Particle Data Group} Collaboration, J.~Beringer {\em et~al.},
  ``{Review of Particle Physics (RPP)},''
\href{http://dx.doi.org/10.1103/PhysRevD.86.010001}{{\em Phys.Rev.} {\bfseries
  D86} (2012) 010001}.
%%CITATION = PHRVA,D86,010001;%%.

\bibitem{Gunion:1989we}
J.~F. Gunion, H.~E. Haber, G.~L. Kane, and S.~Dawson, ``{The Higgs Hunter's
  Guide},''
{\em Front.Phys.} {\bfseries 80} (2000) 1--448.
%%CITATION = FRPHA,80,1;%%.

\bibitem{Hisano:2013sn}
J.~Hisano and K.~Tsumura, ``{Higgs boson mixes with an SU(2) septet
  representation},'' \href{http://dx.doi.org/10.1103/PhysRevD.87.053004}{{\em
  Phys.Rev.} {\bfseries D87} (2013) 053004},
\href{http://arxiv.org/abs/1301.6455}{{\ttfamily arXiv:1301.6455 [hep-ph]}}.
%%CITATION = ARXIV:1301.6455;%%.

\bibitem{Lavoura:1993nq}
L.~Lavoura and L.-F. Li, ``{Making the small oblique parameters large},''
  \href{http://dx.doi.org/10.1103/PhysRevD.49.1409}{{\em Phys.Rev.} {\bfseries
  D49} (1994) 1409--1416},
\href{http://arxiv.org/abs/hep-ph/9309262}{{\ttfamily arXiv:hep-ph/9309262
  [hep-ph]}}.
%%CITATION = HEP-PH/9309262;%%.

\bibitem{Djouadi:2005gj}
A.~Djouadi, ``{The Anatomy of electro-weak symmetry breaking. II. The Higgs
  bosons in the minimal supersymmetric model},''
  \href{http://dx.doi.org/10.1016/j.physrep.2007.10.005}{{\em Phys.Rept.}
  {\bfseries 459} (2008) 1--241},
\href{http://arxiv.org/abs/hep-ph/0503173}{{\ttfamily arXiv:hep-ph/0503173
  [hep-ph]}}.
%%CITATION = HEP-PH/0503173;%%.

\bibitem{Chang:2012ta}
W.-F. Chang, J.~N. Ng, and J.~M. Wu, ``{Constraints on New Scalars from the LHC
  125 GeV Higgs Signal},''
  \href{http://dx.doi.org/10.1103/PhysRevD.86.033003}{{\em Phys.Rev.}
  {\bfseries D86} (2012) 033003},
\href{http://arxiv.org/abs/1206.5047}{{\ttfamily arXiv:1206.5047 [hep-ph]}}.
%%CITATION = ARXIV:1206.5047;%%.

\bibitem{Dorsner:2012pp}
I.~Dorsner, S.~Fajfer, A.~Greljo, and J.~F. Kamenik, ``{Higgs Uncovering Light
  Scalar Remnants of High Scale Matter Unification},''
  \href{http://dx.doi.org/10.1007/JHEP11(2012)130}{{\em JHEP} {\bfseries 1211}
  (2012) 130},
\href{http://arxiv.org/abs/1208.1266}{{\ttfamily arXiv:1208.1266 [hep-ph]}}.
%%CITATION = ARXIV:1208.1266;%%.

\bibitem{LHCcxn}
``Lhc higgs cross section working group.''
\newblock \url{https://twiki.cern.ch/twiki/bin/view/LHCPhysics/LHCHXSWG}.

\bibitem{Chatrchyan:2014tja}
{\bfseries CMS} Collaboration, S.~Chatrchyan {\em et~al.}, ``{Search for
  invisible decays of Higgs bosons in the vector boson fusion and associated ZH
  production modes},''
  \href{http://dx.doi.org/10.1140/epjc/s10052-014-2980-6}{{\em Eur.Phys.J.}
  {\bfseries C74} (2014) 2980},
\href{http://arxiv.org/abs/1404.1344}{{\ttfamily arXiv:1404.1344 [hep-ex]}}.
%%CITATION = ARXIV:1404.1344;%%.

\bibitem{Dorsner:2015mja}
I.~Dorsner, S.~Fajfer, A.~Greljo, J.~F. Kamenik, N.~Kosnik, {\em et~al.},
  ``{New Physics Models Facing Lepton Flavor Violating Higgs Decays at the
  Percent Level},''
\href{http://arxiv.org/abs/1502.07784}{{\ttfamily arXiv:1502.07784 [hep-ph]}}.
%%CITATION = ARXIV:1502.07784;%%.

\bibitem{TheATLAScollaboration:2013lia}
T.~A. collaboration,
``{Search for the bb decay of the Standard Model Higgs boson in associated W/ZH
  production with the ATLAS detector},''.
%%CITATION = ATLAS-CONF-2013-079 ETC.;%%.

\bibitem{Aad:2013wqa}
{\bfseries ATLAS} Collaboration, G.~Aad {\em et~al.}, ``{Measurements of Higgs
  boson production and couplings in diboson final states with the ATLAS
  detector at the LHC},''
  \href{http://dx.doi.org/10.1016/j.physletb.2014.05.011,
  10.1016/j.physletb.2013.08.010}{{\em Phys.Lett.} {\bfseries B726} (2013)
  88--119},
\href{http://arxiv.org/abs/1307.1427}{{\ttfamily arXiv:1307.1427 [hep-ex]}}.
%%CITATION = ARXIV:1307.1427;%%.

\bibitem{ATLASttbb}
T.~A. collaboration,
``{Updated coupling measurements of the Higgs boson with the ATLAS detector
  using up to 25 fb$^{-1}$ of proton-proton collision data},''.
%%CITATION = ATLAS-CONF-2014-009 ETC.;%%.

\bibitem{Aad:2014iia}
{\bfseries ATLAS} Collaboration, G.~Aad {\em et~al.}, ``{Search for Invisible
  Decays of a Higgs Boson Produced in Association with a Z Boson in ATLAS},''
  \href{http://dx.doi.org/10.1103/PhysRevLett.112.201802}{{\em Phys.Rev.Lett.}
  {\bfseries 112} (2014) 201802},
\href{http://arxiv.org/abs/1402.3244}{{\ttfamily arXiv:1402.3244 [hep-ex]}}.
%%CITATION = ARXIV:1402.3244;%%.

\bibitem{CMS:2013jda}
{\bfseries CMS} Collaboration, C.~Collaboration,
``{Higgs to bb in the VBF channel},''.
%%CITATION = CMS-PAS-HIG-13-011 ETC.;%%.

\bibitem{CMS:2013sea}
{\bfseries CMS} Collaboration, C.~Collaboration,
``{Search for Higgs Boson Production in Association with a Top-Quark Pair and
  Decaying to Bottom Quarks or Tau Leptons},''.
%%CITATION = CMS-PAS-HIG-13-019 ETC.;%%.

\bibitem{Chatrchyan:2013iaa}
{\bfseries CMS} Collaboration, S.~Chatrchyan {\em et~al.}, ``{Measurement of
  Higgs boson production and properties in the WW decay channel with leptonic
  final states},'' \href{http://dx.doi.org/10.1007/JHEP01(2014)096}{{\em JHEP}
  {\bfseries 1401} (2014) 096},
\href{http://arxiv.org/abs/1312.1129}{{\ttfamily arXiv:1312.1129 [hep-ex]}}.
%%CITATION = ARXIV:1312.1129;%%.

\bibitem{Chatrchyan:2013mxa}
{\bfseries CMS} Collaboration, S.~Chatrchyan {\em et~al.}, ``{Measurement of
  the properties of a Higgs boson in the four-lepton final state},''
  \href{http://dx.doi.org/10.1103/PhysRevD.89.092007}{{\em Phys.Rev.}
  {\bfseries D89} no.~9, (2014) 092007},
\href{http://arxiv.org/abs/1312.5353}{{\ttfamily arXiv:1312.5353 [hep-ex]}}.
%%CITATION = ARXIV:1312.5353;%%.

\bibitem{CMS:ril}
{\bfseries CMS} Collaboration, S.~Chatrchyan {\em et~al.},
``{Updated measurements of the Higgs boson at 125 GeV in the two photon decay
  channel},''.
%%CITATION = CMS-PAS-HIG-13-001 ETC.;%%.

\bibitem{CMS:2014ala}
{\bfseries CMS} Collaboration, C.~Collaboration,
``{Constraints on the Higgs boson width from off-shell production and decay to
  ZZ to llll and llvv},''.
%%CITATION = CMS-PAS-HIG-14-002 ETC.;%%.

\bibitem{DelNobile:2009st}
E.~Del~Nobile, R.~Franceschini, D.~Pappadopulo, and A.~Strumia, ``{Minimal
  Matter at the Large Hadron Collider},''
  \href{http://dx.doi.org/10.1016/j.nuclphysb.2009.10.004}{{\em Nucl.Phys.}
  {\bfseries B826} (2010) 217--234},
\href{http://arxiv.org/abs/0908.1567}{{\ttfamily arXiv:0908.1567 [hep-ph]}}.
%%CITATION = ARXIV:0908.1567;%%.

\bibitem{Giudice:2014tma}
G.~F. Giudice, G.~Isidori, A.~Salvio, and A.~Strumia, ``{Softened Gravity and
  the Extension of the Standard Model up to Infinite Energy},''
  \href{http://dx.doi.org/10.1007/JHEP02(2015)137}{{\em JHEP} {\bfseries 1502}
  (2015) 137},
\href{http://arxiv.org/abs/1412.2769}{{\ttfamily arXiv:1412.2769 [hep-ph]}}.
%%CITATION = ARXIV:1412.2769;%%.

\bibitem{Barger:2007im}
V.~Barger, P.~Langacker, M.~McCaskey, M.~J. Ramsey-Musolf, and G.~Shaughnessy,
  ``{LHC Phenomenology of an Extended Standard Model with a Real Scalar
  Singlet},'' \href{http://dx.doi.org/10.1103/PhysRevD.77.035005}{{\em
  Phys.Rev.} {\bfseries D77} (2008) 035005},
\href{http://arxiv.org/abs/0706.4311}{{\ttfamily arXiv:0706.4311 [hep-ph]}}.
%%CITATION = ARXIV:0706.4311;%%.

\bibitem{Kleiss:1985gy}
R.~Kleiss, W.~J. Stirling, and S.~Ellis, ``{A New Monte Carlo Treatment of
  Multiparticle Phase Space at High-energies},''
\href{http://dx.doi.org/10.1016/0010-4655(86)90119-0}{{\em Comput.Phys.Commun.}
  {\bfseries 40} (1986) 359}.
%%CITATION = CPHCB,40,359;%%.

\bibitem{Alwall:2011uj}
J.~Alwall, M.~Herquet, F.~Maltoni, O.~Mattelaer, and T.~Stelzer, ``{MadGraph 5
  : Going Beyond},'' \href{http://dx.doi.org/10.1007/JHEP06(2011)128}{{\em
  JHEP} {\bfseries 1106} (2011) 128},
\href{http://arxiv.org/abs/1106.0522}{{\ttfamily arXiv:1106.0522 [hep-ph]}}.
%%CITATION = ARXIV:1106.0522;%%.

\bibitem{Alloul:2013bka}
A.~Alloul, N.~D. Christensen, C.~Degrande, C.~Duhr, and B.~Fuks, ``{FeynRules
  2.0 - A complete toolbox for tree-level phenomenology},''
  \href{http://dx.doi.org/10.1016/j.cpc.2014.04.012}{{\em Comput.Phys.Commun.}
  {\bfseries 185} (2014) 2250--2300},
\href{http://arxiv.org/abs/1310.1921}{{\ttfamily arXiv:1310.1921 [hep-ph]}}.
%%CITATION = ARXIV:1310.1921;%%.

\bibitem{Grober:2015fia}
R.~Grober, M.~Muhlleitner, E.~Popenda, and A.~Wlotzka, ``{Light stop decays
  into $W b \tilde{\chi}_1^0$ near the kinematic threshold},''
\href{http://arxiv.org/abs/1502.05935}{{\ttfamily arXiv:1502.05935 [hep-ph]}}.
%%CITATION = ARXIV:1502.05935;%%.

\bibitem{DNPN}
E.~Del~Nobile, M.~Nardecchia, and P.~Panci, ``{On Decaying Minimal Dark
  Matter},'' {\em work in progress} (2015) .

\bibitem{Kang:2006yd}
J.~Kang, M.~A. Luty, and S.~Nasri, ``{The Relic abundance of long-lived heavy
  colored particles},''
  \href{http://dx.doi.org/10.1088/1126-6708/2008/09/086}{{\em JHEP} {\bfseries
  0809} (2008) 086},
\href{http://arxiv.org/abs/hep-ph/0611322}{{\ttfamily arXiv:hep-ph/0611322
  [hep-ph]}}.
%%CITATION = HEP-PH/0611322;%%.

\bibitem{Fields:2006ga}
B.~Fields and S.~Sarkar, ``{Big-Bang nucleosynthesis (2006 Particle Data Group
  mini-review)},''
\href{http://arxiv.org/abs/astro-ph/0601514}{{\ttfamily arXiv:astro-ph/0601514
  [astro-ph]}}.
%%CITATION = ASTRO-PH/0601514;%%.

\bibitem{Hu:1993gc}
W.~Hu and J.~Silk, ``{Thermalization constraints and spectral distortions for
  massive unstable relic particles},''
\href{http://dx.doi.org/10.1103/PhysRevLett.70.2661}{{\em Phys.Rev.Lett.}
  {\bfseries 70} (1993) 2661--2664}.
%%CITATION = PRLTA,70,2661;%%.

\bibitem{Kribs:1996ac}
G.~D. Kribs and I.~Rothstein, ``{Bounds on longlived relics from diffuse
  gamma-ray observations},'' \href{http://dx.doi.org/10.1103/PhysRevD.56.1822,
  10.1103/PhysRevD.55.4435, 10.1103/PhysRevD.55.4435
  10.1103/PhysRevD.56.1822}{{\em Phys.Rev.} {\bfseries D55} (1997) 4435--4449},
\href{http://arxiv.org/abs/hep-ph/9610468}{{\ttfamily arXiv:hep-ph/9610468
  [hep-ph]}}.
%%CITATION = HEP-PH/9610468;%%.

\bibitem{Ackermann:2012qk}
{\bfseries LAT} Collaboration, M.~Ackermann {\em et~al.}, ``{Fermi LAT Search
  for Dark Matter in Gamma-ray Lines and the Inclusive Photon Spectrum},''
  \href{http://dx.doi.org/10.1103/PhysRevD.86.022002}{{\em Phys.Rev.}
  {\bfseries D86} (2012) 022002},
\href{http://arxiv.org/abs/1205.2739}{{\ttfamily arXiv:1205.2739
  [astro-ph.HE]}}.
%%CITATION = ARXIV:1205.2739;%%.

\bibitem{Burdin:2014xma}
S.~Burdin, M.~Fairbairn, P.~Mermod, D.~Milstead, J.~Pinfold, {\em et~al.},
  ``{Non-collider searches for stable massive particles},''
\href{http://arxiv.org/abs/1410.1374}{{\ttfamily arXiv:1410.1374 [hep-ph]}}.
%%CITATION = ARXIV:1410.1374;%%.

\bibitem{Chuzhoy:2008zy}
L.~Chuzhoy and E.~W. Kolb, ``{Reopening the window on charged dark matter},''
  \href{http://dx.doi.org/10.1088/1475-7516/2009/07/014}{{\em JCAP} {\bfseries
  0907} (2009) 014},
\href{http://arxiv.org/abs/0809.0436}{{\ttfamily arXiv:0809.0436 [astro-ph]}}.
%%CITATION = ARXIV:0809.0436;%%.

\bibitem{Langacker:2011db}
P.~Langacker and G.~Steigman, ``{Requiem for an FCHAMP? Fractionally CHArged,
  Massive Particle},'' \href{http://dx.doi.org/10.1103/PhysRevD.84.065040}{{\em
  Phys.Rev.} {\bfseries D84} (2011) 065040},
\href{http://arxiv.org/abs/1107.3131}{{\ttfamily arXiv:1107.3131 [hep-ph]}}.
%%CITATION = ARXIV:1107.3131;%%.

\bibitem{Perl:2009zz}
M.~L. Perl, E.~R. Lee, and D.~Loomba, ``{Searches for fractionally charged
  particles},''
\href{http://dx.doi.org/10.1146/annurev-nucl-121908-122035}{{\em
  Ann.Rev.Nucl.Part.Sci.} {\bfseries 59} (2009) 47--65}.
%%CITATION = ARNUA,59,47;%%.

\bibitem{Davidson:2000hf}
S.~Davidson, S.~Hannestad, and G.~Raffelt, ``{Updated bounds on millicharged
  particles},'' \href{http://dx.doi.org/10.1088/1126-6708/2000/05/003}{{\em
  JHEP} {\bfseries 0005} (2000) 003},
\href{http://arxiv.org/abs/hep-ph/0001179}{{\ttfamily arXiv:hep-ph/0001179
  [hep-ph]}}.
%%CITATION = HEP-PH/0001179;%%.

\bibitem{Antipin:2015xia}
O.~Antipin, M.~Redi, A.~Strumia, and E.~Vigiani, ``{Accidental Composite Dark
  Matter},''
\href{http://arxiv.org/abs/1503.08749}{{\ttfamily arXiv:1503.08749 [hep-ph]}}.
%%CITATION = ARXIV:1503.08749;%%.

\bibitem{Abbott:1982af}
L.~Abbott and P.~Sikivie, ``{A Cosmological Bound on the Invisible Axion},''
\href{http://dx.doi.org/10.1016/0370-2693(83)90638-X}{{\em Phys.Lett.}
  {\bfseries B120} (1983) 133--136}.
%%CITATION = PHLTA,B120,133;%%.

\bibitem{Dine:1982ah}
M.~Dine and W.~Fischler, ``{The Not So Harmless Axion},''
\href{http://dx.doi.org/10.1016/0370-2693(83)90639-1}{{\em Phys.Lett.}
  {\bfseries B120} (1983) 137--141}.
%%CITATION = PHLTA,B120,137;%%.

\bibitem{Preskill:1982cy}
J.~Preskill, M.~B. Wise, and F.~Wilczek, ``{Cosmology of the Invisible
  Axion},''
\href{http://dx.doi.org/10.1016/0370-2693(83)90637-8}{{\em Phys.Lett.}
  {\bfseries B120} (1983) 127--132}.
%%CITATION = PHLTA,B120,127;%%.

\bibitem{Edsjo:1997bg}
J.~Edsjo and P.~Gondolo, ``{Neutralino relic density including
  coannihilations},'' \href{http://dx.doi.org/10.1103/PhysRevD.56.1879}{{\em
  Phys.Rev.} {\bfseries D56} (1997) 1879--1894},
\href{http://arxiv.org/abs/hep-ph/9704361}{{\ttfamily arXiv:hep-ph/9704361
  [hep-ph]}}.
%%CITATION = HEP-PH/9704361;%%.

\bibitem{Kolb:1990vq}
E.~W. Kolb and M.~S. Turner, ``{The Early Universe},''
{\em Front.Phys.} {\bfseries 69} (1990) 1--547.
%%CITATION = FRPHA,69,1;%%.

\bibitem{Hisano:2006nn}
J.~Hisano, S.~Matsumoto, M.~Nagai, O.~Saito, and M.~Senami, ``{Non-perturbative
  effect on thermal relic abundance of dark matter},''
  \href{http://dx.doi.org/10.1016/j.physletb.2007.01.012}{{\em Phys.Lett.}
  {\bfseries B646} (2007) 34--38},
\href{http://arxiv.org/abs/hep-ph/0610249}{{\ttfamily arXiv:hep-ph/0610249
  [hep-ph]}}.
%%CITATION = HEP-PH/0610249;%%.

\bibitem{Kawasaki:2004qu}
M.~Kawasaki, K.~Kohri, and T.~Moroi, ``{Big-Bang nucleosynthesis and hadronic
  decay of long-lived massive particles},''
  \href{http://dx.doi.org/10.1103/PhysRevD.71.083502}{{\em Phys.Rev.}
  {\bfseries D71} (2005) 083502},
\href{http://arxiv.org/abs/astro-ph/0408426}{{\ttfamily arXiv:astro-ph/0408426
  [astro-ph]}}.
%%CITATION = ASTRO-PH/0408426;%%.

\bibitem{Berger:2008ti}
C.~F. Berger, L.~Covi, S.~Kraml, and F.~Palorini, ``{The Number density of a
  charged relic},'' \href{http://dx.doi.org/10.1088/1475-7516/2008/10/005}{{\em
  JCAP} {\bfseries 0810} (2008) 005},
\href{http://arxiv.org/abs/0807.0211}{{\ttfamily arXiv:0807.0211 [hep-ph]}}.
%%CITATION = ARXIV:0807.0211;%%.

\bibitem{Gondolo:1990dk}
P.~Gondolo and G.~Gelmini, ``{Cosmic abundances of stable particles: Improved
  analysis},''
\href{http://dx.doi.org/10.1016/0550-3213(91)90438-4}{{\em Nucl.Phys.}
  {\bfseries B360} (1991) 145--179}.
%%CITATION = NUPHA,B360,145;%%.

\bibitem{Iocco:2008va}
F.~Iocco, G.~Mangano, G.~Miele, O.~Pisanti, and P.~D. Serpico, ``{Primordial
  Nucleosynthesis: from precision cosmology to fundamental physics},''
  \href{http://dx.doi.org/10.1016/j.physrep.2009.02.002}{{\em Phys.Rept.}
  {\bfseries 472} (2009) 1--76},
\href{http://arxiv.org/abs/0809.0631}{{\ttfamily arXiv:0809.0631 [astro-ph]}}.
%%CITATION = ARXIV:0809.0631;%%.

\bibitem{Lindley:1984bg}
D.~Lindley, ``{Cosmological Constraints on the Lifetime of Massive
  Particles},''
\href{http://dx.doi.org/10.1086/163267}{{\em Astrophys.J.} {\bfseries 294}
  (1985) 1--8}.
%%CITATION = ASJOA,294,1;%%.

\bibitem{Reno:1987qw}
M.~Reno and D.~Seckel, ``{Primordial Nucleosynthesis: The Effects of Injecting
  Hadrons},''
\href{http://dx.doi.org/10.1103/PhysRevD.37.3441}{{\em Phys.Rev.} {\bfseries
  D37} (1988) 3441}.
%%CITATION = PHRVA,D37,3441;%%.

\bibitem{Dimopoulos:1987fz}
S.~Dimopoulos, R.~Esmailzadeh, L.~J. Hall, and G.~Starkman, ``{Is the Universe
  Closed by Baryons? Nucleosynthesis With a Late Decaying Massive Particle},''
\href{http://dx.doi.org/10.1086/166493}{{\em Astrophys.J.} {\bfseries 330}
  (1988) 545}.
%%CITATION = ASJOA,330,545;%%.

\bibitem{Scherrer:1987rr}
R.~J. Scherrer and M.~S. Turner, ``{Primordial Nucleosynthesis with Decaying
  Particles. 1. Entropy Producing Decays. 2. Inert Decays},''
\href{http://dx.doi.org/10.1086/166534}{{\em Astrophys.J.} {\bfseries 331}
  (1988) 19--32}.
%%CITATION = ASJOA,331,19;%%.

\bibitem{Ellis:1990nb}
J.~R. Ellis, G.~Gelmini, J.~L. Lopez, D.~V. Nanopoulos, and S.~Sarkar,
  ``{Astrophysical constraints on massive unstable neutral relic particles},''
\href{http://dx.doi.org/10.1016/0550-3213(92)90438-H}{{\em Nucl.Phys.}
  {\bfseries B373} (1992) 399--437}.
%%CITATION = NUPHA,B373,399;%%.

\bibitem{Pospelov:2006sc}
M.~Pospelov, ``{Particle physics catalysis of thermal Big Bang
  Nucleosynthesis},''
  \href{http://dx.doi.org/10.1103/PhysRevLett.98.231301}{{\em Phys.Rev.Lett.}
  {\bfseries 98} (2007) 231301},
\href{http://arxiv.org/abs/hep-ph/0605215}{{\ttfamily arXiv:hep-ph/0605215
  [hep-ph]}}.
%%CITATION = HEP-PH/0605215;%%.

\bibitem{Kohri:2006cn}
K.~Kohri and F.~Takayama, ``{Big bang nucleosynthesis with long lived charged
  massive particles},''
  \href{http://dx.doi.org/10.1103/PhysRevD.76.063507}{{\em Phys.Rev.}
  {\bfseries D76} (2007) 063507},
\href{http://arxiv.org/abs/hep-ph/0605243}{{\ttfamily arXiv:hep-ph/0605243
  [hep-ph]}}.
%%CITATION = HEP-PH/0605243;%%.

\bibitem{Kaplinghat:2006qr}
M.~Kaplinghat and A.~Rajaraman, ``{Big Bang Nucleosynthesis with Bound States
  of Long-lived Charged Particles},''
  \href{http://dx.doi.org/10.1103/PhysRevD.74.103004}{{\em Phys.Rev.}
  {\bfseries D74} (2006) 103004},
\href{http://arxiv.org/abs/astro-ph/0606209}{{\ttfamily arXiv:astro-ph/0606209
  [astro-ph]}}.
%%CITATION = ASTRO-PH/0606209;%%.

\bibitem{Bird:2007ge}
C.~Bird, K.~Koopmans, and M.~Pospelov, ``{Primordial Lithium Abundance in
  Catalyzed Big Bang Nucleosynthesis},''
  \href{http://dx.doi.org/10.1103/PhysRevD.78.083010}{{\em Phys.Rev.}
  {\bfseries D78} (2008) 083010},
\href{http://arxiv.org/abs/hep-ph/0703096}{{\ttfamily arXiv:hep-ph/0703096
  [hep-ph]}}.
%%CITATION = HEP-PH/0703096;%%.

\bibitem{Jittoh:2007fr}
T.~Jittoh, K.~Kohri, M.~Koike, J.~Sato, T.~Shimomura, {\em et~al.}, ``{Possible
  solution to the Li-7 problem by the long lived stau},''
  \href{http://dx.doi.org/10.1103/PhysRevD.76.125023}{{\em Phys.Rev.}
  {\bfseries D76} (2007) 125023},
\href{http://arxiv.org/abs/0704.2914}{{\ttfamily arXiv:0704.2914 [hep-ph]}}.
%%CITATION = ARXIV:0704.2914;%%.

\bibitem{Jedamzik:2007cp}
K.~Jedamzik, ``{The cosmic Li-6 and Li-7 problems and BBN with long-lived
  charged massive particles},''
  \href{http://dx.doi.org/10.1103/PhysRevD.77.063524}{{\em Phys.Rev.}
  {\bfseries D77} (2008) 063524},
\href{http://arxiv.org/abs/0707.2070}{{\ttfamily arXiv:0707.2070 [astro-ph]}}.
%%CITATION = ARXIV:0707.2070;%%.

\bibitem{Jittoh:2008eq}
T.~Jittoh, K.~Kohri, M.~Koike, J.~Sato, T.~Shimomura, {\em et~al.}, ``{Big-bang
  nucleosynthesis and the relic abundance of dark matter in a stau-neutralino
  coannihilation scenario},''
  \href{http://dx.doi.org/10.1103/PhysRevD.78.055007}{{\em Phys.Rev.}
  {\bfseries D78} (2008) 055007},
\href{http://arxiv.org/abs/0805.3389}{{\ttfamily arXiv:0805.3389 [hep-ph]}}.
%%CITATION = ARXIV:0805.3389;%%.

\bibitem{Cumberbatch:2007me}
D.~Cumberbatch, K.~Ichikawa, M.~Kawasaki, K.~Kohri, J.~Silk, {\em et~al.},
  ``{Solving the cosmic lithium problems with primordial late-decaying
  particles},'' \href{http://dx.doi.org/10.1103/PhysRevD.76.123005}{{\em
  Phys.Rev.} {\bfseries D76} (2007) 123005},
\href{http://arxiv.org/abs/0708.0095}{{\ttfamily arXiv:0708.0095 [astro-ph]}}.
%%CITATION = ARXIV:0708.0095;%%.

\bibitem{Kusakabe:2007fu}
M.~Kusakabe, T.~Kajino, R.~N. Boyd, T.~Yoshida, and G.~J. Mathews, ``{A
  Simultaneous Solution to the 6Li and 7Li Big Bang Nucleosynthesis Problems
  from a Long-Lived Negatively-Charged Leptonic Particle},''
  \href{http://dx.doi.org/10.1103/PhysRevD.76.121302}{{\em Phys.Rev.}
  {\bfseries D76} (2007) 121302},
\href{http://arxiv.org/abs/0711.3854}{{\ttfamily arXiv:0711.3854 [astro-ph]}}.
%%CITATION = ARXIV:0711.3854;%%.

\bibitem{Fields:2011zzb}
B.~D. Fields, ``{The primordial lithium problem},''
  \href{http://dx.doi.org/10.1146/annurev-nucl-102010-130445}{{\em
  Ann.Rev.Nucl.Part.Sci.} {\bfseries 61} (2011) 47--68},
\href{http://arxiv.org/abs/1203.3551}{{\ttfamily arXiv:1203.3551
  [astro-ph.CO]}}.
%%CITATION = ARXIV:1203.3551;%%.

\bibitem{Steigman:2007xt}
G.~Steigman, ``{Primordial Nucleosynthesis in the Precision Cosmology Era},''
  \href{http://dx.doi.org/10.1146/annurev.nucl.56.080805.140437}{{\em
  Ann.Rev.Nucl.Part.Sci.} {\bfseries 57} (2007) 463--491},
\href{http://arxiv.org/abs/0712.1100}{{\ttfamily arXiv:0712.1100 [astro-ph]}}.
%%CITATION = ARXIV:0712.1100;%%.

\bibitem{Lopez:1998vk}
R.~E. Lopez and M.~S. Turner, ``{An Accurate calculation of the big bang
  prediction for the abundance of primordial helium},''
  \href{http://dx.doi.org/10.1103/PhysRevD.59.103502}{{\em Phys.Rev.}
  {\bfseries D59} (1999) 103502},
\href{http://arxiv.org/abs/astro-ph/9807279}{{\ttfamily arXiv:astro-ph/9807279
  [astro-ph]}}.
%%CITATION = ASTRO-PH/9807279;%%.

\bibitem{Mukhanov:2003xs}
V.~F. Mukhanov, ``{Nucleosynthesis without a computer},''
  \href{http://dx.doi.org/10.1023/B:IJTP.0000048169.69609.77}{{\em
  Int.J.Theor.Phys.} {\bfseries 43} (2004) 669--693},
\href{http://arxiv.org/abs/astro-ph/0303073}{{\ttfamily arXiv:astro-ph/0303073
  [astro-ph]}}.
%%CITATION = ASTRO-PH/0303073;%%.

\bibitem{Weinberg:2008zzc}
S.~Weinberg, {\em {Cosmology}}.
\newblock {Oxford University Press},
2008.
\newblock
%%CITATION = ISBN-9780198526827 ETC.;%%.

\bibitem{Pospelov:2010hj}
M.~Pospelov and J.~Pradler, ``{Big Bang Nucleosynthesis as a Probe of New
  Physics},'' \href{http://dx.doi.org/10.1146/annurev.nucl.012809.104521}{{\em
  Ann.Rev.Nucl.Part.Sci.} {\bfseries 60} (2010) 539--568},
\href{http://arxiv.org/abs/1011.1054}{{\ttfamily arXiv:1011.1054 [hep-ph]}}.
%%CITATION = ARXIV:1011.1054;%%.

\bibitem{Englert:2014uua}
C.~Englert, A.~Freitas, M.~Mühlleitner, T.~Plehn, M.~Rauch, {\em et~al.},
  ``{Precision Measurements of Higgs Couplings: Implications for New Physics
  Scales},'' \href{http://dx.doi.org/10.1088/0954-3899/41/11/113001}{{\em
  J.Phys.} {\bfseries G41} (2014) 113001},
\href{http://arxiv.org/abs/1403.7191}{{\ttfamily arXiv:1403.7191 [hep-ph]}}.
%%CITATION = ARXIV:1403.7191;%%.

\bibitem{Carena:2012xa}
M.~Carena, I.~Low, and C.~E. Wagner, ``{Implications of a Modified Higgs to
  Diphoton Decay Width},''
  \href{http://dx.doi.org/10.1007/JHEP08(2012)060}{{\em JHEP} {\bfseries 1208}
  (2012) 060},
\href{http://arxiv.org/abs/1206.1082}{{\ttfamily arXiv:1206.1082 [hep-ph]}}.
%%CITATION = ARXIV:1206.1082;%%.

\bibitem{Picek:2012ei}
I.~Picek and B.~Radovcic, ``{Enhancement of $h \to \gamma \gamma$ by
  seesaw-motivated exotic scalars},''
  \href{http://dx.doi.org/10.1016/j.physletb.2013.01.056}{{\em Phys.Lett.}
  {\bfseries B719} (2013) 404--408},
\href{http://arxiv.org/abs/1210.6449}{{\ttfamily arXiv:1210.6449 [hep-ph]}}.
%%CITATION = ARXIV:1210.6449;%%.

\bibitem{ATLAS:2012hi}
{\bfseries ATLAS} Collaboration, G.~Aad {\em et~al.}, ``{Search for
  doubly-charged Higgs bosons in like-sign dilepton final states at
  $\sqrt{s}=7$ TeV with the ATLAS detector},''
  \href{http://dx.doi.org/10.1140/epjc/s10052-012-2244-2}{{\em Eur.Phys.J.}
  {\bfseries C72} (2012) 2244},
\href{http://arxiv.org/abs/1210.5070}{{\ttfamily arXiv:1210.5070 [hep-ex]}}.
%%CITATION = ARXIV:1210.5070;%%.

\bibitem{Chatrchyan:2012ya}
{\bfseries CMS} Collaboration, S.~Chatrchyan {\em et~al.}, ``{A search for a
  doubly-charged Higgs boson in $pp$ collisions at $\sqrt{s}=7$ TeV},''
  \href{http://dx.doi.org/10.1140/epjc/s10052-012-2189-5}{{\em Eur.Phys.J.}
  {\bfseries C72} (2012) 2189},
\href{http://arxiv.org/abs/1207.2666}{{\ttfamily arXiv:1207.2666 [hep-ex]}}.
%%CITATION = ARXIV:1207.2666;%%.

\bibitem{Melfo:2011nx}
A.~Melfo, M.~Nemevsek, F.~Nesti, G.~Senjanovic, and Y.~Zhang, ``{Type II Seesaw
  at LHC: The Roadmap},''
  \href{http://dx.doi.org/10.1103/PhysRevD.85.055018}{{\em Phys.Rev.}
  {\bfseries D85} (2012) 055018},
\href{http://arxiv.org/abs/1108.4416}{{\ttfamily arXiv:1108.4416 [hep-ph]}}.
%%CITATION = ARXIV:1108.4416;%%.

\bibitem{Ren:2011mh}
B.~Ren, K.~Tsumura, and X.-G. He, ``{A Higgs Quadruplet for Type III Seesaw and
  Implications for $\mu \to e\gamma$ and $\mu - e$ Conversion},''
  \href{http://dx.doi.org/10.1103/PhysRevD.84.073004}{{\em Phys.Rev.}
  {\bfseries D84} (2011) 073004},
\href{http://arxiv.org/abs/1107.5879}{{\ttfamily arXiv:1107.5879 [hep-ph]}}.
%%CITATION = ARXIV:1107.5879;%%.

\bibitem{Babu:2009aq}
K.~Babu, S.~Nandi, and Z.~Tavartkiladze, ``{New Mechanism for Neutrino Mass
  Generation and Triply Charged Higgs Bosons at the LHC},''
  \href{http://dx.doi.org/10.1103/PhysRevD.80.071702}{{\em Phys.Rev.}
  {\bfseries D80} (2009) 071702},
\href{http://arxiv.org/abs/0905.2710}{{\ttfamily arXiv:0905.2710 [hep-ph]}}.
%%CITATION = ARXIV:0905.2710;%%.

\bibitem{Englert:2013wga}
C.~Englert, E.~Re, and M.~Spannowsky, ``{Pinning down Higgs triplets at the
  LHC},'' \href{http://dx.doi.org/10.1103/PhysRevD.88.035024}{{\em Phys.Rev.}
  {\bfseries D88} (2013) 035024},
\href{http://arxiv.org/abs/1306.6228}{{\ttfamily arXiv:1306.6228 [hep-ph]}}.
%%CITATION = ARXIV:1306.6228;%%.

\bibitem{Barate:1997dr}
{\bfseries ALEPH} Collaboration, R.~Barate {\em et~al.}, ``{Search for pair
  production of longlived heavy charged particles in e+ e- annihilation},''
  \href{http://dx.doi.org/10.1016/S0370-2693(97)00715-6}{{\em Phys.Lett.}
  {\bfseries B405} (1997) 379--388},
\href{http://arxiv.org/abs/hep-ex/9706013}{{\ttfamily arXiv:hep-ex/9706013
  [hep-ex]}}.
%%CITATION = HEP-EX/9706013;%%.

\bibitem{Abreu:2000tn}
{\bfseries DELPHI} Collaboration, P.~Abreu {\em et~al.}, ``{Search for heavy
  stable and longlived particles in e+ e- collisions at s**(1/2) = 189-GeV},''
  \href{http://dx.doi.org/10.1016/S0370-2693(00)00265-3}{{\em Phys.Lett.}
  {\bfseries B478} (2000) 65--72},
\href{http://arxiv.org/abs/hep-ex/0103038}{{\ttfamily arXiv:hep-ex/0103038
  [hep-ex]}}.
%%CITATION = HEP-EX/0103038;%%.

\bibitem{Achard:2001qw}
{\bfseries L3} Collaboration, P.~Achard {\em et~al.}, ``{Search for heavy
  neutral and charged leptons in $e^{+} e^{-}$ annihilation at LEP},''
  \href{http://dx.doi.org/10.1016/S0370-2693(01)01005-X}{{\em Phys.Lett.}
  {\bfseries B517} (2001) 75--85},
\href{http://arxiv.org/abs/hep-ex/0107015}{{\ttfamily arXiv:hep-ex/0107015
  [hep-ex]}}.
%%CITATION = HEP-EX/0107015;%%.

\bibitem{Abbiendi:2003yd}
{\bfseries OPAL} Collaboration, G.~Abbiendi {\em et~al.}, ``{Search for stable
  and longlived massive charged particles in e+ e- collisions at s**(1/2) =
  130-GeV to 209-GeV},''
  \href{http://dx.doi.org/10.1016/S0370-2693(03)00639-7}{{\em Phys.Lett.}
  {\bfseries B572} (2003) 8--20},
\href{http://arxiv.org/abs/hep-ex/0305031}{{\ttfamily arXiv:hep-ex/0305031
  [hep-ex]}}.
%%CITATION = HEP-EX/0305031;%%.

\bibitem{Aktas:2004pq}
{\bfseries H1} Collaboration, A.~Aktas {\em et~al.}, ``{Measurement of
  anti-deuteron photoproduction and a search for heavy stable charged particles
  at HERA},'' {\em Eur.Phys.J.} {\bfseries C36} (2004) 413--423,
\href{http://arxiv.org/abs/hep-ex/0403056}{{\ttfamily arXiv:hep-ex/0403056
  [hep-ex]}}.
%%CITATION = HEP-EX/0403056;%%.

\bibitem{Abazov:2008qu}
{\bfseries D0} Collaboration, V.~Abazov {\em et~al.}, ``{Search for Long-Lived
  Charged Massive Particles with the D0 Detector},''
  \href{http://dx.doi.org/10.1103/PhysRevLett.102.161802}{{\em Phys.Rev.Lett.}
  {\bfseries 102} (2009) 161802},
\href{http://arxiv.org/abs/0809.4472}{{\ttfamily arXiv:0809.4472 [hep-ex]}}.
%%CITATION = ARXIV:0809.4472;%%.

\bibitem{Aaltonen:2009kea}
{\bfseries CDF} Collaboration, T.~Aaltonen {\em et~al.}, ``{Search for
  Long-Lived Massive Charged Particles in 1.96 TeV $\bar{p}p$ Collisions},''
  \href{http://dx.doi.org/10.1103/PhysRevLett.103.021802}{{\em Phys.Rev.Lett.}
  {\bfseries 103} (2009) 021802},
\href{http://arxiv.org/abs/0902.1266}{{\ttfamily arXiv:0902.1266 [hep-ex]}}.
%%CITATION = ARXIV:0902.1266;%%.

\bibitem{Abazov:2012ina}
{\bfseries D0} Collaboration, V.~M. Abazov {\em et~al.}, ``{Search for charged
  massive long-lived particles at $\sqrt{s}=1.96$ TeV},''
  \href{http://dx.doi.org/10.1103/PhysRevD.87.052011}{{\em Phys.Rev.}
  {\bfseries D87} no.~5, (2013) 052011},
\href{http://arxiv.org/abs/1211.2466}{{\ttfamily arXiv:1211.2466 [hep-ex]}}.
%%CITATION = ARXIV:1211.2466;%%.

\bibitem{Khachatryan:2011ts}
{\bfseries CMS} Collaboration, V.~Khachatryan {\em et~al.}, ``{Search for Heavy
  Stable Charged Particles in $pp$ collisions at $\sqrt{s}=7$ TeV},''
  \href{http://dx.doi.org/10.1007/JHEP03(2011)024}{{\em JHEP} {\bfseries 1103}
  (2011) 024},
\href{http://arxiv.org/abs/1101.1645}{{\ttfamily arXiv:1101.1645 [hep-ex]}}.
%%CITATION = ARXIV:1101.1645;%%.

\bibitem{Aad:2011mb}
{\bfseries ATLAS} Collaboration, G.~Aad {\em et~al.}, ``{Search for Massive
  Long-lived Highly Ionising Particles with the ATLAS Detector at the LHC},''
  \href{http://dx.doi.org/10.1016/j.physletb.2011.03.033}{{\em Phys.Lett.}
  {\bfseries B698} (2011) 353--370},
\href{http://arxiv.org/abs/1102.0459}{{\ttfamily arXiv:1102.0459 [hep-ex]}}.
%%CITATION = ARXIV:1102.0459;%%.

\bibitem{Aad:2011hz}
{\bfseries ATLAS} Collaboration, G.~Aad {\em et~al.}, ``{Search for Heavy
  Long-Lived Charged Particles with the ATLAS detector in $pp$ collisions at
  $\sqrt{s}=7$ TeV},''
  \href{http://dx.doi.org/10.1016/j.physletb.2011.08.042}{{\em Phys.Lett.}
  {\bfseries B703} (2011) 428--446},
\href{http://arxiv.org/abs/1106.4495}{{\ttfamily arXiv:1106.4495 [hep-ex]}}.
%%CITATION = ARXIV:1106.4495;%%.

\bibitem{Chatrchyan:2012sp}
{\bfseries CMS} Collaboration, S.~Chatrchyan {\em et~al.}, ``{Search for heavy
  long-lived charged particles in $pp$ collisions at $\sqrt{s}=7$ TeV},''
  \href{http://dx.doi.org/10.1016/j.physletb.2012.06.023}{{\em Phys.Lett.}
  {\bfseries B713} (2012) 408--433},
\href{http://arxiv.org/abs/1205.0272}{{\ttfamily arXiv:1205.0272 [hep-ex]}}.
%%CITATION = ARXIV:1205.0272;%%.

\bibitem{Aad:2013pqd}
{\bfseries ATLAS} Collaboration, G.~Aad {\em et~al.}, ``{Search for long-lived,
  multi-charged particles in pp collisions at $\sqrt{s}$=7 TeV using the ATLAS
  detector},'' \href{http://dx.doi.org/10.1016/j.physletb.2013.04.036}{{\em
  Phys.Lett.} {\bfseries B722} (2013) 305--323},
\href{http://arxiv.org/abs/1301.5272}{{\ttfamily arXiv:1301.5272 [hep-ex]}}.
%%CITATION = ARXIV:1301.5272;%%.

\bibitem{Chatrchyan:2013oca}
{\bfseries CMS} Collaboration, S.~Chatrchyan {\em et~al.}, ``{Searches for
  long-lived charged particles in pp collisions at $\sqrt{s}$=7 and 8 TeV},''
  \href{http://dx.doi.org/10.1007/JHEP07(2013)122}{{\em JHEP} {\bfseries 1307}
  (2013) 122},
\href{http://arxiv.org/abs/1305.0491}{{\ttfamily arXiv:1305.0491 [hep-ex]}}.
%%CITATION = ARXIV:1305.0491;%%.

\bibitem{CMS-PAS-EXO-13-006}
{\bfseries CMS} Collaboration, ``{Reinterpreting the results of the search for
  long-lived charged particles in the pMSSM and other BSM scenarios},'' Tech.
  Rep. CMS-PAS-EXO-13-006, CERN, Geneva, 2014.

\bibitem{Martin:2009iq}
A.~Martin, W.~Stirling, R.~Thorne, and G.~Watt, ``{Parton distributions for the
  LHC},'' \href{http://dx.doi.org/10.1140/epjc/s10052-009-1072-5}{{\em
  Eur.Phys.J.} {\bfseries C63} (2009) 189--285},
\href{http://arxiv.org/abs/0901.0002}{{\ttfamily arXiv:0901.0002 [hep-ph]}}.
%%CITATION = ARXIV:0901.0002;%%.

\bibitem{Pinfold:2009oia}
{\bfseries MoEDAL} Collaboration, J.~Pinfold {\em et~al.},
``{Technical Design Report of the MoEDAL Experiment},''.
%%CITATION = CERN-LHCC-2009-006 ETC.;%%.

\bibitem{Acharya:2014nyr}
{\bfseries MoEDAL} Collaboration, B.~Acharya {\em et~al.}, ``{The Physics
  Programme Of The MoEDAL Experiment At The LHC},''
  \href{http://dx.doi.org/10.1142/S0217751X14300506}{{\em Int.J.Mod.Phys.}
  {\bfseries A29} (2014) 1430050},
\href{http://arxiv.org/abs/1405.7662}{{\ttfamily arXiv:1405.7662 [hep-ph]}}.
%%CITATION = ARXIV:1405.7662;%%.

\bibitem{Khachatryan:2014rra}
{\bfseries CMS} Collaboration, V.~Khachatryan {\em et~al.}, ``{Search for dark
  matter, extra dimensions, and unparticles in monojet events in proton-proton
  collisions at $\sqrt{s}$ = 8 TeV},''
\href{http://arxiv.org/abs/1408.3583}{{\ttfamily arXiv:1408.3583 [hep-ex]}}.
%%CITATION = ARXIV:1408.3583;%%.

\bibitem{Zhou:2013fla}
N.~Zhou, D.~Berge, and D.~Whiteson, ``{Mono-everything: combined limits on dark
  matter production at colliders from multiple final states},''
  \href{http://dx.doi.org/10.1103/PhysRevD.87.095013}{{\em Phys.Rev.}
  {\bfseries D87} no.~9, (2013) 095013},
\href{http://arxiv.org/abs/1302.3619}{{\ttfamily arXiv:1302.3619 [hep-ex]}}.
%%CITATION = ARXIV:1302.3619;%%.

\bibitem{Alcaraz:2006mx}
{\bfseries ALEPH, DELPHI, L3, OPAL, LEP Electroweak Working Group}
  Collaboration, J.~Alcaraz {\em et~al.}, ``{A Combination of preliminary
  electroweak measurements and constraints on the standard model},''
\href{http://arxiv.org/abs/hep-ex/0612034}{{\ttfamily arXiv:hep-ex/0612034
  [hep-ex]}}.
%%CITATION = HEP-EX/0612034;%%.

\bibitem{Agashe:2014kda}
{\bfseries Particle Data Group} Collaboration, K.~Olive {\em et~al.}, ``{Review
  of Particle Physics},''
\href{http://dx.doi.org/10.1088/1674-1137/38/9/090001}{{\em Chin.Phys.}
  {\bfseries C38} (2014) 090001}.
%%CITATION = CHPHD,C38,090001;%%.

\bibitem{Aad:2013yna}
{\bfseries ATLAS} Collaboration, G.~Aad {\em et~al.}, ``{Search for charginos
  nearly mass degenerate with the lightest neutralino based on a
  disappearing-track signature in pp collisions at $\sqrt(s)$=8  TeV with
  the ATLAS detector},''
  \href{http://dx.doi.org/10.1103/PhysRevD.88.112006}{{\em Phys.Rev.}
  {\bfseries D88} no.~11, (2013) 112006},
\href{http://arxiv.org/abs/1310.3675}{{\ttfamily arXiv:1310.3675 [hep-ex]}}.
%%CITATION = ARXIV:1310.3675;%%.

\bibitem{Heister:2002mn}
{\bfseries ALEPH} Collaboration, A.~Heister {\em et~al.}, ``{Search for
  charginos nearly mass degenerate with the lightest neutralino in e+ e-
  collisions at center-of-mass energies up to 209-GeV},''
  \href{http://dx.doi.org/10.1016/S0370-2693(02)01584-8}{{\em Phys.Lett.}
  {\bfseries B533} (2002) 223--236},
\href{http://arxiv.org/abs/hep-ex/0203020}{{\ttfamily arXiv:hep-ex/0203020
  [hep-ex]}}.
%%CITATION = HEP-EX/0203020;%%.

\bibitem{Abreu:2000as}
{\bfseries DELPHI} Collaboration, P.~Abreu {\em et~al.}, ``{Update of the
  search for charginos nearly mass-degenerate with the lightest neutralino},''
  \href{http://dx.doi.org/10.1016/S0370-2693(00)00693-6}{{\em Phys.Lett.}
  {\bfseries B485} (2000) 95--106},
\href{http://arxiv.org/abs/hep-ex/0103035}{{\ttfamily arXiv:hep-ex/0103035
  [hep-ex]}}.
%%CITATION = HEP-EX/0103035;%%.

\bibitem{Acciarri:2000wy}
{\bfseries L3} Collaboration, M.~Acciarri {\em et~al.}, ``{Search for charginos
  with a small mass difference with the lightest supersymmetric particle at
  $\sqrt{S}$ = 189-GeV},''
  \href{http://dx.doi.org/10.1016/S0370-2693(00)00488-3}{{\em Phys.Lett.}
  {\bfseries B482} (2000) 31--42},
\href{http://arxiv.org/abs/hep-ex/0002043}{{\ttfamily arXiv:hep-ex/0002043
  [hep-ex]}}.
%%CITATION = HEP-EX/0002043;%%.

\bibitem{Abbiendi:2002vz}
{\bfseries OPAL} Collaboration, G.~Abbiendi {\em et~al.}, ``{Search for nearly
  mass degenerate charginos and neutralinos at LEP},''
  \href{http://dx.doi.org/10.1140/epjc/s2003-01237-x}{{\em Eur.Phys.J.}
  {\bfseries C29} (2003) 479--489},
\href{http://arxiv.org/abs/hep-ex/0210043}{{\ttfamily arXiv:hep-ex/0210043
  [hep-ex]}}.
%%CITATION = HEP-EX/0210043;%%.

\bibitem{Conte:2012fm}
E.~Conte, B.~Fuks, and G.~Serret, ``{MadAnalysis 5, A User-Friendly Framework
  for Collider Phenomenology},''
  \href{http://dx.doi.org/10.1016/j.cpc.2012.09.009}{{\em Comput.Phys.Commun.}
  {\bfseries 184} (2013) 222--256},
\href{http://arxiv.org/abs/1206.1599}{{\ttfamily arXiv:1206.1599 [hep-ph]}}.
%%CITATION = ARXIV:1206.1599;%%.

\bibitem{Abbiendi:2003sc}
{\bfseries OPAL} Collaboration, G.~Abbiendi {\em et~al.}, ``{Search for
  chargino and neutralino production at s**(1/2) = 192-GeV to 209 GeV at
  LEP},'' \href{http://dx.doi.org/10.1140/epjc/s2004-01758-8}{{\em Eur.Phys.J.}
  {\bfseries C35} (2004) 1--20},
\href{http://arxiv.org/abs/hep-ex/0401026}{{\ttfamily arXiv:hep-ex/0401026
  [hep-ex]}}.
%%CITATION = HEP-EX/0401026;%%.

\bibitem{Fairbairn:2006gg}
M.~Fairbairn, A.~Kraan, D.~Milstead, T.~Sjostrand, P.~Z. Skands, {\em et~al.},
  ``{Stable massive particles at colliders},''
  \href{http://dx.doi.org/10.1016/j.physrep.2006.10.002}{{\em Phys.Rept.}
  {\bfseries 438} (2007) 1--63},
\href{http://arxiv.org/abs/hep-ph/0611040}{{\ttfamily arXiv:hep-ph/0611040
  [hep-ph]}}.
%%CITATION = HEP-PH/0611040;%%.

\bibitem{Sjostrand:2006za}
T.~Sjostrand, S.~Mrenna, and P.~Z. Skands, ``{PYTHIA 6.4 Physics and Manual},''
  \href{http://dx.doi.org/10.1088/1126-6708/2006/05/026}{{\em JHEP} {\bfseries
  0605} (2006) 026},
\href{http://arxiv.org/abs/hep-ph/0603175}{{\ttfamily arXiv:hep-ph/0603175
  [hep-ph]}}.
%%CITATION = HEP-PH/0603175;%%.

\bibitem{Moretti:2002eu}
S.~Moretti, K.~Odagiri, P.~Richardson, M.~H. Seymour, and B.~R. Webber,
  ``{Implementation of supersymmetric processes in the HERWIG event
  generator},'' \href{http://dx.doi.org/10.1088/1126-6708/2002/04/028}{{\em
  JHEP} {\bfseries 0204} (2002) 028},
\href{http://arxiv.org/abs/hep-ph/0204123}{{\ttfamily arXiv:hep-ph/0204123
  [hep-ph]}}.
%%CITATION = HEP-PH/0204123;%%.

\bibitem{Bellm:2013lba}
J.~Bellm, S.~Gieseke, D.~Grellscheid, A.~Papaefstathiou, S.~Platzer, {\em
  et~al.}, ``{Herwig++ 2.7 Release Note},''
\href{http://arxiv.org/abs/1310.6877}{{\ttfamily arXiv:1310.6877 [hep-ph]}}.
%%CITATION = ARXIV:1310.6877;%%.

\bibitem{DeRujula:1975ge}
A.~De~Rujula, H.~Georgi, and S.~Glashow, ``{Hadron Masses in a Gauge Theory},''
\href{http://dx.doi.org/10.1103/PhysRevD.12.147}{{\em Phys.Rev.} {\bfseries
  D12} (1975) 147--162}.
%%CITATION = PHRVA,D12,147;%%.

\bibitem{Kraan:2004tz}
A.~C. Kraan, ``{Interactions of heavy stable hadronizing particles},''
  \href{http://dx.doi.org/10.1140/epjc/s2004-01997-7}{{\em Eur.Phys.J.}
  {\bfseries C37} (2004) 91--104},
\href{http://arxiv.org/abs/hep-ex/0404001}{{\ttfamily arXiv:hep-ex/0404001
  [hep-ex]}}.
%%CITATION = HEP-EX/0404001;%%.

\bibitem{Drees:1990yw}
M.~Drees and X.~Tata, ``{Signals for heavy exotics at hadron colliders and
  supercolliders},''
\href{http://dx.doi.org/10.1016/0370-2693(90)90508-4}{{\em Phys.Lett.}
  {\bfseries B252} (1990) 695--702}.
%%CITATION = PHLTA,B252,695;%%.

\bibitem{Mafi:1999dg}
A.~Mafi and S.~Raby, ``{An Analysis of a heavy gluino LSP at CDF: The Heavy
  gluino window},'' \href{http://dx.doi.org/10.1103/PhysRevD.62.035003}{{\em
  Phys.Rev.} {\bfseries D62} (2000) 035003},
\href{http://arxiv.org/abs/hep-ph/9912436}{{\ttfamily arXiv:hep-ph/9912436
  [hep-ph]}}.
%%CITATION = HEP-PH/9912436;%%.

\bibitem{Baer:1998pg}
H.~Baer, K.-m. Cheung, and J.~F. Gunion, ``{A Heavy gluino as the lightest
  supersymmetric particle},''
  \href{http://dx.doi.org/10.1103/PhysRevD.59.075002}{{\em Phys.Rev.}
  {\bfseries D59} (1999) 075002},
\href{http://arxiv.org/abs/hep-ph/9806361}{{\ttfamily arXiv:hep-ph/9806361
  [hep-ph]}}.
%%CITATION = HEP-PH/9806361;%%.

\bibitem{Mackeprang:2006gx}
R.~Mackeprang and A.~Rizzi, ``{Interactions of Coloured Heavy Stable Particles
  in Matter},'' \href{http://dx.doi.org/10.1140/epjc/s10052-007-0252-4}{{\em
  Eur.Phys.J.} {\bfseries C50} (2007) 353--362},
\href{http://arxiv.org/abs/hep-ph/0612161}{{\ttfamily arXiv:hep-ph/0612161
  [hep-ph]}}.
%%CITATION = HEP-PH/0612161;%%.

\bibitem{Mackeprang:2009ad}
R.~Mackeprang and D.~Milstead, ``{An Updated Description of Heavy-Hadron
  Interactions in GEANT-4},''
  \href{http://dx.doi.org/10.1140/epjc/s10052-010-1262-1}{{\em Eur.Phys.J.}
  {\bfseries C66} (2010) 493--501},
\href{http://arxiv.org/abs/0908.1868}{{\ttfamily arXiv:0908.1868 [hep-ph]}}.
%%CITATION = ARXIV:0908.1868;%%.

\bibitem{ATLAS:2014fka}
{\bfseries ATLAS} Collaboration, G.~Aad {\em et~al.}, ``{Searches for heavy
  long-lived charged particles with the ATLAS detector in proton-proton
  collisions at $\sqrt{s}$ = 8 TeV},''
\href{http://arxiv.org/abs/1411.6795}{{\ttfamily arXiv:1411.6795 [hep-ex]}}.
%%CITATION = ARXIV:1411.6795;%%.

\bibitem{Khachatryan:2015jha}
{\bfseries CMS} Collaboration, V.~Khachatryan {\em et~al.}, ``{Search for
  decays of stopped long-lived particles produced in proton-proton collisions
  at $\sqrt{s} = 8$ TeV},''
\href{http://arxiv.org/abs/1501.05603}{{\ttfamily arXiv:1501.05603 [hep-ex]}}.
%%CITATION = ARXIV:1501.05603;%%.

\bibitem{Aad:2013gva}
{\bfseries ATLAS} Collaboration, G.~Aad {\em et~al.}, ``{Search for long-lived
  stopped R-hadrons decaying out-of-time with pp collisions using the ATLAS
  detector},'' \href{http://dx.doi.org/10.1103/PhysRevD.88.112003}{{\em
  Phys.Rev.} {\bfseries D88} no.~11, (2013) 112003},
\href{http://arxiv.org/abs/1310.6584}{{\ttfamily arXiv:1310.6584 [hep-ex]}}.
%%CITATION = ARXIV:1310.6584;%%.

\bibitem{Arvanitaki:2005nq}
A.~Arvanitaki, S.~Dimopoulos, A.~Pierce, S.~Rajendran, and J.~G. Wacker,
  ``{Stopping gluinos},''
  \href{http://dx.doi.org/10.1103/PhysRevD.76.055007}{{\em Phys.Rev.}
  {\bfseries D76} (2007) 055007},
\href{http://arxiv.org/abs/hep-ph/0506242}{{\ttfamily arXiv:hep-ph/0506242
  [hep-ph]}}.
%%CITATION = HEP-PH/0506242;%%.

\bibitem{Mohapatra:1979ia}
R.~N. Mohapatra and G.~Senjanovic, ``{Neutrino Mass and Spontaneous Parity
  Violation},''
\href{http://dx.doi.org/10.1103/PhysRevLett.44.912}{{\em Phys.Rev.Lett.}
  {\bfseries 44} (1980) 912}.
%%CITATION = PRLTA,44,912;%%.

\bibitem{D'Ambrosio:2002ex}
G.~D'Ambrosio, G.~Giudice, G.~Isidori, and A.~Strumia, ``{Minimal flavor
  violation: An Effective field theory approach},''
  \href{http://dx.doi.org/10.1016/S0550-3213(02)00836-2}{{\em Nucl.Phys.}
  {\bfseries B645} (2002) 155--187},
\href{http://arxiv.org/abs/hep-ph/0207036}{{\ttfamily arXiv:hep-ph/0207036
  [hep-ph]}}.
%%CITATION = HEP-PH/0207036;%%.

\bibitem{Abe:2011ts}
K.~Abe, T.~Abe, H.~Aihara, Y.~Fukuda, Y.~Hayato, {\em et~al.}, ``{Letter of
  Intent: The Hyper-Kamiokande Experiment --- Detector Design and Physics
  Potential ---},''
\href{http://arxiv.org/abs/1109.3262}{{\ttfamily arXiv:1109.3262 [hep-ex]}}.
%%CITATION = ARXIV:1109.3262;%%.

\bibitem{Garfagnini:2014nla}
A.~Garfagnini, ``{Neutrinoless Double Beta Decay Experiments},''
\href{http://arxiv.org/abs/1408.2455}{{\ttfamily arXiv:1408.2455 [hep-ex]}}.
%%CITATION = ARXIV:1408.2455;%%.

\bibitem{Machacek:1983tz}
M.~E. Machacek and M.~T. Vaughn, ``{Two Loop Renormalization Group Equations in
  a General Quantum Field Theory. 1. Wave Function Renormalization},''
\href{http://dx.doi.org/10.1016/0550-3213(83)90610-7}{{\em Nucl.Phys.}
  {\bfseries B222} (1983) 83}.
%%CITATION = NUPHA,B222,83;%%.

\bibitem{Mihaila:2012pz}
L.~N. Mihaila, J.~Salomon, and M.~Steinhauser, ``{Renormalization constants and
  beta functions for the gauge couplings of the Standard Model to three-loop
  order},'' \href{http://dx.doi.org/10.1103/PhysRevD.86.096008}{{\em Phys.Rev.}
  {\bfseries D86} (2012) 096008},
\href{http://arxiv.org/abs/1208.3357}{{\ttfamily arXiv:1208.3357 [hep-ph]}}.
%%CITATION = ARXIV:1208.3357;%%.

\end{thebibliography}\endgroup

\end{document}